\documentclass[onecolumn,showkeys,aps,prd]{revtex4}

\usepackage{amsfonts }
\usepackage{latexsym}
\usepackage{tikz}
\usepackage{amsmath,amssymb,amsthm}
\usepackage{mathtools}
\usepackage{lineno}
\usepackage{bbm}
\usepackage{float}      

\usepackage{stackengine,graphicx}

\newcommand\vargtrsim{\mathrel{\ensurestackMath{%
  \stackengine{-.4ex}{>}{\rotatebox{25}{$\sim$}}{U}{l}{F}{T}{S}}}}

\renewcommand{\dj }{\mbox{\kern0.6ex\raise0.8ex\hbox{-}\kern-1.4ex d}}
\newcommand{\Dj }{\mbox{\raise0.3ex\hbox{-}\kern-0.4em D}}

\newcommand{\lc}{\varepsilon}

\newcommand{\nablar}{\stackrel{\rightarrow}{\nabla}\!\!{}}
\newcommand{\nablal}{\stackrel{\leftarrow}{\nabla}\!\!{}}
\newcommand{\rmdr}{\stackrel{\rightarrow}{\mathrm{d}}\!\!{}}
\newcommand{\rmdl}{\stackrel{\leftarrow}{\mathrm{d}}\!\!{}}
\newcommand{\nablalr}{\stackrel{\leftrightarrow}{\nabla}\!\!{}}
\newcommand{\ds}{\displaystyle}

\newcommand{\itGamma}{\mit{\Gamma}}

\newcommand{\diag}{\mathop{\rm diag}\nolimits }

\newcommand{\rmd}{\mathrm{d}}

\newcommand{\celi}{\ensuremath{\mathbb{Z}}}

\newcommand{\realni}{\ensuremath{\mathbb{R}}}
\newcommand{\kompleksni}{\ensuremath{\mathbb{C}}}
\newcommand{\grasmanovi}{\ensuremath{\mathbb{G}}}

\newcommand{\cF}{{\cal F}}
\newcommand{\cG}{{\cal G}}
\newcommand{\cH}{{\cal H}}

\newcommand{\cL}{{\cal L}}
\newcommand{\cM}{{\cal M}}
\newcommand{\cN}{{\cal N}}

\newcommand{\g}{\mathfrak{g}}
\newcommand{\h}{\mathfrak{h}}
\renewcommand{\l}{\mathfrak{l}}

\newcommand{\killing}[2]{ {\langle #1 , #2 \rangle } }

\newcommand{\dual}{\text{\boldmath$\,\star$}}

\newcommand{\osum}{\mathop{\sum\kern-3.1ex\raise-0.27ex\hbox{{\text{\Large$\circ$}}}}\limits}

\newcommand{\dekaedar}{
    \begin{tikzpicture}[scale=0.06, line width=0.14mm]
        \draw (90:2) -- (18:2) -- (306:2) -- (234:2) -- (162:2) -- cycle;
        \draw (90:1.7) -- (306:1.7) -- (162:1.7) -- (18:1.7) -- (234:1.7) -- cycle;
    \end{tikzpicture}
}
\newcommand{\tetraedar}{
    \begin{tikzpicture}[scale=0.06, line width=0.14mm]
        \draw (90:2) -- (210:2) -- (330:2) -- cycle;
        \draw (90:1.8) -- (90:0) -- cycle;
        \draw (210:1.8) -- (210:0) -- cycle;
        \draw (330:1.8) -- (330:0) -- cycle;
    \end{tikzpicture}
}
\newcommand{\trougao}{
    \begin{tikzpicture}[scale=0.06, line width=0.14mm]
        \draw (90:2) -- (210:2) -- (330:2) -- cycle;
    \end{tikzpicture}
}
\newcommand{\dvougao}{
    \begin{tikzpicture}[scale=0.06, line width=0.14mm]
        \draw (90:2) -- (210:2) -- cycle;
    \end{tikzpicture}
}

\begin{document}

\title{A $3BF$ model of quantum gravity coupled to Standard Model matter}

\author{Pavle Stipsi\'c}
 \email{pstipsic@ipb.ac.rs}
\affiliation{Institute of Physics, University of Belgrade, Pregrevica 118, 11080 Belgrade, Serbia}

\author{Marko Vojinovi\'c}
 \email{vmarko@ipb.ac.rs}
 \thanks{corresponding author}
\affiliation{Institute of Physics, University of Belgrade, Pregrevica 118, 11080 Belgrade, Serbia}



\begin{abstract}
We develop an explicit model of quantum gravity coupled to the matter fields of the Standard Model, based on the 3-group structure and the $3BF$ action, within the framework of higher gauge theory. The model is constructed by providing a rigorous definition for the path integral of the theory, achieved by defining the whole theory on a piecewise-flat spacetime manifold. To that end, we develop a method to systematically discretize both the action and the path integral measure by passing from a smooth manifold to a piecewise-flat manifold. Finally, we discuss in some detail the structure of the resulting quantum gravity model, and provide a preliminary analysis of its semiclassical limit.
\end{abstract}

\maketitle

\section{\label{SecI}Introduction}

The construction of a fundamental theory of quantum gravity represents one of the main open problems of modern theoretical physics. Throughout the development of modern physics, there were a number of attempts to tackle this problem, some of which have developed over time into vast research directions such as string theory (ST) \cite{Polchinski1,Polchinski2}, loop quantum gravity (LQG) \cite{Rovelli2004,Thiemann}, causal set theory (CST) \cite{Surya2019}, and others. The main idea behind the LQG approach is to perform a nonperturbative quantization of the gravitational field, initially without matter fields, and then add matter into the model once it becomes developed enough. This approach has than branched into two main frameworks, based on the type of nonperturbative quantization technique. The \emph{canonical} LQG framework \cite{Thiemann} is based on the canonical quantization of gravity, using the foliation of spacetime into space and time, and then imposing the simultaneous commutation relations onto the basic degrees of freedom of the theory. The \emph{covariant} LQG framework \cite{RovelliVidotto2014} is instead focused on providing a rigorous nonperturbative definition of the path integral of the theory, typically by restricting it onto a lattice. Various definitions of the theory and the path integral give rise to the so-called \emph{spinfoam models}, each describing one particular quantum gravity model.

The spinfoam model approach to quantum gravity has been studied for a number of years and is well developed and deeply understood. Unfortunately, the vast majority of spinfoam models have been developed using techniques that exploit particular properties of the gravitational field itself, and are extremely hard to extend to include matter fields as well \cite{SpinfoamFermions}. This represents a drawback, since a model describing pure gravity, without the versatility and dynamics of matter fields, is of limited value. Namely, the most interesting problems one could hope to study using any tentative quantum gravity model are the resolutions of singularities, black hole evaporation and the information problem, gravitational field sourced by matter in spatial superposition, and so on. All these questions involve matter in one way or another, so pure gravity models are hard to apply to these questions.

In recent years, there has been some novel progress on the front of spinfoam model generalizations, with the aim to include matter as well. Based on the recently developed ideas of higher gauge theory \cite{BaezHuerta,BaezDolan,Baez}, an approach to gauge symmetries using category theory generalizations of groups called $n$-groups, a new class of quantum gravity models appeared. Initial development focused on the models based on the algebraic structure of a $2$-group \cite{MikovicVojinovicBook,GirelliPfeifferPopescu2008,FariaMartinsMikovic2011,MikovicOliveira2014,Mikovic2015,Asante2020,MOV2016,MOV2019,Girelli2021}, since it was demonstrated \cite{MikovicVojinovic2012} that such models naturally contain not just the spin connection, but also tetrad fields as the fundamental degrees of freedom in the theory. The presence of tetrad fields proved to be a crucial insight, since tetrads could facilitate a straightforward coupling to matter fields as well, unlike the models that feature spin connection only. Afterwards, the attention has moved to the models based on a $3$-group structure, which proved versatile enough to describe not just gravity, but also matter fields, in particular gauge fields, fermions, and scalar fields (see also \cite{MV2020} for a $4$-group approach).

To every $3$-group one can associate a so-called \emph{topological $3BF$ action}, a classical theory with no propagating degrees of freedom, which can be quantized using the techniques of topological quantum field theory (TQFT) \cite{Witten,Atiyah,Quinn,SozerVirelizier,BaezTQFT,Yetter}. The $3BF$ action could then be deformed into a nontopological theory by deforming it with the so-called \emph{simplicity constraints}, additional terms in the action which give rise to propagating degrees of freedom and nontrivial dynamics. These developments have led to the generalization of the so-called \emph{spinfoam quantization programme} to the level of higher gauge theory. The quantization is performed in three main steps:
\begin{enumerate}
\item Choose a $3$-group structure, and write the classical action as a topological $3BF$ theory deformed by suitable simplicity constraint terms which give the desired dynamics.
\item Use the TQFT techniques to define a path integral (given as an appropriate topological invariant) for the pure topological $3BF$ part of the action.
\item Deform the topological invariant into a path integral for a realistic model of quantum gravity, by imposing the simplicity constraint parts of the action.
\end{enumerate}
The first step of the above programme has been established in \cite{Radenkovic2019}, where a $3BF$ action has been constructed that describes Einstein-Cartan gravity coupled to the whole Standard Model of elementary particle physics. Various additional properties of the resulting $3BF$ action and its variants were studied in \cite{Radenkovic2020proc,Radenkovic2020,Radenkovic2022a,Djordjevic2023,Stipsic2025a}. The second step of the spinfoam quantization programme has been established, for an arbitrary $3$-group, in \cite{Porter,Radenkovic2022b}, giving rise to a novel topological invariant for closed $4$-manifolds. Moreover, in \cite{Radenkovic2025} this topological invariant has been generalized to $4$-manifolds with boundary, giving rise to a full-fledged TQFT.

In this paper, we perform the main, final third step of the spinfoam quantization programme --- we start from a classical theory, described by the topological $3BF$ action deformed by a number of simplicity constraints as was done in \cite{Radenkovic2019}, and then perform the quantization of the whole action, providing a rigorous definition for the expectation value of a generic observable $F$ evaluated via the path integral as:
\begin{equation} \label{eq:AbstractPathIntegral}
\langle F \rangle = \frac{\cN}{Z} \int Dg\,D\phi\; F(g,\phi) \, e^{iS_{3BF}[g,\phi]}\,, \qquad \text{ where } \qquad
Z \equiv \cN \int Dg\,D\phi\; \, e^{iS_{3BF}[g,\phi]}\,.
\end{equation}
Here $g$ and $\phi$ schematically denote the gravitational and matter degrees of freedom, respectively. The quantization itself is based on the following key ingredients and assumptions:
\begin{enumerate}
\item Both gravitational and matter degrees of freedom are treated on an equal footing, and both are described by the underlying algebraic structure of a $3$-group, as a generalization of the notion of a gauge group in the context of higher gauge theory. The $3$-group specifies all dynamical fields that appear in the theory, and since each field is Lie-algebra-valued, the $3$-group defines the domain of integration in the path integral, for each field. We choose in particular the so-called \emph{Standard Model $3$-group}, which describes all degrees of freedom of Einstein-Cartan gravity and the Standard Model of elementary particle physics.
\item The classical $3BF$ action with constraints, which is used as a starting point for the quantization, is written entirely in terms of differential forms, in the sense that all fields it depends on feature in the action as full differential forms, rather than as components of these forms. In particular, the Hodge dual operator is necessarily absent from the action, since it cannot be introduced without explicitly appealing to the components of differential forms. In addition, the covariant exterior derivative is the only differential operator featuring in the action, and it appears at most once in each term of the action. We choose in particular the so-called \emph{Standard Model $3BF$ action}, established in \cite{Radenkovic2019} as a solution of the first step of the spinfoam quantization programme.  It is based on the corresponding Standard Model $3$-group, and fully describes Einstein-Cartan gravity coupled to the Standard Model matter fields.
\item We assume that spacetime is, at the fundamental level, described by a $4$-dimensional piecewise-flat manifold, instead of a smooth manifold. The piecewise-flat structure is assumed to be a \emph{real physical structure}, rather than a mere auxiliary mathematical regulator. It is described as a \emph{triangulation} of a topological manifold, i.e., as a simplicial complex. Moreover, the fields living on this structure are assumed to have \emph{constant value} within each simplex, and change their value only as one moves from one simplex to another. This is an implementation of the ``piecewise-flatness'' idea, providing a physical cutoff at the triangulation scale. The triangulation scale is assumed to be somewhere near the Planck scale, which renders it essentially invisible to resolution of current experiments.
\end{enumerate}
Leveraging the above assumptions enables us to introduce a rigorous definition of the path integral (\ref{eq:AbstractPathIntegral}), thereby performing the quantization of the theory. The key step is to pass from a theory defined on a smooth manifold to a theory defined on a piecewise-flat manifold, using a systematic method of discretization that we develop and then apply both to the classical action and to the path integral itself.

The resulting theory represents a full-fledged model of quantum gravity coupled to the matter of the Standard Model, together with a cosmological constant. It represents one of the first models of its kind. Namely, as mentioned above, most spinfoam models that were proposed so far in the literature are focused on the quantization of the gravitational field only, without matter fields, and the quantization techniques developed for their constructions are not flexible enough to introduce matter fields as well. So most of the models contain no matter at all, or merely one scalar field or such, but nowhere near the matter spectrum of the full Standard Model. Other models, representing various generalizations of spinfoam models, sometimes can and do include matter fields, but these are not treated on an equal footing with gravity, which gives rise to a number of drawbacks in the construction and properties of such models. On the other hand, our model does not suffer from any of these issues, and in this sense represents a qualitatively novel result.

The layout of the paper is as follows. In Section \ref{secII}, we give a short review of the classical theory described by the Standard Model $3BF$ action, as well as the notion of a $3$-group, an algebraic structure that underpins the action, in the context of higher gauge theory. The $3BF$ action describes the Einstein-Cartan gravity coupled to the Standard Model, and it will be the starting point for the construction of the corresponding quantum theory. Sections \ref{secIII} and \ref{secIV} develop the general method of discretizing the action and the path integral. We introduce a key technique how to pass from an action defined on a smooth manifold to an action defined on a piecewise-flat manifold, as well as how to define the path integral measure and domain of integration for Lie-algebra-valued fields living on a piecewise-flat manifold. Section \ref{secV} represents the culmination of our efforts, where we apply the developed discretization method to the Standard Model $3BF$ action, giving rise to a fully defined theory of quantum gravity with matter. We then discuss the structure of the resulting theory, including some major simplifications that can be done under some minimal additional assumptions. In Section \ref{secVI} we give a preliminary analysis of the semiclassical limit of the model, simplifying it even further within the semiclassical approximation scheme. In Section \ref{secVII} we give our concluding remarks and some topics for future research. The Appendices contain some technical results used in the main text.

Our notation and conventions are as follows. Spacetime indices, denoted by the mid-alphabet Greek letters $\mu,\nu,\dots $, are raised and lowered by the spacetime metric $g_{\mu\nu}$, once it is defined. The Lorentz metric is denoted as $\eta_{ab} = \diag (-1,+1,+1,+1)$. The indices that are counting the generators of Lie groups $G$, $H$, and $L$ are denoted with initial Greek letters $\alpha, \beta, \dots $, lowercase initial Latin letters $a, b, c,\dots $, and uppercase Latin indices $A,B,C,\dots $, respectively. The generators themselves are typically denoted as $\tau_\alpha$, $t_a$ and $T_A$, respectively. We work in the natural system of units , defined by $c=\hbar=1$ and $G = l_p^2$, where $l_p$ is the Planck length.

The indices which correspond to the Lorentz group are pairs of indices $ab$ and the quantities that depend on them are antisymmetric with respect to their interchange. This means that all independent components  of these quantities, according to Einstein summation convention, are summed over twice. Because of this, the result of the summation should be divided by two. Alternatively, in order to avoid this problem, one can introduce the notation $[ab]$ which represents  the pair of indices as a single index for which we always assume that $a>b$. Summation over such indices takes into account every independent component precisely once, so it is not necessary to divide the total by two. For example, given some quantity $K^{ab}$, one has
\begin{equation}
K^{[ab]}\sigma_{[ab]}=\frac{1}{2}K^{ab}\sigma_{ab}\,.
\end{equation}
In this work, the square brackets  will exclusively denote the pairs of Lorentz indices, rather than the usual antisymmetrization over those indices.

All additional notation and conventions used throughout the paper are explicitly defined in the text where they first appear.

\section{\label{secII}Short review of the classical $3BF$ action}

We begin with a short review of the classical theory which serves as a starting point for quantization. In order to introduce the relevant mathematical framework and notation, we proceed step by step. First we discuss the topological $3BF$ action, and the corresponding notion of a $3$-group, on which the action is based. Then we discuss the constrained $3BF$ action, which represents the relevant classical theory, combining Einstein-Cartan gravity with the Standard Model of elementary particle physics, along with the cosmological constant (also called the Standard Model $3BF$ action). The quantization of this action is the main focus of the paper. Finally, we briefly introduce the Einstein-Cartan action with spin-spin contact interaction (abreviated as ECC action). The ECC action is classically equivalent to the constrained $3BF$ action on the reduced configuration space, but is much harder to quantize, due to the presence of the Hodge dual operator in some of its terms. Throughout the paper, we compare the terms obtained via quantization of the Standard Model $3BF$ action to various terms in the classical ECC action, in order to gain additional insight into the theory.

\subsection{\label{subsecIIa}The $3$-group and the topological $3BF$ action}

The topological $3BF$ action is based on the underlying algebraic notion of a strict Lie $3$-group, which is a generalization of the notion of a Lie group, stemming from higher category theory. It is equivalent to the algebraic structure called a Lie 2-crossed module --- a triple of Lie groups, $G$, $H$ and $L$, together with two homomorphisms between them,
\begin{equation} \label{eq:jna2}
\partial : H \to G\,, \qquad \delta : L \to H\,,
\end{equation}
the actions of the group $G$ on itself and on the other two groups,
\begin{equation} \label{eq:jna3}
\triangleright : G \times X \to X\,, \qquad X = G, H, L\,,
\end{equation}
and the Peiffer lifting map,
\begin{equation} \label{eq:jna4}
\{ \_ \, , \_\, \}_\mathrm{pf} : H \times H \to L\,.
\end{equation}
All these maps satisfy certain axioms, and taken together they make up a Lie 2-crossed module, denoted as
\begin{equation} \label{eq:jna5}
(\; L\stackrel{\delta}{\to} H \stackrel{\partial}{\to}G \; , \;\triangleright \;, \; \{\_\,,\_\,\}_\mathrm{pf} \;)\,.
\end{equation}
This structure represents  the notion of a $3$-group in the most convenient way for the discussion of the $3BF$ action. An reader interested in further mathematical details may look at additional literature, for example \cite{BaezHuerta,BaezDolan,Radenkovic2019,Radenkovic2022a,Radenkovic2022b,Radenkovic2025,martins2011,SaemannWolf2014b,Wang2014}.

Given the mathematical formulation of a $3$-group, it naturally gives rise to an action, called a $3BF$ action (see also Appendix \ref{AppB} in \cite{Stipsic2025b} for a more detailed explanation of the notation commonly used both in \cite{Stipsic2025b} and in this text). The topological $3BF$ action is defined as:
\begin{equation} \label{eq:3BFaction}
S_{3BF}^\text{top}=\int_{\cM_4} \langle B\wedge {\cal F}\rangle_\mathfrak{g}+\langle C\wedge {\cal G}\rangle_\mathfrak{h}+\langle D\wedge {\cal H}\rangle_\mathfrak{l}.
\end{equation} 
The Lagrange multipliers $B$, $C$ and $D$ are $2$-, $1$- and $0$-forms, valued in Lie algebras $\mathfrak{g}$, $\mathfrak{h}$ and $\mathfrak{l}$, corresponding to the Lie groups $G$, $H$ and $L$, respectively. The field strengths $\cF$, $\cG$ and $\cH$ are $2$-, $3$- and $4$-forms, defined as
\begin{equation} \label{eq:ThreeCurvatureDef}
{\cal F}={\rm d}\alpha+\alpha\wedge \alpha-\partial\beta\,,\qquad {\cal G}={\rm d}\beta+\alpha\wedge^{\triangleright}\beta-\delta\gamma\,,\qquad
{\cal H}={\rm d}\gamma+\alpha\wedge^{\triangleright}\gamma+\{\beta\wedge\beta\}_{\rm pf}\,.
\end{equation}
They are called fake curvatures for the connection $1$-form $\alpha$, connection $2$-form $\beta$ and connection $3$-form $\gamma$, also valued in algebras $\mathfrak{g}$, $\mathfrak{h}$ and $\mathfrak{l}$, respectively. Bilinear forms $\langle\_\, ,\_\rangle_{\mathfrak{g}}$, $\langle\_\, ,\_\rangle_{\mathfrak{h}}$  and $\langle\_\, ,\_\rangle_{\mathfrak{l}}$  are assumed to be symmetric, nondegenerate and $G$-invariant, and they map a pair of algebra elements  into a real number. The structure of the 3-group also allows one to introduce the notion of a covariant derivative as
\begin{equation} \label{eq:jna8}
\nabla = {\rm d} + \alpha \wedge^{\triangleright}
\end{equation}
where the symbol $\wedge^{\triangleright}$ means that, when $\nabla$ acts  for example on the components  $\phi^A$ of the object $\phi \in \l$, the action $\triangleright$ is being applied as the action from the Lie algebra $\g$ to Lie algebra $\l$, giving:
\begin{equation} \label{eq:jna9}
\nabla\phi^A={\rm d}\phi^A+\triangleright_{\alpha B}{}^{A} \,\alpha^{\alpha}\wedge \phi^B\,,
\end{equation}
and similarly for objects  that are elements  of algebras $\g$ and $\h$. Using this notation, one can rewrite the fake curvatures (\ref{eq:ThreeCurvatureDef}) in terms of ordinary curvatures as:
\begin{equation} \label{eq:ThreeCurvatureRewrite}
{\cal F}=\nabla^2 -\partial\beta\,,\qquad {\cal G}=\nabla\beta -\delta\gamma\,,\qquad
{\cal H}=\nabla\gamma +\{\beta\wedge\beta\}_{\rm pf}\,.
\end{equation}
The Appendix \ref{AppB} in \cite{Stipsic2025b} contains more details regarding the above notation, including additional examples.

In order to capture the field content of the Standard Model and Einstein-Cartan gravity, we need to make a specific choice of the 3-group. Called the Standard Model 3-group (see \cite{Radenkovic2019,Radenkovic2020proc,Stipsic2025a} for further details), it is defined as follows. The three Lie groups $G$, $H$ and $L$ are chosen as:
\begin{equation} \label{eq:StandardModel3GroupDef}
G = SO(3,1)\times SU(3)\times SU(2) \times U(1) \,, \qquad H = \mathbb{R}^4\,, \qquad L = \mathbb{C}^4\times\mathbb{G}^{64}\times\mathbb{G}^{64}\times\mathbb{G}^{64}\,.
\end{equation}
This choice has the following physical interpretation. The group $G$ represents  the ordinary Standard Model gauge group, together with the local Lorentz group. The group $H$ are the spacetime translations, while the group $L$ corresponds to matter fields. Specifically, $\mathbb{C}^4$ represents the Higgs sector, while the three Grassmann algebras $\mathbb{G}^{64}$ correspond to the three families of fermions.

In order to complete the definition of the Standard Model $3$-group, one also needs to introduce all relevant maps. In our case, the homomorphisms $\partial$ and $\delta$ are chosen to to be trivial, as well as the Peiffer lifting $\{ \_\, , \_ \,\}_\mathrm{pf}$. The action $\triangleright$ is not trivial, and is defined as follows. The group $G$ can be split into the Lorentz part $SO(3,1)$ (generators counted using the indices $[ab]$) and the internal gauge part $SU(3)\times SU(2) \times U(1)$ (generators counted collectively using indices $\alpha,\beta,\dots $). The action of $G$ on itself is specified by the action of the Lorentz part on itself and on the internal gauge part, as
\begin{equation} \label{eq:actionOfLorentzPartOfG}
\triangleright_{[ab][cd]}{}^{[ef]}\equiv f_{[ab][cd]}{}^{[ef]}=\frac{1}{2}\left(\eta_{[a|c}\delta_{|b]}^{[f|}\delta_{d}^{|e]}-\eta_{[a|d}\delta_{|b]}^{[f|}\delta_{c}^{|e]}\right)\,, \qquad
\triangleright_{[ab] \beta}{}^{\gamma}=0\,,
\end{equation}
while the action of the internal gauge part on itself and on the Lorentz part is given as
\begin{equation} \label{eq:actionOfInternalPartOfG}
\triangleright_{\alpha \beta}{}^{\gamma} = f_{\alpha\beta}{}^{\gamma} \,, \qquad
\triangleright_{\alpha [ab]}{}^{[cd]}=0\,.
\end{equation}
The quantities $f$ are the structure constants for the corresponding subgroups of $G$. The action of $G$ on $H$ is specified assuming that the group $H$ is interpreted as the group of 4-dimensional translations. In that case, the Lorentz part of $G$ acts  in the standard way on translations, while the internal part of $G$ acts  trivially:
\begin{equation}\label{eq:actionGOnH}
\triangleright_{[cd]a}{}^{b}=\frac{1}{2}\eta_{a[d|}\delta_{|c]}^b\,, \qquad
\triangleright_{\alpha a}{}^{b}=0\,.
\end{equation}
Finally, the action of the Lorentz and internal subgroups of $G$ on $L$ is also given in a natural way, namely in line with the transformation properties of various fermions and the Higgs scalar. For example, the action of $G$ on left-isospin fermions is:
\begin{equation}
\triangleright_{[cd]A}{}^{B}=\left(\sigma_{cd}\right)_A{}^{B}\,, \qquad
\triangleright_{\alpha A}{}^{B}=\frac{1}{2}\left(\sigma_{\alpha}\right)_A{}^{B}\,.
\end{equation}
The matrices $\left(\sigma_{\alpha}\right)_A{}^{B}$ are Pauli matrices, and $\left(\sigma_{ab}\right)_A{}^{B}=\frac{1}{4}[\gamma_a,\gamma_b]_A{}^{B}$, where $\gamma_a$ are the standard Dirac matrices, satisfying the anticommutation rule $ \gamma_a\gamma_b + \gamma_b\gamma_a = -2 \eta_{ab}$. Let us also introduce $\gamma_5 \equiv - \gamma_0 \gamma_1 \gamma_2 \gamma_3$. In a similar way, one defines the action of group $G$ for all other fermions and scalars in the group $L$, depending on their particular transformation properties (see \cite{Radenkovic2019} for details).

In addition to the choice of the $3$-group, the action (\ref{eq:3BFaction}) also depends on bilinear forms. For the non-Abelian groups, one naturally chooses the Cartan-Killing form, while for the Abelian groups there is no natural choice, and one is restricted mostly by the requirement of $G$-invariance. Having this in mind, for the Standard Model $3$-group and the action (\ref{eq:3BFaction}) we choose the bilinear forms as follows. For the algebra $\mathfrak{g}$, we have
\begin{equation}\label{snbfg}
g_{[ab][cd]}=\frac{1}{2}\eta_{d[a|}\eta_{|b]c}\,, \qquad
g_{\alpha\beta}=\delta_{\alpha\beta}\,, \qquad
g_{\alpha [ab]}=0\,.
\end{equation}
For the algebra $\mathfrak{h}$, due to the $G$-invariance, we have
\begin{equation}\label{snbfh}
\killing{t_a}{t_b}_\mathfrak{h} \equiv g_{ab}=\eta_{ab}\,.
\end{equation}
Finally, for the algebra $\mathfrak{l}$ the situation is more complicated, due to the anticommutativity of the Grassmann numbers. Namely, for general $A,B\in\mathfrak{l}$, we have
\begin{equation}
\langle A,B \rangle_\mathfrak{l} = A^I B^J g_{IJ}\,, \qquad \langle B,A \rangle_\mathfrak{l} = B^J A^I g_{JI}\,.
\end{equation}
Since the bilinear form must be symmetric, the two expressions must be equal. However, depending on whether the coefficients  $A^I$ and $B^J$ are Grassmann-valued or real-valued, they will either anticommute or commute, and therefore the component matrix $g_{IJ}$ of the bilinear form must be antisymmetric or symmetric, respectively. In our case, the generators $T_A$ of the algebra $\mathfrak{l}$ can be grouped into three classes: $T_{\tilde{A}}$ which belong to the Higgs sector, and a pair $T_{\hat{A}}$, $T^{\hat{A}}$ which belong to the fermion sector. Given this, the components  of the bilinear form can be written as:
\begin{equation}
g_{AB} = \left[
\begin{array}{c|cc}
 \delta_{\tilde{A}\tilde{B}} & 0 & 0 \\ \hline
 0 & 0 & \delta_{\hat{A}}^{\hat{B}} \\
 0 & -\delta_{\hat{A}}^{\hat{B}} & 0 \\
\end{array}
\right]\,.
\end{equation}
Here the upper-left block corresponds to the algebra $\mathbb{C}^4$, while the bottom-right block corresponds to the algebra $\mathbb{G}^{64}\times\mathbb{G}^{64}\times\mathbb{G}^{64}$.

Given the Standard Model $3$-group and the corresponding bilinear forms, one can write the topological $3BF$ action (\ref{eq:3BFaction}) for this particular choice in the following form:
\begin{equation} \label{eq:3BFforStandardModelPrva}
S_{3BF}^\text{top}=\int B_{\alpha}\wedge F^{\alpha}+ B^{[ab]}\wedge R_{[ab]}+e_a\wedge\nabla\beta^{a}+ \phi^{A}(\nabla\tilde{\gamma})_A+ \bar{\psi}_A(\nablar \gamma)^A-(\bar{\gamma} \nablal)_A\psi^A\,.
\end{equation}
Here we have introduced the following new notation. First, $\cF$ is separated into the internal symmetry field strength $2$-form $F^\alpha$, which is a function of the internal symmetry connection $1$-form $\alpha^\alpha$, and the Riemann curvature $2$-form $R_{[ab]}$, which is a function of the spin connection $\omega^{[ab]}$. The Lagrange multiplier $C$ is relabeled as the tetrad field $1$-form $e_a$, while the Lagrange multiplier $D$ is relabeled as a tuple of scalar and fermion fields $(\phi^A, \psi^A, \bar{\psi}_A)$. This change of notation also implicitly suggests  the physical interpretation of the fields present in (\ref{eq:3BFaction}).

\subsection{The Standard Model $3BF$ action}

While the action (\ref{eq:3BFforStandardModelPrva}) is based on the Standard Model $3$-group and features all relevant gravitational, gauge and matter fields, it does not provide the correct classical dynamics. Namely, this action is an example of a {\em topological} theory, and as such it has trivial equations of motion, and no propagating degrees of freedom. In order to fix this, one deforms the action by introducing additional terms, historically called {\em simplicity constraints}. Convenient choices of simplicity constraints allow us to introduce the full classical action representing Einstein-Cartan gravity coupled to the Standard Model, with the correct classical dynamics. Such an action is then commonly called the {\em constrained} $3BF$ action, or more specifically the {\em Standard Model $3BF$} action, and is given as follows:
\begin{equation} \label{eq:RealisticAction}
S_{3BF}=S_{3BF}^\text{top}+S_{\text{grav}}+S_{\text{scal}}+S_{\text{Dirac}}+S_{\text{Yang-Mills}}+S_{\text{Higgs}}+S_{\text{Yukawa}}+S_{\text{spin}}+S_{\text{CC}}\,.
\end{equation}
The contributing terms are given as:
\begin{eqnarray}
S_{3BF}^\text{top}&=&\int B_{\alpha}\wedge F^{\alpha}+ B^{[ab]}\wedge R_{[ab]}+e_a\wedge\nabla\beta^{a}+ \phi^{A}(\nabla\tilde{\gamma})_A+ \bar{\psi}_A(\nablar \gamma)^A-(\bar{\gamma} \nablal)_A\psi^A\,, \label{eq:3BFforStandardModel} \\
S_{\text{grav}}&=&-\int\lambda_{[ab]}\wedge\left(B^{[ab]}-\frac{1}{8\pi l_p^2}\varepsilon^{[ab]cd}e_c\wedge e_d\right)\,, \label{eq:gravConstraint} \\
S_{\text{scal}}&=&\int \tilde{\lambda}^{A}\wedge\left(\tilde{\gamma}_A-H_{abcA}e^a\wedge e^b\wedge e^c\right)+\Lambda^{abA}\wedge\left(H_{abcA}\varepsilon^{cdef}e_d\wedge e_e\wedge e_f-(\nabla\phi)_A\wedge e_a\wedge e_b\right)\,, \label{eq:scalConstraint} \\
S_{\text{Dirac}}&=&\int\bar{\lambda}_A\wedge\left(\gamma^A+\frac{i}{6}\varepsilon_{abcd}e^a\wedge e^b\wedge e^c\left(\gamma^d\psi\right)^{A}\right)-\lambda^{A}\wedge\left(\bar{\gamma}_A-\frac{i}{6}\varepsilon_{abcd}e^a\wedge e^b\wedge e^c \left(\bar{\psi}\gamma^d\right)_A\right)\,, \label{eq:DiracConstraint} \\
S_{\text{Yang-Mills}}&=&\int\lambda^{\alpha}\wedge\left(B_{\alpha}-12{C}_{\alpha\beta}M^{\beta}{}_{ab}e^a\wedge e^b\right)+\zeta_{\alpha}{}^{ab}\left(M^{\alpha}{}_{ab}\varepsilon_{cdef}e^c\wedge e^d \wedge e^e\wedge e^f-F^{\alpha}\wedge e_a\wedge e_b\right)\,, \label{eq:YangMillsConstraintTerm} \\
S_{\text{Higgs}}&=&-\int\frac{2}{4!}\chi\left(\phi^A\phi_A-v^2\right)^2\varepsilon_{abcd}e^a\wedge e^b\wedge e^c\wedge e^d\,, \label{eq:HiggsPotentialConstraint} \\
S_{\text{Yukawa}}&=&-\int \frac{2}{4!}Y_{ABC}\bar{\psi}^A\psi^B\phi^C\varepsilon_{abcd}e^a\wedge e^b\wedge e^c\wedge e^d\,, \label{eq:YukawaConstraint} \\
S_{\text{spin}}&=&\int 2\pi il_p^2\bar{\psi}_A(\gamma_5\gamma^a\psi)^A\varepsilon_{abcd}e^b\wedge e^c\wedge\beta^d\,, \label{eq:spinConstraint} \\
S_{\text{CC}}&=&-\int\frac{1}{96\pi l_p^2}\Lambda\varepsilon_{abcd}e^a\wedge e^b\wedge e^c\wedge e^d\,. \label{eq:CCconstraint}
\end{eqnarray}
While the form of the full action may appear complicated or cryptic, there is the meaning and the purpose of each term in the action, as follows:
\begin{itemize}
\item the topological $3BF$ term (\ref{eq:3BFforStandardModel}) is identical to (\ref{eq:3BFforStandardModelPrva}), tabulating all fields present in the theory (as dictated by the Standard Model $3$-group),
\item the gravitational constraint term (\ref{eq:gravConstraint}) gives rise to the dynamics of the gravitational degrees of freedom,
\item the scalar constraint (\ref{eq:scalConstraint}) gives rise to the dynamics of massless scalar degrees of freedom,
\item the Dirac constraint (\ref{eq:DiracConstraint}) gives rise to the dynamics of massless fermions,
\item the Yang-Mills constraint (\ref{eq:YangMillsConstraintTerm}) gives rise to the dynamics of massless gauge bosons,
\item the Higgs potential constraint (\ref{eq:HiggsPotentialConstraint}) contains the self-interactions and the mass of the Higgs field,
\item the Yukawa constraint (\ref{eq:YukawaConstraint}) contains the interactions between the Higgs field and fermions, as well as fermion mixing matrices,
\item the spin constraint (\ref{eq:spinConstraint}) establishes the appropriate coupling between fermion spins and torsion, and
\item the CC constraint (\ref{eq:CCconstraint}) introduces the cosmological constant.
\end{itemize}
The action features the following free parameters:
\begin{itemize}
\item $l_p$ is the Planck length, featuring in $S_\text{grav}$, $S_\text{spin}$ and $S_\text{CC}$,
\item $C_{\alpha\beta}$ represents  the gauge coupling constant bilinear form, featuring in $S_\text{Yang-Mills}$,
\item $\chi$ is the coupling constant for the quartic self-interaction of the Higgs field, featuring in $S_\text{Higgs}$,
\item $v$ is the vacuum expectation value of the Higgs field, also featuring in $S_\text{Higgs}$,
\item $Y_{ABC}$ represent the Yukawa couplings and fermion mixing matrices, featuring in $S_\text{Yukawa}$, and
\item $\Lambda$ is the cosmological constant, featuring in $S_\text{CC}$.
\end{itemize}
The topological part $S_{3BF}^\text{top}$ and the constraints  $S_\text{scal}$ and $S_\text{Dirac}$ do not contain any free parameters. The configuration space of the theory is defined over the following non-dynamical Lagrange multiplier fields
\begin{equation}\label{nedinpolja3bf}
    M_{\alpha ab}\,,
    \zeta^{\alpha ab}\,,
    \lambda_{\alpha \mu\nu}\,,
    B_{\alpha \mu\nu}\,,
    \lambda_{[ab]\mu\nu}\,,
    B_{[ab]\mu\nu}\,,
    \tilde{\lambda}^A{}_{\mu}\,,
    \tilde{\gamma}^A{}_{\mu\nu\rho}\,,
    H^{abcA}\,,
    \Lambda^{abA}{}_{\mu}\,,
    \gamma^A{}_{\mu\nu\rho}\,,
    \bar{\gamma}_{A\mu\nu\rho}\,,
    \lambda^A{}_{\mu}\,,
    \bar{\lambda}_{A\mu}\,,
    \beta^a{}_{\mu\nu}\,,
    \omega^{[ab]}{}_\mu\,,
\end{equation}
as well as the following dynamical fields
\begin{equation}\label{dinpolja3bf}
e^a{}_\mu\,, \phi^A\,, \psi^A\,, \bar{\psi}_A\,, \alpha^\alpha{}_\mu \,.
\end{equation}
Taken together, (\ref{nedinpolja3bf}) and (\ref{dinpolja3bf}) define the full configuration space of the model. The distinction between dynamical and non-dynamical fields is a consequence of the equations of motion (EoMs), since the EoMs for the Lagrange multiplier fields are algebraic equations, while the EoMs for the dynamical fields are partial differential equations (see Appendix~\ref{AppB} for details). Note that the torsion equation (\ref{spinskakoneksija:jna}) can be explicitly solved for the spin connection as a function of the tetrads and fermion fields, rendering the spin connection as a non-dynamical field as well. This is a well known property of Einstein-Cartan theory \cite{BlagojevicBook,VasilicVojinovic2008}.

A tentative quantum theory corresponding to the action (\ref{eq:RealisticAction}) amounts to a path integral prescription for the expectation value of a generic observable $F$, depending on all fields $\phi_k$ in the configuration space of the theory. It can be schematically written as
\begin{equation}
\langle F \rangle = \frac{\cN}{Z} \int D\phi_k \; F(\phi_k) \; e^{iS_{3BF}[\phi_k]}\,,
\end{equation}
where $Z$ represents the evaluation of the path integral without the kernel $F(\phi_k)$,
\begin{equation}
Z = \cN \int D\phi_k \; e^{iS_{3BF}[\phi_k]}\,,
\end{equation}
and is necessary for the proper normalization of the expectation value. Also, $\cN$ is the overall normalization constant, which is canceled by the same constant present in $Z$. In slightly more detail, the path integral can be written as:
\begin{eqnarray} \label{eq:initialstep}
\nonumber
\langle F\rangle & = & \frac{{\cal N}}{Z} \int D\alpha D\omega D\beta De D\tilde{\gamma} D\gamma D\bar{\gamma} D\phi D\psi D\bar{\psi} DB D\tilde{\lambda} D\lambda D\bar{\lambda} D\Lambda D\zeta DH DM \; F\left(\phi_k\right) \\ 
\nonumber
&& \exp\left[i\int B_{\alpha}\wedge F^{\alpha}+B_{[ab]}\wedge R^{[ab]}+e_a\wedge\left(\nabla\beta\right)^{a}+\phi^A\left(\nabla\tilde{\gamma}\right)_A+\bar{\psi}_A(\nablar\gamma)^A-(\bar{\gamma}\nablal )_A\psi^A\right.\\
\nonumber
&&+\lambda^{\alpha}\wedge\left(B_{\alpha}-12C_{\alpha\beta}M^{\beta}{}_{ab}e^a\wedge e^b\right)-\lambda_{[ab]}\wedge\left(B^{[ab]}-\frac{1}{8\pi l_p^2}\varepsilon^{[ab]cd}e_c\wedge e_d\right)\\
\nonumber
&&+\tilde{\lambda}^A\wedge\left(\tilde{\gamma}_A-H_{abcA}e^a\wedge e^b\wedge e^c\right)+\bar{\lambda}_A\wedge\left(\gamma^A+\frac{i}{6}\varepsilon_{abcd}e^a\wedge e^b\wedge e^c\left(\gamma^d\psi\right)^A\right)\\
\nonumber
&&-\lambda^A\wedge\left(\bar{\gamma}_A-\frac{i}{6}\varepsilon_{abcd}e^a\wedge e^b\wedge e^c\left(\bar{\psi}\gamma^d\right)_A\right)+2\pi i l_p^2 \bar{\psi}_A(\gamma_5\gamma^a\psi)^A
\varepsilon_{abcd}e^b\wedge e^c\wedge \beta^d\\
\nonumber
&&+\zeta_{\alpha}{}^{ab}\left(M^{\alpha}{}_{ab}\varepsilon_{cdef}e^c\wedge e^d\wedge e^e\wedge e^f-F^{\alpha}\wedge e_a\wedge e_b\right)\vphantom{\int}\\
\nonumber
&&+\Lambda^{abA}\wedge\left(H_{abcA}\varepsilon^{cdef}e_d\wedge e_e\wedge e_f-\left(\nabla\phi\right)_A\wedge e_a\wedge e_b\right)\vphantom{\int}\\
&&-\frac{1}{12}\left.\vphantom{\int}\left(\chi\left(\phi^A\phi_A-v^2\right)^2+Y_{ABC}\bar{\psi}^A\psi^B\phi^C+\frac{\Lambda}{8\pi l_p^2}\right)\varepsilon_{abcd}e^a\wedge e^b\wedge e^c\wedge e^d\right]\,.\\
\nonumber
\end{eqnarray}
The abstract form (\ref{eq:initialstep}) of the path integral should be understood as a ``statement of intent''. Namely, the main goal of this paper is to provide a rigorous definition for the expression (\ref{eq:initialstep}).

\subsection{The Einstein-Cartan contact action}

In addition to the Standard Model $3BF$ action, one can also introduce the more traditional, Einstein-Cartan action coupled to Standard Model with the spin-spin contact interaction. The action is given as:
\begin{equation} \label{eq:ECCaction}
\begin{array}{lcl}
S_{ECC} [e,\phi, \psi,\bar{\psi},\alpha] & = & \ds \int \frac{1}{16\pi l_p^2} \varepsilon^{abcd}\, R_{ab}\wedge e_c\wedge e_d -F^{\alpha}\wedge C_{\alpha\beta}\dual F^{\beta}-\left(\nabla\phi\right)^A\wedge \left(\dual\nabla\phi\right)_A \vphantom{\ds\int} \\
&& \ds -\frac{i}{6}\varepsilon_{abcd}\,e^a\wedge e^b\wedge e^c\wedge\left[ \bar{\psi}_A (\nablal\gamma^d\psi)^A-\bar{\psi}_A(\gamma^{d} \nablar\psi )^A\right] \vphantom{\ds\int} +\frac{3\pi l_p^2}{4}s^{a}\wedge \dual s_{a}\\
&& \ds -\frac{1}{12}\vphantom{\int}\left[\chi\left(\phi^A\phi_A-v^2\right)^2+Y_{ABC}\bar{\psi}^A\psi^B\phi^C+\frac{\Lambda}{8\pi l_p^2}\right]\varepsilon_{abcd} \, e^a\wedge e^b\wedge e^c\wedge e^d\,. \vphantom{\ds\int} \\
\end{array}
\end{equation}
In the first row one can recognize the gravitational term, the Yang-Mills kinetic term, and the scalar kinetic term. The second row features the kinetic terms for fermions, and the spin-spin contact interaction term. The third row features the Higgs potential, the Yukawa coupling terms and the cosmological constant. The configuration space of the theory coincides with the configuration space of dynamic fields (\ref{dinpolja3bf}) for the Standard Model $3BF$ action, and the EoMs for (\ref{eq:ECCaction}) are equivalent to the EoMs for (\ref{eq:RealisticAction}), see Appendix~\ref{AppB} for details.

In addition to the classical equivalence of the $3BF$ and ECC actions, the relation between the two theories has been studied also at the abstract quantum level in \cite{Stipsic2025b}. There, it was demonstrated that the expectation values of an arbitrary observable are related between the two theories, in the following way:
\begin{equation} \label{eq:RelationsBFandECC}
\langle F \rangle_{3BF} = \frac{\left\langle \frac{1}{|e|^M} F  \right\rangle_{ECC}}{\left\langle \frac{1}{|e|^M}   \right\rangle_{ECC}} \,, \qquad
\langle F \rangle_{ECC} = \frac{\left\langle |e|^M F  \right\rangle_{3BF}}{\left\langle |e|^M  \right\rangle_{3BF}} \,.
\end{equation}
Here $M$ is an integer constant, evaluated in \cite{Stipsic2025b} to be $M=150$. These relations have been established nonperturbatively using the path integral formalism, and under some very mild and generic assumptions. In particular, the explicit quantization has not been performed for either theory, and instead it was just assumed that the path integral has been defined in the same way for both theories, making them comparable in an abstract sense.

However, in \cite{Stipsic2025b} it was also argued that, if one were to attempt to rigorously define the path integral, this would be easier to do for the $3BF$ theory than for the $ECC$ theory. This is because the ECC action, despite having a smaller configuration space and a simpler action, explicitly features the Hodge dual operator acting on the Yang-Mills field strength, on the scalar field strength, and on the spin $2$-form,
\begin{equation} \label{eq:HodgeDualTetradBasis}
  \dual F^\alpha = \frac{1}{4} F^\alpha{}_{cd} \lc^{abcd} \, e_a \wedge e_b\,, \qquad
  (\dual \nabla \phi)_A = \frac{1}{3!} (\nabla_d \phi)_A \lc^{dabc} \, e_a \wedge e_b \wedge e_c\,, \qquad
  \dual s^a = \frac{1}{4} s^a{}_{bc} \lc^{bcde} \, e_d \wedge e_e\,,
\end{equation}
in contrast to the $3BF$ action which, by design, does not feature the Hodge dual at all. The reason why the presence of the Hodge dual ought to be considered a drawback for the quantization of the theory lies in the fact that it depends on the components of the tetrad fields $e^a{}_\mu$ (and potentially their inverses $e^\mu{}_a$ as well), in addition to the corresponding $1$-forms $e^a$. To see this, note that the components of $\dual F^a$, written in an arbitrary coordinate basis $\rmd x^\mu \wedge \rmd x^\nu$, crucially depend on the metric tensor $g_{\mu\nu} \equiv \eta_{ab} e^a{}_\mu e^b{}_\nu$, and similarly for the components of $(\dual \nabla \phi)_A$ and $\dual s^a$:
\begin{equation} \label{eq:HodgeDualCoordinateBasis}
\begin{array}{c}
\ds  \dual F^\alpha = \frac{1}{4} F^\alpha{}_{\mu\nu} \lc^{\mu\nu\lambda\rho}g_{\lambda\sigma} g_{\rho\tau} \, \rmd x^\sigma \wedge \rmd x^\tau\,, \qquad
  \dual s^a = \frac{1}{4} s^a{}_{\mu\nu} \lc^{\mu\nu\lambda\rho}g_{\lambda\sigma} g_{\rho\tau} \, \rmd x^\sigma \wedge \rmd x^\tau\,, \qquad \vphantom{\ds\int} \\
\ds  (\dual \nabla \phi)_A = \frac{1}{3!} (\nabla_\mu \phi)_A \lc^{\mu\nu\lambda\rho} g_{\nu\sigma} g_{\lambda\tau} g_{\rho\theta} \, \rmd x^\sigma \wedge \rmd x^\tau \wedge \rmd x^\theta\,.  \vphantom{\ds\int} \\
\end{array}
\end{equation}
In other words, regardless of the choice of basis between (\ref{eq:HodgeDualTetradBasis}) and (\ref{eq:HodgeDualCoordinateBasis}), the Hodge dual can be expressed in terms of \emph{components} of forms $F^\alpha$, $(\nabla \phi)_A$ and $e^a$, but it cannot be expressed in terms of those forms \emph{themselves}. This represents a crucial difference between the $3BF$ and ECC actions --- the former is explicitly formulated exclusively in terms of differential forms, while the latter is explicitly formulated not only in terms of differential forms, but also in terms of their components, through the Hodge dual. In this sense, the ECC action is qualitatively no different than the ordinary Einstein-Hilbert action coupled to Standard Model, expressed in textbook-like fashion, in terms of components of general tensors of various ranks.

As we shall demonstrate in Sections \ref{secV} and \ref{secVI}, the exclusive dependence on differential forms indeed represents a decisive advantage of the $3BF$ action, in contrast to the ECC action, precisely because of the presence of the Hodge dual in the latter. Also, in those Sections we explicitly perform the quantization of the $3BF$ action, and as a consequence also of the ECC action, verifying in the process that the relations (\ref{eq:RelationsBFandECC}) indeed hold between the two quantum theories.

\section{\label{secIII}Discretization of the action}

The purpose of this Section is to give a procedure how to define an action on a piecewise-flat manifold. In particular, we give a method to pass from an action defined over a smooth $4$-dimensional spacetime manifold to an action defined over a piecewise-flat manifold. To that end, we will obtain two main results . The first result will be a prescription how to put an integral of a product of differential forms onto a triangulation. The second result will be a prescription how to put differential forms featuring the exterior derivative onto a triangulation. Together, these two prescriptions provide a method to transform any action expressed in terms of differential forms into an action defined on a triangulation.

\subsection{\label{secIIIsubsecA}Preliminaries}

We are interested in discussing piecewise-flat manifolds which are triangulations, in the sense that all flat cells of the manifold are $4$-simplices. Formally, this structure is described as a simplicial complex homeomorfic to the manifold, and it consists  of vertices $v$, edges $\varepsilon$, triangles $\Delta$, tetrahedra $\tau$ and $4$-simplices $\sigma$. Each simplex (except for vertices) has a natural orientation, defined by the order of its  subsimplices. We will discuss the details of the orientation in subsections \ref{secIIIsubsecC} and \ref{secIIIsubsecD}.

Typically, we will encounter various sums over various simplices in the complex. Each sum can be taken over the simplex $a$ which is a subsimplex of $b$, denoted $a \in b$, or a supersimplex of $b$, denoted $a \ni b$. If the domain is not specified, it is assumed that the sum is taken over all simplices in the complex. 

Finally, we will also often discuss vectors that coincide with edges, both in terms of position and orientation. These vectors are formally elements  of a tangent space, defined as a $4$-dimensional flat space $\realni^4$ that contains a given $4$-simplex. The vectors can therefore be expanded into a coordinate basis $\partial_\mu$. Since we only discuss vectors that coincide with edges, the same notation $\varepsilon$ will be used for both.

Throughout the literature, most commonly papers that discuss spinfoam models and their generalizations, one typically discusses the action of the $BF$ type, such as the Plebanski action \cite{Plebanski}. The topological part of the action is of the form
\begin{equation}
\int_\cM B \wedge F \,,
\end{equation}
where $B$ and $F$ are some $2$-forms. When passing from a smooth manifold to a piecewise-flat one, this structure of the integral is often used as a suggestion that the $2$-form $B$ should live on a triangulation, while $F$ should live on its  Poincar\'e dual, or vice versa. This assignment can be considered natural for the actions of the above form, but this is not possible in general. For example, our action (\ref{eq:RealisticAction}) contains (among others) terms of type:
\begin{equation}
\int_\cM \lc_{abcd} \, e^a \wedge e^b \wedge e^c \wedge e^d\,.
\end{equation}
In this case it is completely unnatural to assign any of the differential forms to a triangulation while simultaneously assigning the remainder to the Poincar\'e dual, since all four $1$-forms $e^a$ are of the same nature and should all live on the same structure --- either on a triangulation, specifically on its  edges, or on a dual lattice, specifically on the edges of the dual.

Given this situation, the most natural (and simplest) way to proceed is to assign all differential forms to the simplices on the triangulation, namely each $k$-form to a $k$-simplex ($k \in \{ 0,\dots , 4 \}$). One could alternatively use the Poincar\'e dual and assign all forms to vertices, edges, faces, $3$-polyhedra and $4$-polyhedra of the dual lattice, but in this paper we will opt for triangulation instead, for simplicity.

\subsection{\label{secIIIsubsecB}Combinatorial analysis}

In order to be able to pass from various quantities defined over a smooth manifold to corresponding quantities defined over a piecewise-flat manifold, it will be important to express the resulting quantities in a way that is invariant with respect to the particular details of the specification of a simplex in the triangulation. For example, each tetrahedron can be specified by three of its  edges, or alternatively by one of its  triangles and an additional edge which does not belong to that triangle. There are multiple possible choices of edges and triangles in either of these cases, so it is necessary to count them all in order to take into account all possible ways a tetrahedron can be specified. Similarly, one needs to count all possible ways to specify a triangle, and most importantly, a $4$-simplex, since we are interested in $4$-dimensional manifolds. In addition, one also needs to take into account the notion of orientation for all simplices, since the orientation contributes to the counting in a nontrivial way.

In total, we will be interested in seven particular cases --- one for a triangle, two for tetrahedra, and four for $4$-simplices. They correspond to various ways a simplex can be decomposed into simplices of smaller dimension, and all seven cases are depicted in Figures \ref{SlikaJedanG}, \ref{SlikaDvaG}, \ref{SlikaTriG}, \ref{SlikaCetiriG}, and listed in Table \ref{TabelaJedan}, together with the resulting total number of ways one can specify the simplex. Each count is determined as a product of the distinct number of ways to decompose a simplex with the number of permutations of subsimplices of the same type. The explicit counting procedure for the results displayed in Table \ref{TabelaJedan} is provided in full detail in Appendix \ref{appC}.

\begin{figure}[H]
\begin{center}
  \begin{tikzpicture}[yscale=0.5]
\draw[very thick] (-1,0) -- (1,0) ;
\draw[very thick] (-1,0) -- (0,1) ;
\draw[very thin] (1,0) -- (0,1) ;
\draw[very thin] (0,1) -- (0,3) ;
\draw[very thick] (-1,0) -- (0,3) ;
\draw[very thin] (1,0) -- (0,3) ;
\end{tikzpicture}
,\ 
\begin{tikzpicture}[yscale=0.5]
\draw[very thin] (-1,0) -- (1,0) ;
\draw[very thin] (-1,0) -- (0,1) ;
\draw[very thick] (1,0) -- (0,1) ;
\draw[very thin] (0,1) -- (0,3) ;
\draw[very thick] (-1,0) -- (0,3) ;
\draw[very thick] (1,0) -- (0,3) ;
\end{tikzpicture}
,\ 
\begin{tikzpicture}[yscale=0.5]
\draw[fill=gray] (-1,0) -- (1,0) -- (0,1) -- (-1,0) ;
\draw[very thin] (-1,0) -- (1,0) ;
\draw[very thin] (-1,0) -- (0,1) ;
\draw[very thin] (1,0) -- (0,1) ;
\draw[very thin] (0,1) -- (0,3) ;
\draw[very thick] (-1,0) -- (0,3) ;
\draw[very thin] (1,0) -- (0,3) ;
\end{tikzpicture}
\end{center}
\caption{\label{SlikaJedanG}Decompositions of the tetrahedron. The first two diagrams correspond to the case (2), while the third diagram corresponds to the case (3).}
\end{figure}

\begin{figure}[H]
\begin{center}
\begin{tikzpicture}[scale=1.8,yscale=1.8]
\draw[very thin] (-0.3,0) -- (0.3,0) ;
\draw[very thin] (-0.3,0) -- (0.5,0.3) ;
\draw[very thin] (-0.3,0) -- (0,0.5) ;
\draw[very thick] (-0.3,0) -- (-0.5,0.3) ;
\draw[very thin] (0.5,0.3) -- (0.3,0) ;
\draw[very thin] (0.3,0) -- (0,0.5) ;
\draw[very thick] (-0.5,0.3) -- (0.3,0) ;
\draw[very thin] (0.5,0.3) -- (0,0.5) ;
\draw[very thick] (-0.5,0.3) -- (0.5,0.3) ;
\draw[very thick] (-0.5,0.3) -- (0,0.5) ;
\end{tikzpicture}
,\ 
\begin{tikzpicture}[scale=1.8,yscale=1.8]
\draw[very thick] (-0.3,0) -- (0.3,0) ;
\draw[very thin] (-0.3,0) -- (0.5,0.3) ;
\draw[very thin] (-0.3,0) -- (0,0.5) ;
\draw[very thin] (-0.3,0) -- (-0.5,0.3) ;
\draw[very thin] (0.5,0.3) -- (0.3,0) ;
\draw[very thin] (0.3,0) -- (0,0.5) ;
\draw[very thick] (-0.5,0.3) -- (0.3,0) ;
\draw[very thin] (0.5,0.3) -- (0,0.5) ;
\draw[very thick] (-0.5,0.3) -- (0.5,0.3) ;
\draw[very thick] (-0.5,0.3) -- (0,0.5) ;
\end{tikzpicture}
,\ 
\begin{tikzpicture}[scale=1.8,yscale=1.8]
\draw[very thin] (-0.3,0) -- (0.3,0) ;
\draw[very thin] (-0.3,0) -- (0.5,0.3) ;
\draw[very thin] (-0.3,0) -- (0,0.5) ;
\draw[very thick] (-0.3,0) -- (-0.5,0.3) ;
\draw[very thick] (0.5,0.3) -- (0.3,0) ;
\draw[very thin] (0.3,0) -- (0,0.5) ;
\draw[very thin] (-0.5,0.3) -- (0.3,0) ;
\draw[very thick] (0.5,0.3) -- (0,0.5) ;
\draw[very thin] (-0.5,0.3) -- (0.5,0.3) ;
\draw[very thick] (-0.5,0.3) -- (0,0.5) ;
\end{tikzpicture}
\end{center}
\caption{\label{SlikaDvaG}Decompositions of the $4$-simplex. The three diagrams correspond to the case (4).}
\end{figure}

\begin{figure}[H]
\begin{center}
\begin{tikzpicture}[scale=1.8,yscale=1.8]
\draw[fill=gray] (0,0.5) -- (-0.5,0.3) -- (-0.3,0) -- (0,0.5) ;
\draw[very thick] (-0.3,0) -- (0.3,0) ;
\draw[very thin] (-0.3,0) -- (0.5,0.3) ;
\draw[very thin] (-0.3,0) -- (0,0.5) ;
\draw[very thin] (-0.3,0) -- (-0.5,0.3) ;
\draw[very thick] (0.5,0.3) -- (0.3,0) ;
\draw[very thin] (0.3,0) -- (0,0.5) ;
\draw[very thin] (-0.5,0.3) -- (0.3,0) ;
\draw[very thin] (0.5,0.3) -- (0,0.5) ;
\draw[very thin] (-0.5,0.3) -- (0.5,0.3) ;
\draw[very thin] (-0.5,0.3) -- (0,0.5) ;
\end{tikzpicture}
,\ 
\begin{tikzpicture}[scale=1.8,yscale=1.8]
\draw[fill=gray] (0,0.5) -- (-0.5,0.3) -- (-0.3,0) -- (0,0.5) ;
\draw[very thick] (-0.3,0) -- (0.3,0) ;
\draw[very thin] (-0.3,0) -- (0.5,0.3) ;
\draw[very thin] (-0.3,0) -- (0,0.5) ;
\draw[very thin] (-0.3,0) -- (-0.5,0.3) ;
\draw[very thin] (0.5,0.3) -- (0.3,0) ;
\draw[very thin] (0.3,0) -- (0,0.5) ;
\draw[very thin] (-0.5,0.3) -- (0.3,0) ;
\draw[very thick] (0.5,0.3) -- (0,0.5) ;
\draw[very thin] (-0.5,0.3) -- (0.5,0.3) ;
\draw[very thin] (-0.5,0.3) -- (0,0.5) ;
\end{tikzpicture}
,\ 
\begin{tikzpicture}[scale=1.8,yscale=1.8]
\draw[fill=gray] (0,0.5) -- (-0.5,0.3) -- (-0.3,0) -- (0,0.5) ;
\draw[very thick] (-0.3,0) -- (0.3,0) ;
\draw[very thick] (-0.3,0) -- (0.5,0.3) ;
\draw[very thin] (-0.3,0) -- (0,0.5) ;
\draw[very thin] (-0.3,0) -- (-0.5,0.3) ;
\draw[very thin] (0.5,0.3) -- (0.3,0) ;
\draw[very thin] (0.3,0) -- (0,0.5) ;
\draw[very thin] (-0.5,0.3) -- (0.3,0) ;
\draw[very thin] (0.5,0.3) -- (0,0.5) ;
\draw[very thin] (-0.5,0.3) -- (0.5,0.3) ;
\draw[very thin] (-0.5,0.3) -- (0,0.5) ;
\end{tikzpicture}
\end{center}
\caption{\label{SlikaTriG}Decompositions of the $4$-simplex. The three diagrams correspond to the case (5).}
\end{figure}

\begin{figure}[H]
\begin{center}
\begin{tikzpicture}[scale=1.8,yscale=1.8]
\draw[fill=gray] (-0.5,0.3) -- (-0.3,0) -- (0.3,0) -- (0.5,0.3)  -- (-0.5,0.3) ;
\draw[very thin] (-0.3,0) -- (0.3,0) ;
\draw[very thin] (-0.3,0) -- (0.5,0.3) ;
\draw[very thin] (-0.3,0) -- (0,0.5) ;
\draw[very thin] (-0.3,0) -- (-0.5,0.3) ;
\draw[very thin] (0.5,0.3) -- (0.3,0) ;
\draw[very thin] (0.3,0) -- (0,0.5) ;
\draw[very thin] (-0.5,0.3) -- (0.3,0) ;
\draw[very thin] (0.5,0.3) -- (0,0.5) ;
\draw[very thin] (-0.5,0.3) -- (0.5,0.3) ;
\draw[very thick] (-0.5,0.3) -- (0,0.5) ;
\end{tikzpicture}
,\ 
\begin{tikzpicture}[scale=1.8,yscale=1.8]
\draw[fill=gray] (0,0.5) -- (-0.5,0.3) -- (-0.3,0) -- (0,0.5) ;
\draw[fill=gray] (0,0.5) -- (0.5,0.3) -- (0.3,0) -- (0,0.5) ;
\draw[very thin] (-0.3,0) -- (0.3,0) ;
\draw[very thin] (-0.3,0) -- (0.5,0.3) ;
\draw[very thin] (-0.3,0) -- (0,0.5) ;
\draw[very thin] (-0.3,0) -- (-0.5,0.3) ;
\draw[very thin] (0.5,0.3) -- (0.3,0) ;
\draw[very thin] (0.3,0) -- (0,0.5) ;
\draw[very thin] (-0.5,0.3) -- (0.3,0) ;
\draw[very thin] (0.5,0.3) -- (0,0.5) ;
\draw[very thin] (-0.5,0.3) -- (0.5,0.3) ;
\draw[very thin] (-0.5,0.3) -- (0,0.5) ;
\end{tikzpicture}
\end{center}
\caption{\label{SlikaCetiriG}Decompositions of the $4$-simplex. The left diagram corresponds to the case (6), while the right diagram corresponds to the case (7).}
\end{figure}

\begin{table}[H]
\begin{center}
\begin{tabular}{|c|c|c|c|}
\hline
\ Case\ \ &\ Element\ \ &\ Decomposition\ \ &\ Number of choices\ \ \\
\hline\hline
(1) & $\trougao$ & $2\times\dvougao$ & $3\times 2!=6$\\
\hline
(2) & $\tetraedar$ & $3\times\dvougao$ & $16\times 3!=96$\\
\hline
(3) & $\tetraedar$ & $\dvougao + \trougao$ & $12$\\
\hline
(4) & $\dekaedar$ & $4\times\dvougao$ & $125\times 4!=3000$\\
\hline
(5) & $\dekaedar$ & $2\times\dvougao+\trougao$ & $150\times 2!=300$\\
\hline
(6) & $\dekaedar$ & $\dvougao+\tetraedar$ & $20$\\
\hline
(7) & $\dekaedar$ & $2\times\trougao$ & $15\times 2!=30$\\
\hline
\end{tabular}
\end{center}
\caption{\label{TabelaJedan}Possible decompositions of simplices and the total number of ways each decomposition can be performed.}
\end{table}

This concludes the combinatorial analysis of all possibilities we will encounter below. Specifically, the quantities discussed in the next subsection will be expressed in an invariant way by taking the weighted sum over all possible ways a given simplex can be decomposed into subsimplices.

\subsection{\label{secIIIsubsecC}Discretization of the integral}

Let us now focus on the discretization procedure of the classical action of the theory. In general, as long as the action is written in terms of differential forms, there is a natural way to pass from the action defined over a smooth spacetime manifold $\cM$ to an action defined over a piecewise flat manifold, i.e., a triangulation $T(\cM)$. This is done via the following prescription. Given an action in the form
\begin{equation}
S = \int_\cM \cL\,,
\end{equation}
where $\cL$ is the Lagrangian $4$-form, one can pass to the triangulation $T(\cM)$ by rewriting the action in the form
\begin{equation}
S = \int_{T(\cM)} \cL = \sum_{\sigma\in T(\cM)} \int_\sigma \cL\,,
\end{equation}
where we have split the domain of integration $T(\cM)$ into a sum of its  constituent $4$-simplices $\sigma$. Next, we appeal to the main assumption that the fields do not change their values within any single $4$-simplex. This means that the Lagrangian $4$-form is constant across $\sigma$, so we have
\begin{equation} \label{eq:LagrangianEvaluationPartI}
  \int_\sigma \cL =  \frac{1}{4!} \mathcal{L}_{\mu\nu\rho\lambda}(\sigma) \int_\sigma {\rm d}x^\mu\wedge{\rm d}x^\nu\wedge{\rm d}x^\rho\wedge{\rm d}x^\lambda= \frac{1}{4!} \mathcal{L}_{\mu\nu\rho\lambda}(\sigma) \varepsilon^{\mu\nu\rho\lambda} \frac{1}{e} V(\sigma)\,,
\end{equation}
where $V(\sigma)$ is the $4$-volume of the $4$-simplex,
\begin{equation} \label{eq:FourSimplexFourVolumeDef}
V(\sigma) = \int_\sigma e\; d^4x\,.
\end{equation}
Note that in the above we have applied the elementary identity of differential forms, $ {\rm d}x^\mu\wedge{\rm d}x^\nu\wedge{\rm d}x^\rho\wedge{\rm d}x^\lambda = \varepsilon^{\mu\nu\rho\lambda} \, d^4x$, and that we have assumed that the determinant of the tetrad $e$ is also constant within the $4$-simplex $\sigma$, same as all other fields.

On the other hand, one can obtain the result (\ref{eq:LagrangianEvaluationPartI}) via purely algebraic means. To that end, given a $4$-simplex, one can always associate to it a totally antisymmetric tensor of type $\binom{4}{0}$,
\begin{equation}
\sigma=\frac{1}{4!}\,\varepsilon_1^{\mu}\varepsilon_2^{\nu}\varepsilon_3^\rho\varepsilon_4^\sigma\,\partial_\mu\wedge \partial_\nu\wedge\partial_\rho\wedge\partial_\sigma\,,
\end{equation}
where $\varepsilon_1,\dots ,\varepsilon_4$ are four linearly independent vectors that lie along the four edges of the $4$-simplex. The tensor itself does not depend on the particular choice of these vectors (up to a sign). Namely, any other choice of vectors that lie along some edges of the $4$-simplex must be a linear combination of the vectors $\varepsilon_1,\dots ,\varepsilon_4$, so that the total antisymmetry of the basis $\partial_\mu\wedge \partial_\nu\wedge\partial_\rho\wedge\partial_\sigma$ guarantees that the two choices give the same $\binom{4}{0}$ tensor, up to an overall sign. The result (\ref{eq:LagrangianEvaluationPartI}) can now be obtained by a contraction of the Lagrangian $4$-form $\cL$ and the tensor $\sigma$. Namely, the Lagrangian $4$-form, understood as a tensor of type  $\binom{0}{4}$, is a linear functional over the space of tensors of type $\binom{4}{0}$, due to the elementary biorthogonality relation
\begin{equation} \label{eq:BiorthogonalityRelation}
{\rm d}x^\mu [\partial_\nu] = \delta^\mu_\nu\,,
\end{equation}
which defines the contraction operation. We therefore have:
\begin{equation} \label{eq:LagrangianEvaluationPartII}
\begin{array}{lcl}
\mathcal{L}[\sigma]&=& \ds
\frac{1}{4!}\mathcal{L}_{\mu\nu\rho\lambda}(\sigma)\;\rmd x^{\mu}\wedge \rmd x^{\nu}\wedge \rmd x^{\rho}\wedge \rmd x^{\lambda}[\frac{1}{4!}\varepsilon_{1}^{\mu'}\varepsilon_{2}^{\nu'}\varepsilon_{3}^{\rho'}\varepsilon_{4}^{\lambda'}\;\partial_{\mu'}\wedge\partial_{\nu'}\wedge\partial_{\rho'}\wedge\partial_{\lambda'} ] \vphantom{\ds\int} \\
 & = & \ds
\frac{1}{(4!)^2}\mathcal{L}_{\mu\nu\rho\lambda}(\sigma)\; \varepsilon_{1}^{\mu'}\varepsilon_{2}^{\nu'}\varepsilon_{3}^{\rho'}\varepsilon_{4}^{\lambda'} \; \rmd x^{\mu}\wedge \rmd x^{\nu}\wedge \rmd x^{\rho}\wedge \rmd x^{\lambda}[\partial_{\mu'}\wedge\partial_{\nu'}\wedge\partial_{\rho'}\wedge\partial_{\lambda'} ] \vphantom{\ds\int} \\
 & = & \ds
\frac{1}{(4!)^2}\mathcal{L}_{\mu\nu\rho\lambda}(\sigma)\; \varepsilon_{1}^{\mu'}\varepsilon_{2}^{\nu'}\varepsilon_{3}^{\rho'}\varepsilon_{4}^{\lambda'} \; (-1) \varepsilon^{\mu\nu\rho\lambda} \varepsilon_{\mu'\nu'\rho'\lambda'} \vphantom{\ds\int} \\
 & = & \ds
\frac{1}{4!}\mathcal{L}_{\mu\nu\rho\lambda}(\sigma) \varepsilon^{\mu\nu\rho\lambda} \frac{1}{e} \left(- \frac{1}{4!} \, e\, \varepsilon_{1}^{\mu'}\varepsilon_{2}^{\nu'}\varepsilon_{3}^{\rho'}\varepsilon_{4}^{\lambda'} \varepsilon_{\mu'\nu'\rho'\lambda'} \right) \vphantom{\ds\int} \\
 & = & \ds
\frac{1}{4!}\mathcal{L}_{\mu\nu\rho\lambda}(\sigma) \varepsilon^{\mu\nu\rho\lambda} \frac{1}{e} V(\sigma)\,. \vphantom{\ds\int} \\
\end{array}
\end{equation}
In the final step above, we have recognized the expression in the parentheses as the $4$-volume (\ref{eq:FourSimplexFourVolumeDef}) of the $4$-simplex. Comparing (\ref{eq:LagrangianEvaluationPartI}) with (\ref{eq:LagrangianEvaluationPartII}), we obtain the following main result:
\begin{equation} \label{eq:DiscretizationMainResult}
\int_{T(\cM)} \cL = \sum_{\sigma\in T(\cM)} \cL[\sigma]\,.
\end{equation}

Let us summarize the analysis given above. If we start from an action which can be written as an integral of a Lagrangian $4$-form $\cL$, and if the fields in the action are constant within each $4$-simplex $\sigma$ of the triangulation $T(\cM)$, then the action can be rewritten in an algebraic way, namely as a sum of contractions $\cL[\sigma]$ over all simplices. The main benefit of (\ref{eq:DiscretizationMainResult}) lies in the fact that the contractions $\cL[\sigma]$ are readily expressible in an algebraic manner, without performing any integrals. Thus the formula (\ref{eq:DiscretizationMainResult}) enables us to pass from an action over a smooth manifold to an action over a piecewise flat manifold in a unique and convenient way.

In a $4$-dimensional spacetime, there are four nontrivial types of contractions $\cL[\sigma]$, corresponding to four classes of integrals. For the purpose of convenient description of these classes, let us introduce the following notation:
\begin{equation} \label{eq:AllFormsNotation}
\text{ 0-forms:} \quad \phi_a \,, \qquad 
\text{ 1-forms:}\quad \alpha_a\,, \qquad 
\text{ 2-forms:}\quad \beta_a\,, \qquad 
\text{ 3-forms:}\quad \gamma_a\,, \qquad 
\text{ 4-forms:}\quad \delta_a\,,
\end{equation}
where the index $a$ is counting the forms within some given set. The four classes of integrals that can appear in the action are then of the following types:
\begin{equation} \label{eq:FourTypesOfIntegrals}
  \int \alpha_a\wedge\alpha_b\wedge\alpha_c\wedge\alpha_d\,, \qquad
  \int\alpha_a\wedge\alpha_b\wedge \beta_c\,, \qquad
  \int \beta_a\wedge \beta_b\,, \qquad
  \int\alpha_a\wedge\gamma_b\,.
\end{equation}
Note that these four types of integrals correspond to the decompositions of a $4$-simplex in Table \ref{TabelaJedan}, specifically to the cases (4), (5), (7) and (6), respectively. Now we proceed to evaluate them in turn on a triangulation, using the formula (\ref{eq:DiscretizationMainResult}). In the case of the first class, we have:
\begin{eqnarray} \label{eq:fourAlphaIntegral}
\nonumber
\int \alpha_a\wedge\alpha_b\wedge\alpha_c\wedge\alpha_d&=&\sum_\sigma\alpha_a\wedge\alpha_b\wedge\alpha_c\wedge\alpha_d[\sigma]\\
\nonumber
&=&\sum_\sigma\alpha_{a\mu}\alpha_{b\nu}\alpha_{c\rho}\alpha_{d\sigma}\;\frac{1}{4!}\varepsilon_1^{\lambda}\varepsilon_2^{\chi}\varepsilon_3^{\varphi}\varepsilon_4^{\xi}\;{\rm d}x^{\mu}\wedge {\rm d}x^{\nu}\wedge {\rm d}x^{\rho}\wedge {\rm d}x^{\sigma}\left[\partial_\lambda\wedge\partial_\chi\wedge\partial_\varphi\wedge\partial_\xi\right]\\
\nonumber
&=&\frac{1}{4!}\sum_\sigma\alpha_{a\mu}\alpha_{b\nu}\alpha_{c\rho}\alpha_{d\sigma}\;\varepsilon_1^{\lambda}\varepsilon_2^{\chi}\varepsilon_3^{\varphi}\varepsilon_4^{\xi}\left(\delta_\lambda^\mu\delta_\chi^\nu\delta_\varphi^\rho\delta_\xi^\sigma+(23)\right)\\
&=&\frac{1}{3000}\sum_\sigma\sum_{\varepsilon_1,...,\varepsilon_4\in \sigma}\alpha_a[\varepsilon_1]\alpha_b[\varepsilon_2]\alpha_c[\varepsilon_3]\alpha_d[\varepsilon_4]\;W[\varepsilon_1,\varepsilon_2,\varepsilon_3,\varepsilon_4]\,.
\end{eqnarray}
Several comments  regarding the above calculation are in order. First, the notation ``$+(23)$'' in the third row denotes a shorthand for the remaining $23$ terms in the parentheses, rendering the first term, i.e., the product of Kronener symbols, totally antisymmetric.

Second, the function $W$ is valued in the set $\{-1, 0,1\}$, and keeps track of the overall sign. It is defined in detail below.

Third, it is important to discuss the steps of the calculation. In the first step we have applied the formula (\ref{eq:DiscretizationMainResult}). The second step amounts  to expanding everything with respect to a basis. The third step is the contraction of basis vectors and basis $1$-forms, using the biorthogonality relation (\ref{eq:BiorthogonalityRelation}). Finally, the fourth step is nontrivial. In addition to expressing the result in terms of contractions $\alpha[\epsilon]$ of four $1$-forms and four vectors, we have introduced a sum over all possible choices one can make for the four vectors $\varepsilon_1,\dots ,\varepsilon_4$. The idea is to render the resulting expression manifestly invariant with respect to these choices. To that end, the latter sum has to be normalized with the factor $3000$, in correspondence with the combinatorial analysis of the previous subsection, specifically the case (4) of Table \ref{TabelaJedan}.

Next we discuss the second class of integrals in (\ref{eq:FourTypesOfIntegrals}). The calculation is similar to the previous case above:
\begin{eqnarray} \label{eq:TwoAlphaBetaIntergal}
\nonumber
\int\alpha_a\wedge\alpha_b\wedge \beta_c&=&\sum_\sigma\alpha_a\wedge\alpha_b\wedge \beta_c[\sigma] \\
\nonumber
 &=&\sum_\sigma\alpha_{a\mu}\alpha_{b\nu}\frac{1}{2}\beta_{c\rho\sigma}\frac{1}{4!}\varepsilon_1^{\lambda}\varepsilon_2^{\chi}\varepsilon_3^{\varphi}\varepsilon_4^{\xi}\left(\delta_\lambda^\mu\delta_\chi^\nu\delta_\varphi^\rho\delta_\xi^\sigma+(23)\right)\\
&=&\frac{1}{300}\sum_\sigma\sum_{\varepsilon_1,\varepsilon_2,\Delta\in\sigma}\alpha_a[\varepsilon_1]\alpha_b[\varepsilon_2]\beta_{c}[\Delta]W[\varepsilon_1,\varepsilon_2,\Delta]\,.
\end{eqnarray}
As before, the first step is the application of our formula (\ref{eq:DiscretizationMainResult}), and the second step is the expansion in terms of a basis and application of the biorthogonality relation (\ref{eq:BiorthogonalityRelation}). The third step is again the introduction of the summation over all possible choices of four vectors $\varepsilon_1,\dots ,\varepsilon_4$, together with appropriate normalization. However, since the $2$-form $\beta_c$ corresponds to a triangle, two of the four vectors $\varepsilon$ have to form a triangle $\Delta$. Therefore, instead of the sum over four vectors, the $4$-simplex is decomposed into two vectors and a triangle, so we need to sum over those, and normalize with a factor of $300$. This corresponds to the case (5) of the combinatorial analysis (see Table \ref{TabelaJedan} and Figure \ref{SlikaTri}).

The third class of integrals in (\ref{eq:FourTypesOfIntegrals}) is evaluated as follows:
\begin{eqnarray} \label{eq:BetaBetaIntegral}
\nonumber
\int \beta_a\wedge \beta_b&=& \sum_\sigma \beta_a\wedge \beta_b[\sigma] \\
\nonumber
&=&\sum_\sigma\frac{1}{2}\beta_{a\mu\nu}\frac{1}{2}\beta_{b\rho\sigma}\frac{1}{4!}\varepsilon_1^{\lambda}\varepsilon_2^{\chi}\varepsilon_3^{\varphi}\varepsilon_4^{\xi}\left(\delta_\lambda^\mu\delta_\chi^\nu\delta_\varphi^\rho\delta_\xi^\sigma+(23)\right)\\
&=&\frac{1}{30}\sum_\sigma\sum_{\Delta_1,\Delta_2\in\sigma}\beta_a[\Delta_1]\beta_b[\Delta_2]W[\Delta_1,\Delta_2]\,.
\end{eqnarray}
In this case, after applying the formula (\ref{eq:DiscretizationMainResult}) and the biorthogonality relation (\ref{eq:BiorthogonalityRelation}), we consider the decomposition of the $4$-simplex into two triangles $\Delta_1,\Delta_2$, which correspond to the two $2$-forms $\beta_a$ and $\beta_b$. The sum over all possible choices of the two triangles, and the normalization factor $30$, correspond to the case (7) of the combinatorial analysis (see Table \ref{TabelaJedan} and the second diagram in Figure \ref{SlikaCetiri}).

Finally, the fourth class of integrals in (\ref{eq:FourTypesOfIntegrals}) is evaluated in a similar way:
\begin{eqnarray} \label{eq:AlphaGammaIntegral}
  \nonumber
  \int\alpha_a\wedge\gamma_b &=& \sum_\sigma \alpha_a\wedge\gamma_b[\sigma] \\
\nonumber
&=&\sum_\sigma\alpha_{a\mu}\frac{1}{3!}\gamma_{b\nu\rho\sigma}\frac{1}{4!}\varepsilon_1^{\lambda}\varepsilon_2^{\chi}\varepsilon_3^{\varphi}\varepsilon_4^{\xi}\left(\delta_\lambda^\mu\delta_\chi^\nu\delta_\varphi^\rho\delta_\xi^\sigma+(23)\right) \\
&=&\frac{1}{20}\sum_\sigma\sum_{\varepsilon,\tau\in\sigma}\alpha_a[\varepsilon]\gamma_b[\tau]W[\varepsilon,\tau]\,.
\end{eqnarray}
This class corresponds to the decomposition of a $4$-simplex into a tetrahedron and one vector, in line with the case (6) of the combinatorial analysis (see Table \ref{TabelaJedan} and the first diagram in Figure \ref{SlikaCetiri}).

In addition to the classes of integrals (\ref{eq:FourTypesOfIntegrals}), an action may feature one or more scalar fields as well. The scalar fields correspond to $0$-forms, which are naturally associated to vertices of the triangulation, which means that the intergation of a scalar field over the vertex is trivial. Namely, the value of the scalar field at a vertex is formally equal to the contraction of a $0$-form and a tensor of type $\binom{0}{0}$. Therefore, the evaluation of the scalar fields on a triangulation is independent of the integration, so we have the general equation for $n$ scalar fields $\phi_a$ ($a\in \{ 1,\dots ,n \}$):
\begin{equation} \label{eq:ScalarTypeIntegral}
\int \phi_1\dots \phi_n \, \delta_a=\sum_\sigma \delta_a[\sigma]\frac{1}{5^n}\sum_{v_1,\dots ,v_n\in\sigma}\phi_1[v_1]\dots \phi_n[v_n]\,.
\end{equation}
Here $\delta_a$ is some given $4$-form, and the normalization factor $5^n$ accounts  for the fact that every scalar field can be evaluated at every vertex.

Let us now give a definition for the functions $W$ which keep track of the signs in the four integrals (\ref{eq:fourAlphaIntegral}), (\ref{eq:TwoAlphaBetaIntergal}), (\ref{eq:BetaBetaIntegral}) and (\ref{eq:AlphaGammaIntegral}). In order to do so, we need an invariant way to evaluate the 4-volume (\ref{eq:FourSimplexFourVolumeDef}) of a 4-simplex, expressed in terms of the tetrads:
\begin{equation} \label{eq:IntegralOverFourTetrads}
\int_\sigma \frac{1}{4!}\varepsilon_{abcd}e^a\wedge e^b\wedge e^c\wedge e^d= - \int_\sigma e\, d^4x = -V(\sigma)\,.
\end{equation}
The minus sign is a consequence of the Minkowski signature. Also, note that in spaces with Minkowski signature there are two different notions of a $4$-volume, namely the real-valued $4$-volume (\ref{eq:FourSimplexFourVolumeDef}), and the complex-valued $4$-volume ${}^{(4)}V(\sigma)$, defined in terms of the square root of the appropriate Cayley-Menger determinant (see Appendix \ref{AppA}). The relation between them is ${}^{(4)}V(\sigma) = i V(\sigma)$.

On the other hand, the integral of the four tetrads can be evaluated explicitly, in terms of a choice of edge vectors $\varepsilon_1,\dots , \varepsilon_4$ for a given $4$-simplex. Namely, the tetrad $1$-forms have one special feature. Remembering that the tetrads are $\mathfrak{h}$-valued, evaluating the tensor product of two tetrad $1$-forms using the bilinear form $\killing{\_\,}{\_}_\mathfrak{h}$, we have:
\begin{equation}
\langle e \otimes e \rangle_\mathfrak{h} = e^a{}_\mu e^b{}_\nu \, \rmd x^\mu \otimes \rmd x^\nu \, \killing{t_a}{t_b}_\mathfrak{h} = \eta_{ab} e^a{}_\mu e^b{}_\nu \, \rmd x^\mu \otimes \rmd x^\nu =  g_{\mu\nu} \, \rmd x^\mu \otimes \rmd x^\nu = \rmd s^2\,.
\end{equation}
Here we have used the fact that $\killing{t_a}{t_b}_\mathfrak{h} = \eta_{ab}$, see equation (\ref{snbfh}), and the standard relation between the components  of tetrads and of the metric tensor, $g_{\mu\nu} = \eta_{ab} e^a{}_\mu e^b{}_\nu$. One can now contract this with $\varepsilon \otimes \varepsilon$, for a given vector $\varepsilon$. Using the biorthogonality relation (\ref{eq:BiorthogonalityRelation}) one can easily see that
\begin{equation}
\langle e \otimes e \rangle_\mathfrak{h} [\varepsilon \otimes \varepsilon] = g_{\mu\nu} \varepsilon^\mu \varepsilon^\nu = \varepsilon^2\,,
\end{equation}
establishing the relationship between the tetrad fields and the lengths of the edges of a $4$-simplex. Moreover, one can also contract a single tetrad $1$-form with the vector $\varepsilon$,
\begin{equation} \label{eq:tetradAsAtransformationMatrix}
e^{a}[\varepsilon]=e^a{}_{\mu}\varepsilon^\mu=\Gamma^{a}{}_\mu\varepsilon^{\mu}=\varepsilon^a\,.
\end{equation}
Here, the components  $\varepsilon^a $ of the vector are expressed in the locally inertial reference frame, because the components  of the tetrad $e^a{}_\mu$ can be simultaneously understood both as components  of an $\mathfrak{h}$-valued $1$-form, and as the vector representation $\Gamma^{a}{}_\mu$ of the element of $GL(4,\realni)$ group, so that $\Gamma^{a}{}_\mu$ represents  a transformation matrix from a coordinate basis $\partial_\mu$ to the basis of the locally inertial frame. The equation (\ref{eq:tetradAsAtransformationMatrix}) in fact represents  the definition of a locally inertial frame, and the $G$-invariance of the bilinear form $\killing{\_\,}{\_}_\mathfrak{h}$ guarantees that the locally inertial frame is defined up to the action of the Lorentz group $SO(3,1)$.

Coming back to the integral (\ref{eq:IntegralOverFourTetrads}) of the four tetrads, we can now employ (\ref{eq:tetradAsAtransformationMatrix}) to evaluate it in terms of the four linearly independent vectors $\varepsilon_1,\dots , \varepsilon_4$ that span a $4$-simplex $\sigma$. Expressing the vectors in the basis of a locally inertial frame, the integral is evaluated to give the $4$-volume of $\sigma$ in the form
\begin{equation} \label{eq:EquationForW}
\int_\sigma \frac{1}{4!}\varepsilon_{abcd}e^a\wedge e^b\wedge e^c\wedge e^d=\frac{1}{4!}\varepsilon_{abcd}\varepsilon_1^a\varepsilon_2^b\varepsilon_3^c\varepsilon_4^d \, W[\varepsilon_1,\varepsilon_2,\varepsilon_3,\varepsilon_4]\,,
\end{equation}
where the function $W$ keeps track of the order of vectors $\varepsilon_1,\dots , \varepsilon_4$ so that the sign of the $4$-volume is always the same. Thus, comparing (\ref{eq:EquationForW}) with (\ref{eq:IntegralOverFourTetrads}), one obtains the explicit expression for $W$ in the form:
\begin{equation} \label{eq:Wdef1}
W[\varepsilon_1,\varepsilon_2,\varepsilon_3,\varepsilon_4] = -\frac{1}{4!V(\sigma)}\varepsilon_{abcd}\varepsilon_1^a\varepsilon_2^b\varepsilon_3^c\varepsilon_4^d\,.
\end{equation}
One can see that the possible values of $W$ are always from the set $\{ -1,0,1 \}$, where $0$ is obtained for the degenerate case when the vectors $\varepsilon_1,\dots , \varepsilon_4$ are not linearly independent.

Equation (\ref{eq:Wdef1}) represents  the definition of the $W$ function which appears in the integral (\ref{eq:fourAlphaIntegral}) discussed above. The remaining three integrals (\ref{eq:TwoAlphaBetaIntergal}), (\ref{eq:BetaBetaIntegral}) and (\ref{eq:AlphaGammaIntegral}) depend on similar $W$ functions that are defined in terms of (\ref{eq:Wdef1}) as:
\begin{eqnarray}
W[\Delta,\varepsilon_1,\varepsilon_2] = W[\varepsilon_1,\Delta,\varepsilon_2] = W[\varepsilon_1,\varepsilon_2,\Delta] &=& W[\varepsilon_{\Delta1},\varepsilon_{\Delta2},\varepsilon_1,\varepsilon_2]\,,\\
W[\Delta,\tilde{\Delta}]&=& W[\varepsilon_{\Delta1},\varepsilon_{\Delta2},\varepsilon_{\tilde{\Delta}1},\varepsilon_{\tilde{\Delta}2}]\,,\\
W[\varepsilon,\tau] = - W[\tau,\varepsilon] &=& W[\varepsilon,\varepsilon_{\tau1},\varepsilon_{\tau2},\varepsilon_{\tau3}]\,.
\end{eqnarray}
They naturally depend on the arguments  corresponding to a triangle and two edges, or two triangles, or a tetrahedron and one edge, respectively, in line with the combinatorial classification discussed in the previous subsection.

\subsection{\label{secIIIsubsecD}Discretization of the derivative}

In the previous subsection, we have derived and studied the formula (\ref{eq:DiscretizationMainResult}) that prescribes the way to pass from an action defined over a smooth manifold to an action defined over a piecewise-flat manifold. We have applied this formula to the four classes of integrals (\ref{eq:FourTypesOfIntegrals}) with the addition of scalar fields (\ref{eq:ScalarTypeIntegral}). However, the action may also depend on derivatives of fields, which was not covered above. Specifically, the notion of a derivative intrinsically relies on the notion of smoothness, since it follows the elementary definition of a limit of finite differences. Therefore, the question how the derivative passes from a smooth manifold to a piecewise-flat manifold is nontrivial. Generically, in the literature there are many different prescriptions to introduce the notion of a derivative on a lattice structure. Each prescription has its  own advantages and disadvantages, and there is no consensus on a preferred choice. This problem is most obvious in numerical treatments  of differential and partial differential equations \cite{Smith1985,Shashkov1996,MortonMayers2005,Lemeshevsky2016}.

Nevertheless, in the special case in which the fields that enter the action can all be written as differential forms, it turns out that there is a unique and very natural way to define the derivative on a piecewise-flat manifold. Specifically, under the assumption that all fields are explicitly written as differential forms, the only type of derivative that can appear in the Lagrangian $4$-form is the exterior derivative, and due to its  fundamental nilpotency property $\rmd \rmd \equiv 0$ (Poincar\'e lemma), in each term in the Lagrangian it can appear only once. Given all this, one can employ another fundamental result of differential geometry to define the exterior derivative on a piecewice-flat manifold --- the Stokes theorem.

To see how this can be done, first note that a triangulation of a $4$-dimensional manifold consists  of vertices, edges, triangles, tetrahedra and $4$-simplices. Therefore, the derivative of a field that lives on a $k$-simplex ($k \in \{ 1,2,3,4 \}$) can be defined in terms of the field itself, the latter living on $(k-1)$-simplices which constitute the boundary of the $k$-simplex. Since any $k$-simplex $\sigma_k$ is itself a smooth manifold, the Stokes theorem can be written as
\begin{equation} \label{eq:StokesTheorem}
\int_{\sigma_k} {\rm d} \omega =\oint_{\partial\sigma_k} \omega\,,
\end{equation}
where $\omega$ is a $(k-1)$-form. The boundary $\partial\sigma_k$ is then a triangulation of a $(k-1)$-sphere, and one can apply our formula (\ref{eq:DiscretizationMainResult}) to evaluate the right-hand side of (\ref{eq:StokesTheorem}) as:
\begin{equation}
\oint_{\partial\sigma_k} \omega  = \osum_{\sigma_{k-1}\in\partial\sigma_k} \omega[\sigma_{k-1}]\,.
\end{equation}
Since we want the Stokes theorem to hold on a single $k$-simplex, we can \emph{define} the left-hand side of (\ref{eq:StokesTheorem}) as a contraction of a $k$-form $\rmd \omega$ with $\sigma_k$, namely:
\begin{equation}
\int_{\sigma_k}{\rm d} \omega \equiv {\rm d} \omega[\sigma_k]\,.
\end{equation}
Therefore, combining the left-hand side of the Stokes theorem with the right-hand side, we obtain the definition of the contraction of the exterior derivative with $\sigma_k$ as:
\begin{equation} \label{eq:DiscretizationSecondMainResultTemp}
{\rm d} \omega[\sigma_k] = \osum_{\sigma_{k-1}\in\partial\sigma_k} \omega[\sigma_{k-1}]\,.
\end{equation}
The circled sum symbol denotes the sum of elements  over an oriented boundary, so we have to keep track both of the orientation of $\sigma_k$ and the orientation of $\partial\sigma_k$ relative to $\sigma_k$. For that purpose, we introduce another sign function, denoted $z[\sigma_k,\sigma_{k-1}]$. This function is valued in the set $\{ -1,0,1 \}$, where the value $0$ is obtained whenever $\sigma_{k-1} \notin \partial \sigma_k$, and its  explicit definition is given below. Using this, we can rewrite (\ref{eq:DiscretizationSecondMainResultTemp}) into its  final form:
\begin{equation} \label{eq:DiscretizationSecondMainResult}
{\rm d} \omega[\sigma_k] = \sum_{\sigma_{k-1}\in T(\cM)} \omega[\sigma_{k-1}] \, z[\sigma_k,\sigma_{k-1}]\,.
\end{equation}

The formula (\ref{eq:DiscretizationSecondMainResult}) is the second main result of our paper, extending the applicability of (\ref{eq:DiscretizationMainResult}). Namely, using (\ref{eq:DiscretizationSecondMainResult}), we can now discuss additional four classes of integrals, of the following form:
\begin{equation} \label{eq:FourTypesOfIntegralsWithDerivatives}
  \int \gamma \wedge \rmd \phi\,, \qquad \int \beta \wedge \rmd \alpha \,, \qquad
  \int \alpha \wedge \rmd \beta\,, \qquad \int \rmd \gamma\,.
\end{equation}
The integrals can be evaluated by applying the integrals (\ref{eq:fourAlphaIntegral}), (\ref{eq:TwoAlphaBetaIntergal}), (\ref{eq:BetaBetaIntegral}) and (\ref{eq:AlphaGammaIntegral}), and then applying the formula (\ref{eq:DiscretizationSecondMainResult}) in their solutions, at all places where the derivative appears. Specifically, for the first integral in (\ref{eq:FourTypesOfIntegralsWithDerivatives}), we first apply (\ref{eq:AlphaGammaIntegral}) to obtain:
\begin{equation}
\int \gamma \wedge \rmd \phi = \frac{1}{20}\sum_\sigma\sum_{\varepsilon,\tau\in\sigma}\gamma[\tau] \rmd \phi[\varepsilon] W[\tau,\varepsilon]\,.
\end{equation}
Then we apply (\ref{eq:DiscretizationSecondMainResult}) onto the term $\rmd \phi[\varepsilon]$, for the case $\sigma_k \equiv \varepsilon$ and $\sigma_{k-1} \equiv v$, to obtain:
\begin{equation} \label{eq:GammaDPhiIntegral}
\int \gamma \wedge \rmd \phi = \frac{1}{20}\sum_\sigma\sum_{\tau,\varepsilon,v\in\sigma}\gamma[\tau]\phi[v]z[\varepsilon,v]W[\tau,\varepsilon]\,.
\end{equation}
Using this procedure, we have managed to evaluate the integral containing a derivative on a piecewise-flat manifold, in a way consistent with the Stokes theorem. Similarly, we can evaluate the remaining three classes of integrals in (\ref{eq:FourTypesOfIntegralsWithDerivatives}):
\begin{eqnarray}
\int \beta\wedge{\rm d}\alpha&=&\frac{1}{30}\sum_{\sigma}\sum_{\Delta_1,\Delta_2,\varepsilon\in\sigma}\beta[\Delta_1]\alpha[\varepsilon]z[\Delta_2,\varepsilon]W[\Delta_1,\Delta_2]\,,\\
\int \alpha\wedge {\rm d}\beta&=&\frac{1}{20}\sum_{\sigma}\sum_{\varepsilon,\tau,\Delta\in\sigma}\alpha[\varepsilon]\beta[\Delta]z[\tau,\Delta]W[\varepsilon,\tau]\,,\\
\int {\rm d}\gamma&=&\sum_{\sigma}\sum_{\tau\in\sigma}\gamma[\tau]z[\sigma,\tau]\,. \label{eq:dGammaIntegral}
\end{eqnarray}

At this point, the only remaining task is to provide a detailed definition of the sign function $z[\sigma_k,\sigma_{k-1}]$. This is done for each dimension $k \in \{1,2,3,4 \}$ separately. Specifically, for $k=1$ we have $\sigma_1 \equiv \varepsilon$ and $\sigma_0 \equiv v$. Every oriented edge can be represented via its  two boundary vertices $\varepsilon=(v_{\varepsilon1},v_{\varepsilon2})$, where by convention the edge is oriented from the second towards the first vertex. The sign function is then defined as
\begin{equation}
z[\varepsilon,v]=\delta_{v_{\varepsilon1},v}-\delta_{v_{\varepsilon2},v}\,.
\end{equation}
Here $\delta_{vv'}$ is the Kronecker symbol, equal to $1$ if $v \equiv v'$ and $0$ otherwise. We can observe that the sign function is equal to $\pm 1$ if $v$ is an element of $\partial\varepsilon$, while it is zero otherwise.

In case $k=2$, we have $\sigma_2 \equiv \Delta$, and $\sigma_1 \equiv \varepsilon$. Every oriented triangle can be represented via one pair of its  edges $\Delta = (\varepsilon_{\Delta 1}, \varepsilon_{\Delta 2})$, emanating from the same vertex. By convention, the orientation of the triangle is specified by the order of the two edges. The sign function is then defined as:
\begin{equation}
z[\Delta,\varepsilon]=\frac{\varepsilon_{\Delta1}^a\varepsilon_a}{\varepsilon^a\varepsilon_a}\delta_{\varepsilon_{\Delta1},\varepsilon}-\frac{\varepsilon_{\Delta2}^a\varepsilon_a}{\varepsilon^a\varepsilon_a}\delta_{\varepsilon_{\Delta2},\varepsilon}+\frac{(\varepsilon_{\Delta2}-\varepsilon_{\Delta1})^a\varepsilon_a}{\varepsilon^a\varepsilon_a}\delta_{\varepsilon_{\Delta2}-\varepsilon_{\Delta1},\varepsilon}\,.
\end{equation}
The Kronecker symbol for the edges is defined in the same way as for the vertices, ignoring their orientation.

In case $k=3$, we have $\sigma_3 \equiv \tau$, and $\sigma_2 \equiv \Delta$. Every oriented tetrahedron can be represented via a triple of its  edges emanating from the same vertex, or a pair of its  triangles sharing one common edge, $\tau=(\varepsilon_{\tau 1}\varepsilon_{\tau 2},\varepsilon_{\tau 3})=(\Delta_{\tau 1},\Delta_{\tau 2})$, such that $\Delta_{\tau1}=(\varepsilon_{\tau1},\varepsilon_{\tau2})$ and $\Delta_{\tau2}=(\varepsilon_{\tau2},\varepsilon_{\tau3})$. In addition, for notational simplicity, we also introduce the remaining two triangles as $\Delta_{\tau3}=(\varepsilon_{\tau3},\varepsilon_{\tau1})$ and $\Delta_{\tau4}=(\varepsilon_{\tau1}-\varepsilon_{\tau2},\varepsilon_{\tau3}-\varepsilon_{\tau2})$. By convention, the orientation of the tetrahedron is specified by the order of the two triangles $\Delta_{\tau1},\Delta_{\tau 2}$. The sign function is then defined as
\begin{equation}
z[\tau,\Delta]=\sum_{j=1}^4\frac{ \Delta_{\tau j}^{ab} \Delta_{ab}}{\Delta^{ab}\Delta_{ab}}\delta_{\Delta_{\tau j},\Delta}\,, 
\end{equation}
where
\begin{equation}
\Delta^{ab} \equiv \frac{1}{2}\left(\varepsilon_{\Delta 1}^a\varepsilon_{\Delta 2}^b-\varepsilon_{\Delta 1}^b\varepsilon_{\Delta 2}^a\right)\,.
\end{equation}

Finally, in the case $k=4$, we have $\sigma_4 \equiv \sigma$, and $\sigma_3 \equiv \tau$. Every oriented $4$-simplex can be represented via four edges emanating from the same vertex, $\sigma = (\varepsilon_{\sigma 1},\varepsilon_{\sigma 2},\varepsilon_{\sigma 3},\varepsilon_{\sigma 4})$. It contains the following five positively oriented tetrahedra: $\tau_{\sigma 1}=(\varepsilon_{\sigma 1},\varepsilon_{\sigma 2},\varepsilon_{\sigma3})$, $\tau_{\sigma 2}=(\varepsilon_{\sigma1},\varepsilon_{\sigma4},\varepsilon_{\sigma2})$, $\tau_{\sigma 3}=(\varepsilon_{\sigma1},\varepsilon_{\sigma3},\varepsilon_{\sigma4})$, $\tau_{\sigma 4}=(\varepsilon_{\sigma2},\varepsilon_{\sigma4},\varepsilon_{\sigma3})$ and $\tau_{\sigma 5}=(\varepsilon_{\sigma1}-\varepsilon_{\sigma2},\varepsilon_{\sigma2}-\varepsilon_{\sigma3},\varepsilon_{\sigma4}-\varepsilon_{\sigma3})$. The sign function is then defined as
\begin{equation}
z[\sigma,\tau]=\sum_{j=1}^5\frac{\tau_{\sigma j}^{abc} \tau_{abc}}{\tau^{abc}\tau_{abc}}\delta_{\tau_{\sigma j},\tau}\,,
\end{equation} 
where
\begin{equation}
\tau^{abc}=\frac{1}{3!}\varepsilon^{abcd}\varepsilon_{defg}\varepsilon_{\tau 1}^e\varepsilon_{\tau 2}^f\varepsilon_{\tau 3}^g\,.
\end{equation}
This concludes the definition of the sign function.

Note also that every $k$-simplex can be uniquely represented via the $(k+1)$-tuple of its  vertices, $\sigma_k = (v_1,\dots , v_{k+1})$. Given two such $k$-simplices, $A$ and $B$, one can define the general case of the Kronecker symbol $\delta_{A,B}$ as
\begin{eqnarray}
\delta_{A,B}={\rm perm}\left(\delta_{v_{Ai},v_{Bj}}\right)\,,
\end{eqnarray}
where $\rm perm $ denotes the permanent of the $(k+1)\times (k+1) $ matrix constructed from the Kronecker symbols between the vertices of $A$ and $B$. Due to the properties of the permanent, the Kronecker symbol will be automatically zero if one row or column in the matrix is zero. This means that the Kronecker symbol will give a nonzero result only if both $k$-simplices are spanned over the same set of vertices, i.e., if they coincide, irrespective of their relative orientation.

\section{\label{secIV}Discretization of the path integral}

After the discussion of the method to discretize the action given in the previous Section, we now turn to the question of the discretization of the path integral, specifically its  measure.

When passing from a smooth manifold to a piecewise-flat manifold, a crucial assumption was that within any given simplex all fields are constant. This means that the number of degrees of freedom, that the path integral should be taken over, is drastically reduced. Specifically, instead of integrating a given field at every point in spacetime, one should integrate its value only on its corresponding simplices. In contrast to the uncountably infinitely many spacetime points, there are only countably many simplices in a triangulation. Therefore, passing from a smooth manifold to a piecewise-flat manifold reduces the path integral to a countable number of ordinary integrals.

In practice, the measure of the path integral is therefore defined as a product of measures of ordinary integrals. Recalling the notation (\ref{eq:AllFormsNotation}), and noting that $0$-forms live on vertices, $1$-forms live on edges, and so on, we can write the path integral measure as follows:
\begin{eqnarray}\label{defmere}
D\phi=\prod_v d\phi[v]\,,\quad D\alpha=\prod_\varepsilon d\alpha[\varepsilon]\,,\quad
D\beta=\prod_\Delta d\beta[\Delta]\,,\quad
D\gamma=\prod_\tau d\gamma[\tau]\,,\quad
D\delta=\prod_\sigma d\delta[\sigma]\,.
\end{eqnarray}
In this way, the path integral reduces to either a finite or countably infinite number of ordinary integrals.

The above definition is of course not unique, but can be regarded as a natural choice. An alternative choice can be constructed as follows. Given an arbitrary $k$-form $F$, one can always expand it into a basis corresponding to locally inertial coordinate frame,
\begin{equation}
F=\frac{1}{k!}F_{a_1 \dots  a_k}e^{a_1}\wedge \dots  \wedge e^{a_k}\,,
\end{equation}
where $e^a$ are the tetrad $1$-forms defining the locally inertial coordinate frame. One can also introduce a tensor $\sigma_k$ of the type $\binom{k}{0}$ corresponding to the $k$-simplex, and contract it with the basis $k$-form $e^{a_1}\wedge \dots  \wedge e^{a_k}$ to obtain:
\begin{equation}
e^{a_1}\wedge \dots  \wedge e^{a_k}[\sigma_k]=V(\sigma_k) E^{a_1 \dots  a_k}\,,
\end{equation}
where $V(\sigma_k)$ denotes a $k$-dimensional volume of the $k$-simplex described by the tensor $\sigma_k$, while $E^{a_1 \dots  a_k}$ encodes the details of the orientation of the basis with respect to the $k$-simplex. One can exploit this to evaluate the contraction $F[\sigma_k]$ as
\begin{equation}
F[\sigma_k]= \frac{1}{k!}F_{a_1...a_k}(\sigma_k)E^{a_1...a_k}V(\sigma_k)=\tilde{F}_{\sigma_k}V(\sigma_k)\,.
\end{equation}
The purpose of this is to rewrite the contraction $F[\sigma_k]$ as a product of a scale-independent scalar quantity $\tilde{F}_{\sigma_k}$ and the $k$-volume of the $k$-simplex, carrying the information about the scale. Applying this transformation to the measure (\ref{defmere}) we obtain a new definition of the path integral measure:
\begin{equation} \label{eq:defdrugemere}
D\phi=\prod_v d\tilde{\phi}_v\,,\quad D\alpha=\prod_\varepsilon l(\varepsilon)d\tilde{\alpha}_\varepsilon\,,\quad
D\beta=\prod_\Delta A(\Delta)d\tilde{\beta}_\Delta\,,
\quad
D\gamma=\prod_\tau V(\tau)d\tilde{\gamma}_\tau\,,\quad
D\delta=\prod_\sigma {}^{(4)}V(\sigma)d\tilde{\delta}_\sigma\,.
\end{equation}
In this form, the measure has explicit dependence on the scales of all simplices in the triangulation, which depend only on the metric of the manifold (see Appendix \ref{AppA} for explicit definitions). This property can be considered useful, for example in numerical evaluations of the path integral. Nevertheless, given that the relation between (\ref{eq:defdrugemere}) and (\ref{defmere}) is just a change of variables, in the remainder of the paper we will use the original definition (\ref{defmere}).

The next important thing to note regarding the measure (\ref{defmere}) is that the integration is performed over Lie algebra valued quantities, since in the case of the action (\ref{eq:RealisticAction}) all differential forms are algebra-valued, in line with the choice of the $3$-group (\ref{eq:StandardModel3GroupDef}). Therefore, the path integral itself should consist of countably many integrals, each of them being an integral over the Lie algebra corresponding to the integration variable. Since algebra integrals are naturally related to group integrals, let us briefly recall the definition and basic properties of integration over a group.

Given a Lie group $G$, an integral of a function $F: G \to \kompleksni$ over the group is denoted as:
\begin{equation}
I_G[F]=\int_G F(g)\, d\mu(g)\,,
\end{equation}
where $d\mu(g)$ is the Haar measure on the Lie group $G$. In order to explicitly evaluate this integral, one introduces the parameters $\theta^i$ ($i\in \{ 1,\dots ,n \}$, where $n$ is the dimension of the group $G$) and writes the integral in the form:
\begin{equation} \label{eq:ParametrizationOfGroupIntegral}
I_G[F]=\int_{\mathbb{F}^n} F(g(\theta)) \, \Pi(\theta) \, \mu(g(\theta)) \, d^n\theta\,.
\end{equation}
Here, $\mathbb{F}$ is the space of parameters $\theta$, and is typically a set of real numbers $\realni$, a set of complex numbers $\kompleksni$, or a set of Grassmann numbers $\grasmanovi$. The function $\Pi(\theta)$ is the rectangular window-function playing the role of the compact support of the domain of parameters. The function $\mu(g(\theta))$ is the weight function of the Haar measure. Below we will tabulate the explicit choices of $\Pi$ and $\mu$ functions, as well as the set of parameters $\theta$, for each Lie group which contributes to the 3-group
 (\ref{eq:StandardModel3GroupDef}). In addition, we will tabulate the explicit expressions for the Dirac delta function defined on a group. This function satisfies the identity
\begin{equation}
\int_G F(g)\,\delta(gg_0^{-1})\, d\mu(g)=F(g_0)\,,
\end{equation}
for any function $F$, and for any $g_0\in G$. Specifically, choosing this function to be a map from the group to its representation $D$, we can apply the above identity to obtain:
\begin{equation}
D\left(g_0^{-1}\right)=\int_G D(g)\,\delta(gg_0)\,d\mu(g)\,.
\end{equation}
Using (\ref{eq:ParametrizationOfGroupIntegral}), this integral can be rewritten as
\begin{equation} \label{eq:RepInverseElement}
D\left(g_0^{-1}\right)=\int_{\mathbb{F}^n} D(g(\theta)) \, \delta(g(\theta)g_0) \, \Pi(\theta) \, \mu(g(\theta)) \, d^n\theta\,. 
\end{equation}
For the case of compact Lie groups, or Lie groups which contain a compact Lie subgroup, the domain of the parameters is nontrivial, since in general it is possible to choose multiple values of the parameters which correspond to a single group element. This means that the representation of the inverse element (\ref{eq:RepInverseElement}) can be multivalued, unless the domain of integration is limited to a specific subset of $\mathbb{F}^n$. The purpose of the window-function $\Pi(\theta)$ is precisely to restrict the domain of integration to this specific subset.

In addition to the group integral, one can similarly introduce the corresponding algebra integral. For any function $f:\mathfrak{g} \to \kompleksni$, the integral over the algebra is denoted as:
\begin{equation} \label{eq:AlgebraIntegralDef}
I_\mathfrak{g}[f] = \int_\mathfrak{g} f(\underline{g}) d\mu_\g(\underline{g})\,,
\end{equation}
where we use the underlined notation $\underline{g} \in \g$ to distinguish the element of the algebra from the element of the group. Here $d\mu_\g(\underline{g})$ is the measure on the algebra $\g$. One can always introduce the parametrization $\underline{g} = \theta^i t_i$, where $\theta^i$ are $\mathbb{F}$-valued parameters specifying the element $\underline{g} \in \g$, while $t_i$ are the generators of the algebra. In this parametrization, the measure $d\mu_\g(\underline{g})$ becomes trivial, so the integral (\ref{eq:AlgebraIntegralDef}) can be explicitly evaluated as
\begin{equation}
I_\mathfrak{g}[f] = \int_{\mathbb{F}^n} f(\theta^i t_i) \, d^n\theta \,.
\end{equation}

For a certain class of functions on the group and on the algebra, one can establish a relation between the group integral and the algebra integral. Namely, given the functions $F: G \to \kompleksni$ and $f: \mathfrak{g} \to \kompleksni$, we say that they correspond to each other if the following equation holds:
\begin{equation} \label{eq:RelationFandf}
F(e^{\theta^it_i}) \, \mu(e^{\theta^it_i}) = f(\theta^i t_i)\,.
\end{equation}
Here we have used the exponential map parametrization $g(\theta) = e^{\theta^it_i}$ to denote the group element in the function $F$ and in the weight $\mu$ of the Haar measure. Given this, one can establish the relation between the algebra and group integrals as follows. Start from the algebra integral $I_\g[f]$, and substitute the relation (\ref{eq:RelationFandf}):
\begin{equation}
I_\mathfrak{g}[f] = \int_{\mathbb{F}^n} f(\theta^i t_i) \, d^n\theta = \int_{\mathbb{F}^n} F(e^{\theta^it_i}) \, \mu(e^{\theta^it_i}) \, d^n\theta\,.
\end{equation}
Next, note that the domain of integration $\mathbb{F}^n$ can be split into disjoint subsets, defined by the window-function $\Pi(\theta)$ of the group $G$ and by the periodicity of the integrand $F(e^{\theta^it_i}) \, \mu(e^{\theta^it_i}) $ on the right hand side. This means that the right hand side can be integrated to give the same result in each of the $\cN$ disjoint subsets. The main disjoint subset is the support of the window-function $\Pi(\theta)$, which gives us
\begin{equation}
\int_{\mathbb{F}^n} F(e^{\theta^it_i}) \, \mu(e^{\theta^it_i}) \, d^n\theta = \cN \int_{\mathbb{F}^n} \Pi(\theta) F(e^{\theta^it_i}) \, \mu(e^{\theta^it_i}) \, d^n\theta\,.
\end{equation}
On the other hand, the latter integral is nothing but the parametrized version of the integral $I_G[F]$ of $F$ over the group $G$. Therefore, we obtain the following relation between the algebra and group integrals:
\begin{equation} \label{eq:RelationAlgebraGroupInt}
I_\mathfrak{g}[f] \equiv \int_\g f(\underline{g}) \, d\mu_\g(\underline{g})
= \cN \int_G F(g) \, d\mu(g) 
\equiv \cN I_G[F] \,.
\end{equation}
Under the assumption (\ref{eq:RelationFandf}), this property allows us to freely switch between the two types of integrals, which is extremely useful in practical calculations. The overall factor $\cN$ is immaterial, since the expectation value of an observable is evaluated as a ratio of the integrals.

Before we proceed to tabulate the explicit definitions of integrals over the relevant groups, it is important to comment on one more theoretical aspect of these integrals. In the context of finite groups (groups with finite number of elements, denoted $|G|$), one can define the set of orthogonal functions on the group, called characters $\chi^{(\Lambda)}(g)$, labeled with the irreducible representations $\Lambda$ of the group. Also, one can define the Dirac delta function $\delta(g)$, also called the regular representation of the group, as:
\begin{equation}
\delta(g)=\left\{\begin{matrix}
0 &\quad g\neq e\,,\\
|G| &\quad g=e\,.
\end{matrix}\right.
\end{equation}
The decomposition of the function $\delta(g)$ into the basis of characters can be written as
\begin{equation} \label{eq:DiracDeltaDecomposition}
\delta(g)=\sum_\Lambda {\rm dim}\Lambda \,\chi^{(\Lambda)}(g)\,.
\end{equation}
In the context of compact Lie groups, all the above notions can be generalized by virtue of the Peter-Weyl theorem.

Various constructions of spinfoam models present in the literature leverage the Peter-Weyl theorem to perform the decomposition (\ref{eq:DiracDeltaDecomposition}) in order to explicitly evaluate the path integral in each irreducible subspace separately, leading to the notion of a state sum over irreducible representations. This procedure can be explicitly done for the topological $BF$ theory, after which the nontopological parts of the action are implemented as simplicity constraints, deforming the state sum from a topological invariant to a physical state sum. In contrast, when discussing the $3BF$ theory, which is based on the notion of a Lie $3$-group rather than an ordinary Lie group, one cannot repeat this procedure, since the representation theory for $3$-groups is scarcely developed, and the analog of the Peter-Weyl theorem is not known to exist.

In this sense, the explicit evaluation of the path integral in our case cannot be performed in analogy to the spinfoam model constructions. However, given that a $3$-group can be understood as an ordered triple of three groups $(G,H,L)$, we can try to evaluate the path integral using the Peter-Weyl theorem for each of the three groups separately. Unfortunately, this alternative strategy is of limited value. Namely, we are ultimately interested in evaluating expectation values of observables, see (\ref{eq:initialstep}). Given that the observable is a generic function of fields, in order to evaluate the path integral it would be necessary to decompose both this function and the exponent of the action into irreducible components. In practice, though, this is not feasible, so the path integral cannot be explicitly transformed into a state sum over irreducible representations, in contrast to spinfoam models. In such a situation the best course of action is to simply refrain from trying to explicitly evaluate the path integral, so in this paper the resulting quantum gravity model (described in detail in the next Section) is not expressed as a state sum.

Finally, let us turn to the particular choice (\ref{eq:StandardModel3GroupDef}) of the $3$-group, and the particular action (\ref{eq:RealisticAction}) of our model. The fields that feature in the theory can be split into two distinct classes. The first class consists of fields which are valued in only one of the Lie algebras $\mathfrak{g}$, $\mathfrak{h}$ and $\mathfrak{l}$. The second class consists of fields which are valued in various products of subalgebras of $\mathfrak{g}$, $\mathfrak{h}$, $\mathfrak{l}$. To each of those algebras we now assign the corresponding Lie group. Note, however, that since the theory features fermionic fields, the Lorentz group $SO(3,1)$ must be substituted with $SL(2,\kompleksni)$. The fields themselves, the groups they correspond to, and the simplices of the triangulation they live on, are displayed in Tables \ref{TabelaDva} and \ref{TabelaTri}. Specifically, Table \ref{TabelaDva} contains fields belonging to the first class, while Table \ref{TabelaTri} contains fields belonging to the second class.

\begin{table}[H]
\begin{center}
\begin{tabular}{|c|c|c|c|c|c|}
\hline
 & \multicolumn{2}{c|}{$\g$} & $\h$ & \multicolumn{2}{c|}{$\l$} \\
\hline
\ \ simplex\ \ \ & $\quad SL(2,{\mathbb{C}})\quad$ & $\quad SU(3)\times SU(2)\times U(1)\quad$ & $\quad\mathbb{R}^4\quad$ & $\quad\mathbb{C}^4\quad$ & $\quad\mathbb{G}^{64}\times\mathbb{G}^{64}\times\mathbb{G}^{64}\quad$\\
\hline
$\cdot$ & & & & $\phi^A$ & $\psi^A$, $\bar{\psi}_A$\\
\hline
$\dvougao$ & $\omega^{[ab]}$ & $\alpha^\alpha$ & $e^a$ & $\tilde{\lambda}^A$ & $\lambda^A$, $\bar{\lambda}_A$\\
\hline
$\trougao$ & $\lambda_{[ab]}$, $B_{[ab]}$ & $\lambda_\alpha$, $B_\alpha$ & $\beta^a$ & &\\
\hline
$\tetraedar$ & & & & $\tilde{\gamma}^A$ & $\gamma^A$, $\bar{\gamma}_A$\\
\hline
$\dekaedar$ & & & & &\\
\hline
\end{tabular}
\end{center}
\caption{\label{TabelaDva}Distribution of fields from the first class with respect to Lie groups and simplices of the triangulation. Each field is valued in only one of the Lie algebras $\mathfrak{g}$, $\mathfrak{h}$, $\mathfrak{l}$.}
\end{table}

\begin{table}[H]
\begin{center}
\begin{tabular}{|c|c|}
\hline
$\quad SU(3)\times SU(2)\times U(1)\times \mathbb{R}^4\times \mathbb{R}^4\quad$ & $\quad M_{\alpha ab}$, $\zeta^{\alpha ab}\quad$\\
\hline
$\mathbb{R}^4\times\mathbb{R}^4\times\mathbb{C}^4$ & $\Lambda^{abA}$\\
\hline
$\mathbb{R}^4\times\mathbb{R}^4\times\mathbb{R}^4\times\mathbb{C}^4$ & $H^{abcA}$\\
\hline
\end{tabular}
\end{center}
\caption{\label{TabelaTri}Distribution of the fields from the second class with respect to the corresponding Lie groups. Each field lives on the vertices of the triangulation.}
\end{table}

Looking at the list of Lie groups in the Table \ref{TabelaDva}, we can observe that all Lie groups are products of the following elementary groups:
\begin{equation} \label{eq:BuildingBlockGroups}
U(1)\,, \qquad SU(2)\,, \qquad SU(3)\,, \qquad SL(2,\kompleksni)\,, \qquad \realni\,, \qquad \kompleksni\,, \qquad \grasmanovi\,.
\end{equation}
As a final step, let us give explicit definitions of the window-function $\Pi(\theta)$, weight of the Haar measure $\mu(g(\theta))$, the expression for the Dirac delta function, as well as the parametrization choice, for each of these groups:
\newpage
\begin{itemize}
\item Group $U(1)$

  The algebra is parametrized with a single parameter $\theta$. The weight of the Haar measure is constant:
\begin{equation} \label{eq:GroupPropertiesBeginning}
d\mu(g)=\frac{1}{2\pi}d\theta\,.
\end{equation}
The window-function is defined as:
\begin{equation}
\Pi(\theta)={\rm H}(\theta+\pi){\rm H}(\pi-\theta)\,,
\end{equation}
where ${\rm H}(x)$ denotes the Heaviside step function.
The Dirac delta function is:
\begin{equation}
\delta(g)=2\pi\sum_{k\in\celi}\delta(\theta+2k\pi)\,.
\end{equation}

\item Group $SU(2)$

  The algebra is parametrized with three parameters $\theta^\alpha$, but the weight, window-function and the Dirac delta depend only on the invariant parameter that corresponds to the radial spherical coordinate $r=\sqrt{\theta_\alpha\theta^\alpha}$. The weight of the Haar measure, the window-function and the Dirac delta are given as:
\begin{equation}
d\mu(g)=\frac{1}{2\pi}\sin^2\left(\frac{r}{2}\right)dr\,,
\qquad
\Pi(r)={\rm H}(r){\rm H}(2\pi-r)\,,
\qquad
\delta(g)=\frac{4\pi}{\sin^2\left(\frac{r}{2}\right)}\sum_{k\in\celi}\delta(r+4k\pi)\,.
\end{equation}


\item Group $SU(3)$

  Group $SU(3)$ has eight parameters which live in a space of topology $S^3\times S^5$, so the parameters are chosen in correspondence with this topology \cite{radzaSU3}. However, the weight, window-function and Dirac delta depend only on the following two invariant parameters:
\begin{equation}
\varphi=\frac{2}{\sqrt{3}}\sqrt{\sigma_1}\sin\left(\frac{1}{3}{\rm arctg}\left(\frac{\sqrt{3}\sigma_2}{\sqrt{\sigma_1^3-3\sigma_2^2}}\right)\right)\,, \qquad\xi=\sqrt{\sigma_1}\cos\left(\frac{1}{3}{\rm arctg}\left(\frac{\sqrt{3}\sigma_2}{\sqrt{\sigma_1^3-3\sigma_2^2}}\right)\right)\,,
\end{equation}
where
\begin{equation}
\sigma_1=\theta_\alpha \theta^{\alpha}\,,\qquad \sigma_2=d_{\alpha\beta\gamma}\theta^{\alpha}\theta^{\beta}\theta^{\gamma}\,.
\end{equation}
The coefficients $d_{\alpha\beta\gamma}$ feature in the anticommutation relations:
\begin{equation}
\{\tau_\alpha,\tau_\beta\}=d_{\alpha\beta\gamma}\tau^{\gamma}\,.
\end{equation}
The weight, window-function and Dirac delta are given as:
\begin{equation}
d\mu(g)=\frac{2}{3\pi^2}\sin^2\left(\frac{\varphi}{2}\right)\sin^2\left(\frac{\varphi+3\xi}{4}\right)\sin^2\left(\frac{\varphi-3\xi}{4}\right)d\varphi d\xi\,,
\end{equation}
\begin{equation}
\Pi(\xi,\varphi)={\rm H}(\xi+2\pi){\rm H}(2\pi-\xi){\rm H}(\varphi+2\pi){\rm H}(2\pi-\varphi)\,,
\end{equation}
\begin{equation}
\delta(g)=\frac{3\pi^2}{2\sin^2\left(\frac{\varphi}{2}\right)\sin^2\left(\frac{\varphi+3\xi}{4}\right)\sin^2\left(\frac{\varphi-3\xi}{4}\right)}\sum_{k,l\in\celi}\delta(\xi+2k\pi)\delta(\xi+\varphi+4l\pi)\,.
\end{equation}

\item Group $SL(2,{\mathbb{C}})$

The algebra is parametrized with six parameters $\omega_{[ab]}$ (here $[ab] \in \{ 01,\dots,23 \}$), and it can be identified with the complexified $\mathfrak{su}(2)$ algebra. The latter algebra can be decomposed into a direct sum  of two $\mathfrak{su}(2)$ algebras parametrized with $\theta_i+i\rho_i$ and $\theta_i-i\rho_i$, where $\theta_i=\varepsilon_{ijk}\omega^{[jk]}$ and $\rho_i=\omega_{[0i]}$ (here $i,j,k\in\{ 1,2,3 \}$). The weight, window-function and the Dirac delta are functions of invariant parameters
\begin{equation}  
\sqrt{a+ib} \,, \qquad \sqrt{a-ib}\,, \qquad\text{ where }\qquad a\equiv\theta_i\theta^i-\rho_i\rho^i\,,\qquad b\equiv 2\theta_i\rho^i\,.
\end{equation}
The expressions for the weight, window-function and Dirac delta become:
\begin{equation}
d\mu(g)=\frac{\left|\sin^2\left(\frac{1}{2}\sqrt{a-ib}\right)\right|^2}{8\pi^2\sqrt{a^2+b^2}}dadb\,,
\end{equation}
\begin{equation}
\Pi(a,b)={\rm H}\left(2\pi-\left|{\rm Re}\left(\sqrt{a-ib}\right)\right|\right)\,,
\end{equation}
\begin{equation}
\delta(g)=4\pi^2\frac{\delta\left({\rm Im}\left(\sqrt{a-ib}\right)\right)}{\left|\sin^2\left(\frac{1}{2}\sqrt{a-ib}\right)\right|^2}\sum_{k\in\celi}\delta\left({\rm Re}\left(\sqrt{a-ib}\right)+4k\pi\right)\,.
\end{equation}

\item Groups ${\mathbb{R}}$, ${\mathbb{C}}$ and ${\mathbb{G}}$

First let us note that these spaces are considered to be groups with respect to addition rather than multiplication, and that the group and its algebra coincide. They are parametrized with a single real-valued, complex-valued and Grassmann-valued parameter $\theta$, respectively. The weight, window-function and Dirac delta are given as:
\begin{equation} \label{eq:GroupPropertiesEnd}
d\mu(g)=d\theta\,,\qquad \Pi(\theta)=1\,,\qquad \delta(g)=\delta(\theta)\,.
\end{equation}
Note that in the Grassmann case the basic integration rules are given via the Berezin integrals:
\begin{equation}
\int_\mathbb{G} d\theta=0\,,\qquad\int_\mathbb{G}\theta d\theta=1\,.
\end{equation}
Given this, the Dirac delta function must be equal to its argument, $\delta(\theta)=\theta$, since we require it to satisfy the usual definition via the integral:
\begin{equation}
\qquad\int_\mathbb{G} \delta(\theta)\, d\theta=1\,.
\end{equation}
\end{itemize}

This concludes the analysis of the discretization of the path integral measure. We have passed from a traditional (ill-defined) path integral measure featuring uncountably infinitely many integrals to a discretized (well-defined) path integral measure (\ref{defmere}) featuring only finitely many or at most countably infinitely many integrals. This is one of the main benefits of introducing a triangulation, i.e., working on a piecewise-flat spacetime manifold. Moreover, as a feature of the approach based on a $3$-group, each field that appears in the path integral measure is algebra-valued for some Lie algebra. In cases when one can freely switch between the integral over the algebra and the integral over its corresponding group, one can define the domains of integration for all fields in the path integral measure to be a corresponding Lie group. Finally, all integrals for all Lie groups (\ref{eq:BuildingBlockGroups}) appearing in the Standard Model $3$-group (\ref{eq:StandardModel3GroupDef}) have been explicitly defined. Taken together, these results represent a method to provide a rigorous definition for the path integral.

\section{\label{secV}Application to the Standard Model 3BF action}

After the development of the method to define both the classical action and the path integral over a piecewise-flat manifold discussed in the previous two Sections, we now finally put all pieces together, and apply the method to the case of the Standard Model $3BF$ action. The action is specified in detail by (\ref{eq:RealisticAction}) and (\ref{eq:3BFforStandardModel})-(\ref{eq:CCconstraint}), and we apply our two main results (\ref{eq:DiscretizationMainResult}) and (\ref{eq:DiscretizationSecondMainResult}) to define it on a triangulation. All terms in the action fall into the classes of integrals (\ref{eq:FourTypesOfIntegralsWithDerivatives}) and (\ref{eq:FourTypesOfIntegrals}), depending on whether they contain an exterior derivative or not, respectively. Therefore, one can apply equations (\ref{eq:fourAlphaIntegral})-(\ref{eq:ScalarTypeIntegral}) and (\ref{eq:GammaDPhiIntegral})-(\ref{eq:dGammaIntegral}), and pass from the smooth manifold to the piecewise-flat manifold for all terms in the action.

Additionally, the path integral (\ref{eq:initialstep}) can be defined precisely on a piecewise-flat manifold by introducing the measure (\ref{defmere}), where the domain of each integration is a combination of Lie groups (\ref{eq:BuildingBlockGroups}). The explicit choices for the parametrization, weight function, window-function, and Dirac delta are listed in (\ref{eq:GroupPropertiesBeginning})-(\ref{eq:GroupPropertiesEnd}). Alternatively, we can employ the relation (\ref{eq:RelationAlgebraGroupInt}) and simplify the integration by passing from group integrals to the corresponding algebra integrals.

Thus, as the third main result of our paper, we finally obtain the explicit definition for the expectation value of an observable in the model of quantum gravity based on the Standard Model $3$-group and the corresponding constrained $3BF$ action:
{\small
\begin{eqnarray} 
\nonumber
\langle F\rangle & = & \frac{{\cal N}}{Z}\Bigg(\prod_v \int d\phi[v] d\psi[v] d\bar{\psi}[v] d\Lambda[v] d\zeta[v] dH[v] dM[v]
\prod_{\varepsilon}\int d\alpha[\varepsilon] d\omega[\varepsilon] de[\varepsilon] d\tilde{\lambda}[\varepsilon] d\lambda[\varepsilon] d\bar{\lambda}[\varepsilon]\\
\nonumber 
&&\qquad\qquad\prod_{\Delta}\int dB[\Delta] d\beta[\Delta] d\lambda[\Delta]\prod_\tau\int d\tilde{\gamma}[\tau] d\gamma[\tau] d\bar{\gamma}[\tau]\Bigg)  \prod_\sigma F\Big(\phi_k[\sigma]\Big)\\
\nonumber
&&\prod_\sigma\left\{\exp\Big[\frac{i}{30}\sum_{\Delta_1\in\sigma}B_\alpha[\Delta_1]\Big(\sum_{\Delta_2,\varepsilon\in\sigma}\alpha^{\alpha}[\varepsilon]z[\Delta_2,\varepsilon]W[\Delta_1,\Delta_2]+\frac{1}{10}\sum_{\varepsilon_1,\varepsilon_2\in\sigma}\alpha^{\beta}[\varepsilon_1]\alpha^{\gamma}[\varepsilon_2]f_{\beta\gamma}{}^\alpha W[\Delta_1,\varepsilon_1,\varepsilon_2]\Big)\Big]\right.\\
\nonumber
&&\exp\Big[\frac{i}{30}\sum_{\Delta_1\in\sigma}B_{[ab]}[\Delta_1]\Big(\sum_{\Delta_2,\varepsilon\in\sigma}\omega^{[ab]}[\varepsilon]z[\Delta_2,\varepsilon]W[\Delta_1,\Delta_2]+\frac{1}{10}\sum_{\varepsilon_1,\varepsilon_2\in\sigma}\omega^{[cd]}[\varepsilon_1]\omega^{[ef]}[\varepsilon_2]f_{[cd][ef]}{}^{[ab]}W[\Delta_1,\varepsilon_1,\varepsilon_2]\Big)\Big]\\
\nonumber
&&\exp\Big[\frac{i}{20}\sum_{\varepsilon_1\in\sigma}\varepsilon_{1a}\Big(\sum_{\tau,\Delta\in\sigma}\beta^a[\Delta]z[\tau,\Delta]W[\varepsilon_1,\tau]+\frac{1}{15}\sum_{\varepsilon_2,\Delta\in\sigma}\omega^{[cd]}[\varepsilon_2]\beta^b[\Delta]\triangleright_{[cd]b}{}^a W[\varepsilon_1,\varepsilon_2,\Delta]\Big)\Big]\\
\nonumber
&&\exp\Big[\frac{i}{5}\sum_{v\in\sigma}\phi^A[v]\Big(\sum_{\tau\in\sigma}\tilde{\gamma}_A[\tau]z[\sigma,\tau]+\frac{1}{20}\sum_{\varepsilon,\tau\in\sigma}W[\varepsilon,\tau]\tilde{\gamma}^B[\tau]\alpha^{\alpha}[\varepsilon]\triangleright_{\alpha BA}\Big)\Big]\\
\nonumber
&&\exp\Big[\frac{i}{5}\sum_{v\in\sigma}\bar{\psi}_A[v]\Big(\sum_{\tau\in\sigma}\gamma^A[\tau]z[\sigma,\tau]+\frac{1}{20}\sum_{\varepsilon,\tau\in\sigma}W[\varepsilon,\tau]\gamma^B[\tau]\Big(\alpha^{\alpha}[\varepsilon]\triangleright_{\alpha B}{}^{A}+\omega^{[ab]}[\varepsilon]\triangleright_{[ab]B}{}^{A}\Big)\Big)\Big]\\
\nonumber
&&\exp\Big[\frac{i}{5}\Big(\sum_{\tau\in\sigma}\bar{\gamma}_A[\tau]z[\sigma,\tau]-\frac{1}{20}\sum_{\varepsilon,\tau\in\sigma}W[\varepsilon,\tau]\bar{\gamma}^B[\tau]\Big(\alpha^{\alpha}[\varepsilon]\triangleright_{\alpha BA}+\omega^{[ab]}[\varepsilon]\triangleright_{[ab]BA}\Big)\Big)\sum_{v\in\sigma}\psi^A[v]\Big]\\
\nonumber
&&\exp\Big[\frac{i}{30}\sum_{\Delta_1\in\sigma}\lambda^\alpha[\Delta_1]\Big(\sum_{\Delta_2\in\sigma}B_\alpha[\Delta_2]W[\Delta_1,\Delta_2]-\frac{12}{50}C_{\alpha\beta}\sum_{v\in\sigma}M^{\beta}{}_{ab}[v]\sum_{\varepsilon_1,\varepsilon_2\in\sigma}\varepsilon_1^a\varepsilon_2^b W[\Delta_1,\varepsilon_1,\varepsilon_2]\Big)\Big]\\
\nonumber
&&\exp\Big[\frac{i}{30}\sum_{\Delta_1\in\sigma}\lambda^{[ab]}[\Delta_1]\Big(\frac{1}{80\pi l_p^2}\varepsilon_{[ab]cd} \sum_{\varepsilon_1,\varepsilon_2\in\sigma}\varepsilon_1^c\varepsilon_2^d W[\Delta_1,\varepsilon_1,\varepsilon_2]-\sum_{\Delta_2\in\sigma}B_{[ab]}[\Delta_2]W[\Delta_1,\Delta_2]\Big)\Big]\\
\nonumber
&&\exp\Big[\frac{i}{20}\sum_{\varepsilon_1\in\sigma}\tilde{\lambda}^A[\varepsilon_1]\Big(\sum_{\tau\in\sigma}\tilde{\gamma}_A[\tau]W[\varepsilon_1,\tau]-\frac{1}{750}\sum_{v\in\sigma}H_{abcA}[v]\sum_{\varepsilon_2,\varepsilon_3,\varepsilon_4\in\sigma}\varepsilon_2^a\varepsilon_3^b\varepsilon_4^c W[\varepsilon_1,\varepsilon_2,\varepsilon_3,\varepsilon_4]\Big)\Big]\\
\nonumber
&&\exp\Big[\frac{i}{20}\sum_{\varepsilon_1\in\sigma}\bar{\lambda}_A[\varepsilon_1]\Big(\sum_{\tau\in\sigma}\gamma^A[\tau]W[\varepsilon_1,\tau]+\frac{i}{4500}\varepsilon_{abcd}\sum_{v\in\sigma}\sum_{\varepsilon_2,\varepsilon_3,\varepsilon_4\in\sigma}\varepsilon_2^a\varepsilon_3^b\varepsilon_4^c(\gamma^d\psi[v])^A W[\varepsilon_1,\varepsilon_2,\varepsilon_3,\varepsilon_4]\Big)\Big]\\
\nonumber
&&\exp\Big[-\frac{i}{20}\sum_{\varepsilon_4\in\sigma}\Big(\sum_{\tau\in\sigma}\bar{\gamma}_A[\tau]W[\tau,\varepsilon_4]-\frac{i}{4500}\varepsilon_{abcd}\sum_{v\in\sigma}\sum_{\varepsilon_1,\varepsilon_2,\varepsilon_3\in\sigma}\varepsilon_1^a\varepsilon_2^b\varepsilon_3^c(\bar{\psi}[v]\gamma^d)_A W[\varepsilon_1,\varepsilon_2,\varepsilon_3,\varepsilon_4]\Big)\lambda^A[\varepsilon_4]\Big]\\
\nonumber
&&\exp\Big[-\frac{2\pi l_p^2}{7500}\varepsilon_{abcd}\sum_{v_1,v_2,\varepsilon_1,\varepsilon_2,\Delta\in\sigma}\bar{\psi}_A[v_1](\gamma_5\gamma^a\psi)^A[v_2]\varepsilon_1^b\varepsilon_2^c\beta^d[\Delta]W[\varepsilon_1,\varepsilon_2,\Delta]\Big]\\
\nonumber
&&\exp\Big[\frac{i}{1500}\sum_{v_1\in\sigma}\zeta_\alpha{}^{ab}[v_1]\sum_{\varepsilon_1,\varepsilon_2\in\sigma}\varepsilon_1^c\varepsilon_2^d\Big(\sum_{\varepsilon_3,\varepsilon_4\in\sigma}W[\varepsilon_1,\varepsilon_2,\varepsilon_3,\varepsilon_4]\Big(\frac{1}{50}\varepsilon_{cdef}\varepsilon_3^e\varepsilon_4^f\sum_{v_2\in\sigma}M^\alpha{}_{ab}[v_2]-\frac{1}{10}f_{\beta\gamma}{}^{\alpha}\alpha^{\beta}[\varepsilon_3]\alpha^{\gamma}[\varepsilon_4]\Big)\\
\nonumber
&&\qquad\qquad-\sum_{\Delta,\varepsilon_3\in\sigma}\alpha^{\alpha}[\varepsilon_3]z[\Delta,\varepsilon_3]W[\Delta,\varepsilon_1,\varepsilon_2]\eta_{ac}\eta_{bd}\Big)\Big]\\
\nonumber
&&\exp\Big[\frac{i}{3000}\sum_{\varepsilon_1,\varepsilon_2,\varepsilon_3,\varepsilon_4,v\in\sigma}\Lambda^{abA}[\varepsilon_1]W[\varepsilon_1,\varepsilon_2,\varepsilon_3,\varepsilon_4]\Big(\frac{1}{5}H_{abcA}[v]\varepsilon^{cdef}\varepsilon_{2d}\varepsilon_{3e}\varepsilon_{4f}\\
\nonumber
&&\qquad\qquad+\Big(\phi_A[v]z[\varepsilon_2,v]+\frac{1}{5}\phi^B[v]\alpha^{\alpha}[\varepsilon_2]\triangleright_{\alpha BA}\Big)\varepsilon_{3a}\varepsilon_{4b}\Big)\Big]\\
\nonumber
&&\exp\Big[-\frac{i}{36000}\sum_{\varepsilon_1,\varepsilon_2,\varepsilon_3,\varepsilon_4\in\sigma}\varepsilon_{abcd}\varepsilon_1^a\varepsilon_2^b\varepsilon_3^c\varepsilon_4^d\Big(\frac{\Lambda}{8\pi l_p^2}+\chi v^4-\frac{2\chi v^2}{25}\sum_{v_1,v_2\in\sigma}\phi_A[v_1]\phi^A[v_2]\\
  &&\qquad\qquad+\left.\frac{1}{125}Y_{ABC}\sum_{v_1,v_2,v_3\in\sigma}\bar{\psi}^A[v_1]\psi^B[v_2]\phi^C[v_3]+\frac{\chi}{625}\sum_{v_1,v_2,v_3,v_4\in\sigma}\phi_A[v_1]\phi^A[v_2]\phi_B[v_3]\phi^B[v_4]\Big)\right\}\,. \label{eq:ThirdMainResult}
\end{eqnarray}
}%

Equation (\ref{eq:ThirdMainResult}), while representing the full model of quantum gravity, may look intimidating. Nevertheless, there is a certain amount of structure within the equation. This structure is partly inherited from the corresponding structure of the classical action (\ref{eq:RealisticAction}), while partly it is a consequence of the combinatorial structure of the underlying triangulation. Let us discuss both of these in turn.

\subsection{General structure of the model}

In the programme of constructing spinfoam models, one typically starts from a topological $BF$ action and performs the quantization first by constructing the corresponding topological invariant, and afterwards by deforming this invariant in order to implement the simplicity constraints, which encode the nontopological part of the action. In our case, the situation is similar --- the action (\ref{eq:RealisticAction}) features a topological $3BF$ part, as well as a number of simplicity constraints. The quantization of the topological part has been performed in \cite{Radenkovic2022b} and gives rise to a genuine topological invariant of $4$-manifolds based on a $3$-group. Moreover, it also gives rise to a proper TQFT, constructed in \cite{Radenkovic2025}. However, in our case, it is not possible to directly deform this invariant in order to implement simplicity constraints. The reason is that the invariant has been constructed by explicit integration of the $3BF$ action over some variables (Lagrange multipliers) which correspond to the dynamical fields in our model. In particular, the construction of the topological invariant along the lines of \cite{Radenkovic2022b} starting from (\ref{eq:3BFforStandardModel}) would imply the integration over the tetrad, scalar and fermion fields. This is unacceptable, given that these fields explicitly feature not just in the simplicity constraint terms, but also in the observable whose expectation value is being evaluated.

To circumvent this issue, instead of assuming that the physical observables can depend on all fields from the full configuration space (given by the union of (\ref{nedinpolja3bf}) and (\ref{dinpolja3bf})), we can restrict the observables to depend only on the dynamical part (\ref{dinpolja3bf}) of the configuration space of the theory. Given this restriction, one can indeed perform a construction of a topological invariant similar to the construction performed in \cite{Radenkovic2022b}, but in our case the choice of the variables to be integrated over will be different. Technically, the action (\ref{eq:3BFaction}) can be integrated either over the variables $B$, $C$ and $D$, as was done in \cite{Radenkovic2022b}, or alternatively over $B$, $\beta$ and $\gamma$, as we opt to do in this paper. This will again lead to the same topological invariant, but now expressed in terms of more convenient set of variables.

Specifically, let us perform the explicit integration in (\ref{eq:ThirdMainResult}) over the variables $B_\alpha[\Delta]$, $B_{[ab]}[\Delta]$, $\beta^a[\Delta]$, $\tilde{\gamma}^A[\tau]$, $\bar{\gamma}_A[\tau]$, $\gamma^A[\tau]$, $\Lambda^{abA}[\varepsilon]$ and $\zeta_{\alpha ab}[v]$, with the assumption that the observable $F$ does not depend on them. To that end, we apply the following identity:
\begin{equation}
\int\mathrm{exp}\Big(i\sum_\sigma\sum_{k\in\sigma}f[k]F[k,\sigma]\Big)\prod_k df[k]
= \prod_k \int df[k]\, \mathrm{exp}\Big(i f[k] \sum_{\sigma \ni k} F[k,\sigma]\Big)
= \mathcal{N}\prod_{k}\delta\Big(\sum_{\sigma\ni k}F[k,\sigma]\Big)\,.
\end{equation}
The identity is applied by identifying the integration variable $f[k]$ with each of the above variables in turn, and by identifying $F[k,\sigma]$ with the expression that is being multiplied by $f[k]$ in (\ref{eq:ThirdMainResult}), for all vertices, edges, triangles and tetrahedra ($k \in \{v,\varepsilon,\Delta,\tau \}$). As a result, the full expression (\ref{eq:ThirdMainResult}) can be rewritten in the following form, emphasizing its general structure:
\begin{eqnarray}\label{oblikF}
  \langle F\rangle&=&\frac{{\cal N}}{Z}\int D\phi_k\,F(\phi_k)\,
  \mathrm{TI}\Big[\mathcal{F}[\phi_k]-\lambda[\phi_k]\Big]\, \mathrm{INT}\Big[\lambda[\phi_k]-G[\phi_k]\Big]\,\mathrm{SC}\Big[\phi_k\Big]\,.
\end{eqnarray}
Here the $\rm TI[\dots ]$ term represents the topological invariant term, analogous to the invariant defined in \cite{Radenkovic2022b}, but now keeping the dynamical variables:
{\small
\begin{eqnarray}
\nonumber
&& {\rm TI}\Big[\mathcal{F}[\phi_k]-\lambda[\phi_k]\Big] = \\
\nonumber
&& \qquad   \prod_{\Delta_1}\delta\Big[\sum_{\sigma\ni\Delta_1}\Big(\sum_{\Delta_2,\varepsilon\in\sigma}\alpha^\alpha[\varepsilon]z[\Delta_2,\varepsilon]W[\Delta_1,\Delta_2]+\frac{1}{10}\sum_{\varepsilon_1,\varepsilon_2\in\sigma}\alpha^{\beta}[\varepsilon_1]\alpha^{\gamma}[\varepsilon_2]f_{\beta\gamma}{}^{\alpha}W[\Delta_1,\varepsilon_1,\varepsilon_2]+\sum_{\Delta_2\in\sigma}\lambda^\alpha[\Delta_2]W[\Delta_1,\Delta_2]\Big)\Big]\\
\nonumber
&& \qquad\prod_{\Delta_1}\delta\Big[\sum_{\sigma\ni\Delta_1}\Big(\sum_{\Delta_2,\varepsilon\in\sigma}\omega^{[ab]}[\varepsilon]z[\Delta_2,\varepsilon]W[\Delta_1,\Delta_2]-\sum_{\Delta_2\in\sigma}\lambda^{[ab]}[\Delta_2]W[\Delta_1,\Delta_2]\\
\nonumber
&& \qquad\qquad\qquad+\frac{1}{10}\sum_{\varepsilon_1,\varepsilon_2\in\sigma}\omega^{[cd]}[\varepsilon_1]\omega^{[ef]}[\varepsilon_2]f_{[cd][ef]}{}^{[ab]}W[\Delta_1,\varepsilon_1,\varepsilon_2]\Big)\Big]\\
\nonumber
&& \qquad\prod_{\Delta}\delta\Big[\sum_{\sigma\ni\Delta}\Big(\sum_{\varepsilon,\tau\in\sigma}\varepsilon_a z[\tau,\Delta]W[\varepsilon,\tau]+\frac{1}{15}\hspace{-1mm}\sum_{\varepsilon_1,\varepsilon_2\in\sigma}\hspace{-1mm}\varepsilon_{1b} \omega^{[cd]}[\varepsilon_2]\triangleright_{[cd]a}{}^{b}W[\varepsilon_1,\varepsilon_2,\Delta]\\
\nonumber
&& \qquad\qquad\qquad+\frac{4\pi i l_p^2}{750}\sum_{v_1,v_2,\varepsilon_1,\varepsilon_2\in\sigma}\bar{\psi}_A[v_1](\gamma_5\gamma^d\psi)^A[v_2]\varepsilon_1^b\varepsilon_2^c\varepsilon_{dbca}W[\varepsilon_1,\varepsilon_2,\Delta]\Big)\Big]\\
\nonumber
&& \qquad\prod_{\tau}\delta\Big[\sum_{\sigma\ni\tau}\Big(\sum_{v\in\sigma}\phi^A[v]\Big(\delta^B_Az[\sigma,\tau]+\frac{1}{20}\sum_{\varepsilon\in\sigma}W[\varepsilon,\tau]\alpha^{\alpha}[\varepsilon]\triangleright_{\alpha}{}^B{}_A\Big) +\frac{1}{4}\sum_{\varepsilon\in\sigma}\tilde{\lambda}^B[\varepsilon]W[\varepsilon,\tau]\Big)\Big]\\
\nonumber
&& \qquad\prod_{\tau}\delta\Big[\sum_{\sigma\ni\tau}\Big(\sum_{v\in\sigma}\bar{\psi}_B[v]\Big(\delta^B_Az[\sigma,\tau]+\frac{1}{20}\sum_{\varepsilon\in\sigma}W[\varepsilon,\tau]\Big(\alpha^{\alpha}[\varepsilon]\triangleright_{\alpha A}{}^B+\omega^{[ab]}[\varepsilon]\triangleright_{[ab]A}{}^B\Big)\vphantom{\frac{1}{20}\sum_{\varepsilon\in\sigma}}\Big)+\frac{1}{4}\sum_{\varepsilon\in\sigma}\bar{\lambda}_A[\varepsilon]W[\varepsilon,\tau]\Big)\Big]\\
&& \qquad\prod_{\tau}\delta\Big[\sum_{\sigma\ni\tau}\Big(\Big(\delta^A_Bz[\sigma,\tau]+\frac{1}{20}\sum_{\varepsilon\in\sigma}W[\varepsilon,\tau]\Big(\alpha^{\alpha}[\varepsilon]\triangleright_{\alpha B}{}^A+\omega^{[ab]}[\varepsilon]\triangleright_{[ab]B}{}^A\Big)\vphantom{\frac{1}{20}\sum_{\varepsilon\in\sigma}}\Big)\sum_{v\in\sigma}\psi^B[v]-\frac{1}{4}\sum_{\varepsilon\in\sigma}\lambda^A[\varepsilon]W[\varepsilon,\tau]\Big)\Big]\,.
\end{eqnarray}
}%
The terms $\rm INT[\dots ]$ and $\rm SC[\dots ]$ correspond to the simplicity constraint terms in the action. The $\rm INT[\dots ]$ term describes the interactions, and is given as:
{\small
\begin{eqnarray}
\nonumber
\mathrm{INT}\Big[\lambda[\phi_k]-G[\phi_k]\Big] & = & \prod_\sigma\Big[\exp \Big(-\frac{i}{125}\sum_{\Delta,\varepsilon_1,\varepsilon_2,v\in\sigma}C_{\alpha\beta}\lambda^{\alpha}[\Delta]M^{\beta}{}_{ab}[v]\varepsilon_1^a\varepsilon_2^b W[\Delta,\varepsilon_1,\varepsilon_2]\Big)\\
\nonumber
&& \exp \Big(\frac{i}{2400\pi l_p^2}\sum_{\Delta,\varepsilon_1,\varepsilon_2\in\sigma}\varepsilon_{[ab]cd}\lambda^{[ab]}[\Delta]\varepsilon_1^c\varepsilon_2^d W[\Delta,\varepsilon_1,\varepsilon_2]\Big)\\
\nonumber
&& \exp \Big(-\frac{i}{15000}\sum_{\varepsilon_1,\varepsilon_2,\varepsilon_3,\varepsilon_4,v\in\sigma}\tilde{\lambda}^A[\varepsilon_1]H_{abcA}[v]\varepsilon_2^a\varepsilon_3^b\varepsilon_4^c W[\varepsilon_1,\varepsilon_2,\varepsilon_3,\varepsilon_4]\Big)\\
\nonumber
&& \exp \Big(\frac{1}{90000}\sum_{\varepsilon_1,\varepsilon_2,\varepsilon_3,\varepsilon_4,v\in\sigma}\bar{\lambda}_A[\varepsilon_1]\Big(\gamma^d\psi[v]\Big)^A\varepsilon_2^a\varepsilon_3^b\varepsilon_4^c\varepsilon_{abcd} W[\varepsilon_1,\varepsilon_2,\varepsilon_3,\varepsilon_4]\Big)\\
\nonumber
&& \exp \Big(\frac{1}{90000}\sum_{\varepsilon_1,\varepsilon_2,\varepsilon_3,\varepsilon_4,v\in\sigma}\bar{\psi}_A[v]\Big(\gamma^d\lambda[\varepsilon_4]\Big)^A\varepsilon_1^a\varepsilon_2^b\varepsilon_3^c\varepsilon_{abcd} W[\varepsilon_1,\varepsilon_2,\varepsilon_3,\varepsilon_4]\Big)\\
\nonumber
&& \exp \Big(-\frac{i}{36000}\sum_{\varepsilon_1,\varepsilon_2,\varepsilon_3,\varepsilon_4\in\sigma}\varepsilon_{abcd}\varepsilon_1^a\varepsilon_2^b\varepsilon_3^c\varepsilon_4^d\Big(\frac{\Lambda}{8\pi l_p^2}+\chi v^4-\frac{2\chi v^2}{25}\sum_{v_1,v_2\in\sigma}\phi_A[v_1]\phi^A[v_2]\\
&& +\frac{1}{125}Y_{ABC}\sum_{v_1,v_2,v_3\in\sigma}\bar{\psi}^A[v_1]\psi^B[v_2]\phi^C[v_3] +\frac{\chi}{625}\sum_{v_1,v_2,v_3,v_4\in\sigma}\phi_A[v_1]\phi^A[v_2]\phi_B[v_3]\phi^B[v_4]\Big)\Big)\Big]\,.
\end{eqnarray}
}%
The remaining $\rm SC[\dots ]$ term describes the strong constraints, and is given as:
{\small
\begin{eqnarray}
\nonumber
&& \mathrm{SC}\Big[\phi_k\Big]= \\
\nonumber
&&  \qquad\prod_{\varepsilon_1}\delta\Big[\sum_{\sigma\ni\varepsilon_1}\Big(\sum_{v,\varepsilon_2,\varepsilon_3,\varepsilon_4\in\sigma}W[\varepsilon_1,\varepsilon_2,\varepsilon_3,\varepsilon_4]\Big(\frac{1}{5}H_{abcA}[v]\varepsilon^{cdef}\varepsilon_{2d}\varepsilon_{3e}\varepsilon_{4f}+\Big(\phi_A[v]z[\varepsilon_2,v]+\frac{1}{5}\phi^B[v]\alpha^\alpha[\varepsilon_2]\triangleright_{\alpha BA}\Big)\varepsilon_{3a}\varepsilon_{4b}\Big)\Big)\Big]\\
\nonumber
&& \qquad\prod_{v}\delta\Big[\sum_{\sigma\ni v}\Big(\sum_{\varepsilon_1,\varepsilon_2,\varepsilon_3\in\sigma}\varepsilon_1^c\varepsilon_2^d\Big(\sum_{\varepsilon_4\in\sigma}W[\varepsilon_1,\varepsilon_2,\varepsilon_3,\varepsilon_4]\Big(\frac{1}{50}\varepsilon_{cdef}\varepsilon_3^e\varepsilon_4^f\sum_{\tilde{v}\in\sigma}M^\alpha{}_{ab}[\tilde{v}]-\frac{1}{10}f_{\beta\gamma}{}^{\alpha}\alpha^{\beta}[\varepsilon_3]\alpha^{\gamma}[\varepsilon_4]\eta_{ac}\eta_{bd}\Big)\\
&& \qquad\qquad\qquad-\sum_{\Delta\in\sigma}\alpha^{\alpha}[\varepsilon_3]z[\Delta,\varepsilon_3]W[\Delta,\varepsilon_1,\varepsilon_2]\eta_{ac}\eta_{bd}\Big)\Big)\Big]\,.
\end{eqnarray}
}%

It is important to note the natural distinction between the interaction term and the strong constraints term. Namely, unlike the former, the latter is expressed as a product of Dirac delta functions, enforcing their arguments to be interpreted as strong constraints. Aside from the Dirac delta functions which appear in the topological term $\rm TI[\dots ]$, the only additional Dirac deltas are present in $\rm SC[\dots ]$. In contrast, the terms in the $\rm INT[\dots ]$ term cannot be written as Dirac delta functions, and therefore can rather be understood as weakly enforced constraints.

The second aspect of the structure (\ref{oblikF}) of our model (\ref{eq:ThirdMainResult}) lies in the combinatorial nature of the underlying triangulation. Namely, one could attempt to simplify the structure (\ref{oblikF}) further, by integrating out the Dirac delta functions appearing in either the topological or the strong constraints term. In a theory defined over a smooth manifold this can be explicitly performed, as was demonstrated in \cite{Stipsic2025b}. However, in addition to the topological structure of a smooth manifold, a simplicial complex also contains combinatorial structure, which must be taken into account in our model. In principle, one can (for example using numerical techniques) perform the integration over all Dirac delta functions, but the result depends on the combinatorial structure of the triangulation. Therefore, the result of the integration cannot be expressed in closed form in general. Nevertheless, for a large class of triangluations, there are substantial further simplifications that one can perform, along the lines of transformations demonstrated in the theory on a smooth manifold \cite{Stipsic2025b}.

Those further simplifications of (\ref{oblikF}) rely on the following identity:
\begin{equation} \label{eq:DeltaSumToProduct}
\prod_k \delta\Big( \sum_{\sigma \ni k} F(\sigma) \Big) = \prod_k \prod_{\sigma \ni k} \delta (F(\sigma))\,,
\end{equation}
where $k\in\{v,\varepsilon,\Delta,\tau \}$. The above identity does not hold for all triangulations, but holds for a rather large class of them. Namely, if the number of $4$-simplices in any given trianguation is less or equal to the number of $k$-simplies, $|\sigma|\leqslant |k|$, then the Dirac delta functions on the left-hand side are nonzero only if their arguments form a homogeneous linear system of $|k|$ equations for $|\sigma|$ quantities $F(\sigma)$, which has only a trivial solution, $F(\sigma) = 0$. The latter is equivalent to the product of Dirac delta functions on the right-hand side of (\ref{eq:DeltaSumToProduct}), establishing the above identity. Unfortunately, if the number of $4$-simplices is greater than the number of $k$-simplices, $|\sigma| > |k|$, the equality between the left-hand side and the right-hand side does not hold. However, the case $|\sigma| > |k|$ represents a rather atypical triangulation, since every $4$-simplex consists of five vertices and tetrahedra, and ten edges and triangles. Therefore, one ought to expect that for a generic triangulation, the number of $4$-simplices should be smaller than the number of subsimplices, so that (\ref{eq:DeltaSumToProduct}) can be considered a rigorous identity for all such triangulations. In the remainder of the paper, we will therefore restrict ourseleves to the class of triangulations which satisfy the criterion $|\sigma|\leqslant |k|$.

Applying (\ref{eq:DeltaSumToProduct}) to our path integral (\ref{eq:ThirdMainResult}), and writing it in the form (\ref{oblikF}), the strong constraint term becomes
{\small
\begin{eqnarray}
\nonumber
&& \mathrm{SC}\Big[\phi_k\Big]= \\
\nonumber
&&\qquad\prod_{\sigma}\prod_{\varepsilon_1\in\sigma}\delta\Big[\sum_{v,\varepsilon_2,\varepsilon_3,\varepsilon_4\in \sigma}W[\varepsilon_1,\varepsilon_2,\varepsilon_3,\varepsilon_4]\Big(\frac{1}{5}H_{abcA}[v]\varepsilon^{cdef}\varepsilon_{2d}\varepsilon_{3e}\varepsilon_{4f}+\Big(\phi_A[v]z[\varepsilon_2,v]+\frac{1}{5}\phi^B[v]\alpha^\alpha[\varepsilon_2]\triangleright_{\alpha BA}\Big)\varepsilon_{3a}\varepsilon_{4b}\Big)\Big]\\
\nonumber
&& \qquad \prod_{\sigma}\delta\Big[\sum_{\varepsilon_1,\varepsilon_2,\varepsilon_3\in\sigma}\varepsilon_1^c\varepsilon_2^d\Big(\sum_{\varepsilon_4\in\sigma}W[\varepsilon_1,\varepsilon_2,\varepsilon_3,\varepsilon_4]\Big(\frac{1}{50}\varepsilon_{cdef}\varepsilon_3^e\varepsilon_4^f\sum_{v\in\sigma}M^\alpha{}_{ab}[v]-\frac{1}{10}f_{\beta\gamma}{}^{\alpha}\alpha^{\beta}[\varepsilon_3]\alpha^{\gamma}[\varepsilon_4]\eta_{ac}\eta_{bd}\Big)\\
&&\qquad\qquad\qquad-\sum_{\Delta\in\sigma}\alpha^{\alpha}[\varepsilon_3]z[\Delta,\varepsilon_3]W[\Delta,\varepsilon_1,\varepsilon_2]\eta_{ac}\eta_{bd}\Big)\Big]\,,
\end{eqnarray}
}%
while the topological invariant term becomes
{\small
\begin{eqnarray}
\nonumber
&& {\rm TI}\Big[\mathcal{F}[\phi_k]-\lambda[\phi_k]\Big]=\\
\nonumber
&& \qquad\prod_\sigma\Big[\prod_{\Delta_1\in\sigma}\delta\Big(\sum_{\Delta_2,\varepsilon\in\sigma}\alpha^\alpha[\varepsilon]z[\Delta_2,\varepsilon]W[\Delta_1,\Delta_2]+\frac{1}{10}\sum_{\varepsilon_1,\varepsilon_2\in\sigma}\alpha^{\beta}[\varepsilon_1]\alpha^{\gamma}[\varepsilon_2]f_{\beta\gamma}{}^{\alpha}W[\Delta_1,\varepsilon_1,\varepsilon_2]\\
\nonumber
&& \qquad\qquad\qquad\qquad +\sum_{\Delta_2\in\sigma}\lambda^\alpha[\Delta_2]W[\Delta_1,\Delta_2]\Big)\\
\nonumber
&& \qquad\qquad \prod_{\Delta_1\in\sigma}\delta\Big(\sum_{\Delta_2,\varepsilon\in\sigma}\omega^{[ab]}[\varepsilon]z[\Delta_2,\varepsilon]W[\Delta_1,\Delta_2]+\frac{1}{10}\sum_{\varepsilon_1,\varepsilon_2\in\sigma}\omega^{[cd]}[\varepsilon_1]\omega^{[ef]}[\varepsilon_2]f_{[cd][ef]}{}^{[ab]}W[\Delta_1,\varepsilon_1,\varepsilon_2]\\
\nonumber
&& \qquad\qquad\qquad\qquad -\sum_{\Delta_2\in\sigma}\lambda^{[ab]}[\Delta_2]W[\Delta_1,\Delta_2]\Big)\\
\nonumber
&& \qquad\qquad \prod_{\Delta\in\sigma}\delta\Big(\sum_{\varepsilon,\tau\in\sigma}\varepsilon_a z[\tau,\Delta]W[\varepsilon,\tau]+\frac{1}{15}\sum_{\varepsilon_1,\varepsilon_2\in\sigma}\varepsilon_{1b} \omega^{[cd]}[\varepsilon_2]\triangleright_{[cd]a}{}^{b}W[\varepsilon_1,\varepsilon_2,\Delta]\\
\nonumber
&& \qquad\qquad\qquad\qquad +\frac{4\pi i l_p^2}{750}\sum_{v_1,v_2,\varepsilon_1,\varepsilon_2\in\sigma}\bar{\psi}_A[v_1](\gamma_5\gamma^d\psi)^A[v_2]\varepsilon_1^b\varepsilon_2^c\varepsilon_{dbca}W[\varepsilon_1,\varepsilon_2,\Delta]\Big)\\
\nonumber
&& \qquad\qquad \prod_{\tau\in\sigma}\delta\Big(\sum_{v\in\sigma}\phi^A[v]\Big(\delta^B_Az[\sigma,\tau]+\frac{1}{20}\sum_{\varepsilon\in\sigma}W[\varepsilon,\tau]\alpha^{\alpha}[\varepsilon]\triangleright_{\alpha}{}^B{}_A\Big)+\frac{1}{4}\sum_{\varepsilon\in\sigma}\tilde{\lambda}^B[\varepsilon]W[\varepsilon,\tau]\Big)\\
\nonumber
&& \qquad\qquad \prod_{\tau\in\sigma}\delta\Big(\sum_{v\in\sigma}\bar{\psi}_B[v]\Big(\delta^B_Az[\sigma,\tau]+\frac{1}{20}\sum_{\varepsilon\in\sigma}W[\varepsilon,\tau]\Big(\alpha^{\alpha}[\varepsilon]\triangleright_{\alpha A}{}^B+\omega^{[ab]}[\varepsilon]\triangleright_{[ab]A}{}^B\Big)\Big)+\frac{1}{4}\sum_{\varepsilon\in\sigma}\bar{\lambda}_A[\varepsilon]W[\varepsilon,\tau]\Big)\\
&& \qquad\qquad \prod_{\tau\in\sigma}\delta\Big(\Big(\delta^A_Bz[\sigma,\tau]+\frac{1}{20}\sum_{\varepsilon\in\sigma}W[\varepsilon,\tau]\Big(\alpha^{\alpha}[\varepsilon]\triangleright_{\alpha B}{}^A+\omega^{[ab]}[\varepsilon]\triangleright_{[ab]B}{}^A\Big)\Big)\sum_{v\in\sigma}\psi^B[v]-\frac{1}{4}\sum_{\varepsilon\in\sigma}\lambda^A[\varepsilon]W[\varepsilon,\tau]\Big)\Big]\,.
\end{eqnarray}
}%
The purpose of using the identity (\ref{eq:DeltaSumToProduct}) to perform the above transformations lies in the fact that the terms $\rm SC[\dots ]$ and $\rm TI[\dots ]$ are now expressed as products over all $4$-simplices in the triangulation. The next step is the following change of variables for the $\rm SC[\dots ]$ term,
\begin{equation} \label{eq:ChangeOfVarsSC}
\Xi^{\alpha}{}_{ab}[\sigma]=\frac{1}{5}\sum_{v\in\sigma}M^{\alpha}{}_{ab}[v]\,,\qquad\Theta_{abcA}[\sigma]=\frac{1}{5}\sum_{v\in\sigma}H_{abcA}[v]\,,
\end{equation}
and for the $\rm TI[\dots ]$ term,
\begin{equation} \label{eq:ChangeOfVarsTI}
\begin{array}{c}
\ds  Q^{[ab]}[\Delta,\sigma]=\frac{1}{3}\sum_{\tilde{\Delta}\in\sigma}\lambda^{[ab]}[\tilde{\Delta}]W[\Delta,\tilde{\Delta}]\,,\qquad Q^{\alpha}[\Delta,\sigma]=\frac{1}{3}\sum_{\tilde{\Delta}\in\sigma}\lambda^{\alpha}[\tilde{\Delta}]W[\Delta,\tilde{\Delta}]\,, \\
\ds \tilde{Q}^A[\tau,\sigma]=\frac{1}{4}\sum_{\varepsilon\in\sigma}\tilde{\lambda}^A[\varepsilon]W[\varepsilon,\tau]\,, \qquad
\bar{Q}_A[\tau,\sigma]=\frac{1}{4}\sum_{\varepsilon\in\sigma}\bar{\lambda}_A[\varepsilon]W[\varepsilon,\tau]\,, \qquad
Q^A[\tau,\sigma]=\frac{1}{4}\sum_{\varepsilon\in\sigma}\lambda^A[\varepsilon]W[\varepsilon,\tau]\,. \\
\end{array}
\end{equation}
Namely, in any given $4$-simplex, we have five fields $M^\alpha{}_{ab}[v]$, one for each vertex $v\in\sigma$. The change of variables (\ref{eq:ChangeOfVarsSC}) substitutes one of these five with a new field $\Xi^\alpha{}_{ab}[\sigma]$ living on the $4$-simplex $\sigma$, while the other four remain unchanged. This represents a 1-to-1 map between the old and new variables, within a single simplex. However, note that some of the fields $M^\alpha{}_{ab}[v]$ are common to more than one $4$-simplex, so a priori there may be a clash of changes of variables across neighboring $4$-simplices. Nevertheless, this can be avoided for our class of triangulations, since the total number of $4$-simplices in the triangulation is always smaller than the total number of vertices in the triangulation, $|\sigma|\leqslant |v|$. Therefore, the total number of new fields $\Xi^\alpha{}_{ab}[\sigma]$ is small enough so that the triangulation can always accommodate the change of variables. An analogous argument holds for all other substitutions in (\ref{eq:ChangeOfVarsSC}) and (\ref{eq:ChangeOfVarsTI}).

In addition to the above, each $Q$-type field defined in (\ref{eq:ChangeOfVarsTI}) depends not only on the $k$-simplex it lives on, but also on the $4$-simplex $\sigma$ used to evaluate the corresponding sum. This is compatible with our application of the identity (\ref{eq:DeltaSumToProduct}) which has increased the overall number of Dirac delta functions by expressing them as products over all $4$-simplices in the triangulation. In this sense, the dependence of $Q$-type fields on the $4$-simplex $\sigma$ precisely matches the corresponding Dirac delta functions.

The above change of variables should be applied to the $\rm INT[\dots ]$ term as well, in addition to the $\rm SC[\dots ]$ and $\rm TI[\dots ]$ terms. To that end, let us first note that the following identities hold:
\begin{eqnarray}
\frac{1}{300}\sum_{\tilde{\Delta},\varepsilon_1,\varepsilon_2\in\sigma}\lambda^{[cd]}[\tilde{\Delta}]W[\tilde{\Delta},\varepsilon_1,\varepsilon_2]\varepsilon_1^a\varepsilon_2^b
&=&\frac{1}{10}\sum_{\Delta\in\sigma}Q^{[cd]}[\Delta,\sigma]\Delta^{ab}\,, \\
\frac{1}{300}\sum_{\tilde{\Delta},\varepsilon_1,\varepsilon_2\in\sigma}\lambda^\alpha[\tilde{\Delta}]W[\tilde{\Delta},\varepsilon_1,\varepsilon_2]\varepsilon_1^a\varepsilon_2^b
&=&\frac{1}{10}\sum_{\Delta\in\sigma}Q^\alpha[\Delta,\sigma]\Delta^{ab}\,, \\
\frac{1}{3000}\sum_{\varepsilon_1,\varepsilon_2,\varepsilon_3,\varepsilon_4\in\sigma}\tilde{\lambda}^A[\varepsilon_1]\varepsilon^a_2\varepsilon^b_3\varepsilon_4^c W[\varepsilon_1,\varepsilon_2,\varepsilon_3,\varepsilon_4]
& = &\frac{1}{5}\sum_{\tau\in\sigma}\tilde{Q}^A[\tau,\sigma]\tau^{abc}\,, \\
\frac{1}{3000}\sum_{\varepsilon_1,\varepsilon_2,\varepsilon_3,\varepsilon_4\in\sigma}\bar{\lambda}_A[\varepsilon_1]\varepsilon^a_2\varepsilon^b_3\varepsilon_4^c W[\varepsilon_1,\varepsilon_2,\varepsilon_3,\varepsilon_4]
& = &\frac{1}{5}\sum_{\tau\in\sigma}\bar{Q}_A[\tau,\sigma]\tau^{abc}\,, \\
\frac{1}{3000}\sum_{\varepsilon_1,\varepsilon_2,\varepsilon_3,\varepsilon_4\in\sigma}\lambda^A[\varepsilon_1]\varepsilon^a_2\varepsilon^b_3\varepsilon_4^c W[\varepsilon_1,\varepsilon_2,\varepsilon_3,\varepsilon_4]
& = &\frac{1}{5}\sum_{\tau\in\sigma}Q^A[\tau,\sigma]\tau^{abc}\,.
\end{eqnarray}
Each identity is straightforward to prove by substituting to its right-hand side the appropriate $Q$-type field from (\ref{eq:ChangeOfVarsTI}). Next, we note that the $\rm INT[\dots ]$ term in fact depends on all $\lambda$-type fields only via the expressions identical to the ones on the left-hand sides of the above identities. Moreover, it depends on $M$-fields and $H$-fields only via the expressions identical to the ones on the right-hand sides of (\ref{eq:ChangeOfVarsSC}). In other words, the $\rm INT[\dots ]$ term in fact depends only on the newly introduced fields in our change of variables, while its dependence on $M$, $H$ and $\lambda$ fields is eliminated.

Once the change of variables is introduced, the three terms take the following forms.  The $\rm TI[\dots ]$ term becomes:
{\small
\begin{eqnarray} \label{eq:TItermBeforeIntegration}
\nonumber
&& {\rm TI}\Big[\mathcal{F}[\phi_k]-\lambda[\phi_k]\Big]= \\
\nonumber
&& \qquad  \prod_\sigma\Big[\prod_{\Delta_1\in\sigma}\delta\Big(\sum_{\Delta_2,\varepsilon\in\sigma}\alpha^\alpha[\varepsilon]z[\Delta_2,\varepsilon]W[\Delta_1,\Delta_2]+\frac{1}{10}\sum_{\varepsilon_1,\varepsilon_2\in\sigma}\alpha^{\beta}[\varepsilon_1]\alpha^{\gamma}[\varepsilon_2]f_{\beta\gamma}{}^{\alpha}W[\Delta_1,\varepsilon_1,\varepsilon_2]+3Q^\alpha[\Delta_1,\sigma]\vphantom{\prod_{\Delta_1\in\sigma}}\Big)\\
\nonumber
&& \qquad  \qquad \prod_{\Delta_1\in\sigma}\delta\Big(\sum_{\Delta_2,\varepsilon\in\sigma}\omega^{[ab]}[\varepsilon]z[\Delta_2,\varepsilon]W[\Delta_1,\Delta_2]+\frac{1}{10}\sum_{\varepsilon_1,\varepsilon_2\in\sigma}\omega^{[cd]}[\varepsilon_1]\omega^{[ef]}[\varepsilon_2]f_{[cd][ef]}{}^{[ab]}W[\Delta_1,\varepsilon_1,\varepsilon_2]-3Q^{[ab]}[\Delta_1,\sigma]\Big)\\
\nonumber
&& \qquad\qquad \prod_{\Delta\in\sigma}\delta\Big(\sum_{\varepsilon,\tau\in\sigma}\varepsilon_a z[\tau,\Delta]W[\varepsilon,\tau]+\frac{1}{15}\sum_{\varepsilon_1,\varepsilon_2\in\sigma}\varepsilon_{1b} \omega^{[cd]}[\varepsilon_2]\triangleright_{[cd]a}{}^{b}W[\varepsilon_1,\varepsilon_2,\Delta]\\
\nonumber
&& \qquad\qquad\qquad\qquad\qquad\qquad +\frac{4\pi i l_p^2}{750}\sum_{v_1,v_2,\varepsilon_1,\varepsilon_2\in\sigma}\bar{\psi}_A[v_1](\gamma_5\gamma^d\psi)^A[v_2]\varepsilon_1^b\varepsilon_2^c\varepsilon_{dbca}W[\varepsilon_1,\varepsilon_2,\Delta]\Big)\\
\nonumber
&& \qquad  \qquad \prod_{\tau\in\sigma}\delta\Big(\sum_{v\in\sigma}\phi^A[v]\Big(\delta^B_Az[\sigma,\tau]+\frac{1}{20}\sum_{\varepsilon\in\sigma}W[\varepsilon,\tau]\alpha^{\alpha}[\varepsilon]\triangleright_{\alpha}{}^B{}_A\Big)+\tilde{Q}^B[\tau,\sigma]\Big)\\
\nonumber
&& \qquad  \qquad \prod_{\tau\in\sigma}\delta\Big(\sum_{v\in\sigma}\bar{\psi}_B[v]\Big(\delta^B_Az[\sigma,\tau]+\frac{1}{20}\sum_{\varepsilon\in\sigma}W[\varepsilon,\tau]\Big(\alpha^{\alpha}[\varepsilon]\triangleright_{\alpha A}{}^B+\omega^{[ab]}[\varepsilon]\triangleright_{[ab]A}{}^B\Big)\Big)+\bar{Q}_A[\tau,\sigma]\Big)\\
&& \qquad  \qquad \prod_{\tau\in\sigma}\delta\Big(\Big(\delta^A_Bz[\sigma,\tau]+\frac{1}{20}\sum_{\varepsilon\in\sigma}W[\varepsilon,\tau]\Big(\alpha^{\alpha}[\varepsilon]\triangleright_{\alpha B}{}^A+\omega^{[ab]}[\varepsilon]\triangleright_{[ab]B}{}^A\Big)\Big)\sum_{v\in\sigma}\psi^B[v]-Q^A[\tau,\sigma]\Big)\Big]\,.
\end{eqnarray}
}%
The $\rm SC[\dots ]$ term becomes:
{\small
\begin{eqnarray} \label{eq:SCtermBeforeIntegration}
\nonumber
\mathrm{SC}\Big[\phi_k\Big] &=&  \prod_{\sigma}\Big[
  \prod_{\varepsilon_1\in\sigma}\delta\Big(\sum_{\varepsilon_2,\varepsilon_3,\varepsilon_4\in \sigma}W[\varepsilon_1,\varepsilon_2,\varepsilon_3,\varepsilon_4]\Big(\Theta_{abcA}[\sigma]\varepsilon^{cdef}\varepsilon_{2d}\varepsilon_{3e}\varepsilon_{4f} \\
\nonumber
&& \qquad\qquad\qquad +\sum_{v\in \sigma}\Big(\phi_A[v]z[\varepsilon_2,v]+\frac{1}{5}\phi^B[v]\alpha^\alpha[\varepsilon_2]\triangleright_{\alpha BA}\Big)\varepsilon_{3a}\varepsilon_{4b}\Big)\Big)\\
\nonumber
&&\qquad
\frac{1}{\big|V(\sigma)\big|^{|\alpha||[ab]|}}\delta\Big(\Xi^\alpha{}_{ab}[\sigma]+\frac{1}{4!V(\sigma)}\Big(\frac{1}{300}\sum_{\varepsilon_1,\varepsilon_2,\varepsilon_3,\Delta\in\sigma}\alpha^{\alpha}[\varepsilon_3]z[\Delta,\varepsilon_3]W[\Delta,\varepsilon_1,\varepsilon_2]\varepsilon_{1a}\varepsilon_{2b}\\
&&\qquad\qquad\qquad+\frac{1}{3000}\sum_{\varepsilon_1,\varepsilon_2,\varepsilon_3,\varepsilon_4\in\sigma}f_{\beta\gamma}{}^{\alpha}\alpha^{\beta}[\varepsilon_3]\alpha^{\gamma}[\varepsilon_4]\varepsilon_{1a}\varepsilon_{2b}W[\varepsilon_1,\varepsilon_2,\varepsilon_3,\varepsilon_4]\Big)\Big)\Big]\,.
\end{eqnarray}
}%
Finally, the $\rm INT[\dots ]$ term takes the form:
{\small
\begin{eqnarray} \label{eq:INTtermBeforeIntegration}
\nonumber
&& \mathrm{INT}\Big[\lambda[\phi_k]-G[\phi_k]\Big]=\\
\nonumber
&& \qquad \prod_\sigma\Big[\exp \Big(-\frac{6i}{5}\sum_{\tilde{\Delta},\Delta\in\sigma}C_{\alpha\beta}Q^{\alpha}[\tilde{\Delta},\sigma]W[\Delta,\tilde{\Delta}]\Xi^{\beta}{}_{ab}[\sigma]\Delta^{ab}\Big)\\
\nonumber
&& \qquad\qquad \exp \Big(\frac{i}{80\pi l_p^2}\sum_{\Delta,\tilde{\Delta}\in\sigma}\varepsilon_{[ab]cd}Q^{[ab]}[\tilde{\Delta},\sigma]W[\Delta,\tilde{\Delta}]\Delta^{cd}\Big)\\
\nonumber
&& \qquad\qquad \exp \Big(-\frac{i}{5}\sum_{\tau\in\sigma}\tilde{Q}^A[\tau,\sigma]\Theta_{abcA}[\sigma]\tau^{abc}\Big)\\
\nonumber
&& \qquad\qquad \exp \Big(\frac{1}{150}\sum_{v,\tau\in\sigma}\bar{Q}_A[\tau,\sigma]\Big(\gamma^d\psi[v]\Big)^A\varepsilon_{abcd}\tau^{abc}\Big)\\
\nonumber
&& \qquad\qquad \exp \Big(\frac{1}{150}\sum_{v,\tau\in\sigma}\bar{\psi}_A[v]\Big(\gamma^d Q[\tau,\sigma]\Big)^A\varepsilon_{abcd}\tau^{abc}\Big)\\
\nonumber
&& \qquad\qquad \exp \Big(-\frac{i}{36000}\sum_{\varepsilon_1,\varepsilon_2,\varepsilon_3,\varepsilon_4\in\sigma}\varepsilon_{abcd}\varepsilon_1^a\varepsilon_2^b\varepsilon_3^c\varepsilon_4^d\Big(\frac{\Lambda}{8\pi l_p^2}+\chi v^4-\frac{2\chi v^2}{25}\sum_{v_1,v_2\in\sigma}\phi_A[v_1]\phi^A[v_2]\\
&& \qquad\qquad\qquad +\frac{1}{125}Y_{ABC}\sum_{v_1,v_2,v_3\in\sigma}\bar{\psi}^A[v_1]\psi^B[v_2]\phi^C[v_3] +\frac{\chi}{625}\sum_{v_1,v_2,v_3,v_4\in\sigma}\phi_A[v_1]\phi^A[v_2]\phi_B[v_3]\phi^B[v_4]\Big)\Big)\Big]\,.
\end{eqnarray}
}%
At this point, we are ready to perform the integration over all $Q$-type fields, and over the $\Xi^{\alpha}{}_{ab}[\sigma]$ field, as well as over any remaining $H$-, $M$- and $\lambda$-type fields, namely those that were not eliminated in favor of $Q$, $\Xi$ and $\Theta$, if there are any. Since the terms $\rm SC[\dots ]$, $\rm TI[\dots ]$ and $\rm INT[\dots ]$ do not explicitly depend any more on $H$-, $M$- and $\lambda$-type fields, integration over the latter will amount only to a constant, which can be absorbed into the overall normalization constant $\cN$. Next, looking at the $\rm SC[\dots ]$ and $\rm TI[\dots ]$ terms, one can see that each $Q$ and $\Xi$ field appears precisely in one Dirac delta function, ready to be integrated. Therefore, the integration over these fields amounts to substituting the corresponding expressions within Dirac delta functions in the place of $Q$ and $\Xi$ fields withing the $\rm INT[\dots ]$ term. After the integration, the $\rm SC[\dots ]$ and $\rm TI[\dots ]$ terms reduce to the following form, each term depending only on the remaining Dirac delta functions that were not integrated over:
{\small
\begin{eqnarray} \label{eq:RemainingDiracDeltaFunctions}
\nonumber
\mathrm{SC}\Big[\phi_k\Big] &=&  \prod_{\sigma}
  \prod_{\varepsilon_1\in\sigma}\delta\Big(\sum_{\varepsilon_2,\varepsilon_3,\varepsilon_4\in \sigma}W[\varepsilon_1,\varepsilon_2,\varepsilon_3,\varepsilon_4]\Big(\Theta_{abcA}[\sigma]\varepsilon^{cdef}\varepsilon_{2d}\varepsilon_{3e}\varepsilon_{4f} \\
\nonumber
&& \qquad\qquad\qquad +\sum_{v\in \sigma}\Big(\phi_A[v]z[\varepsilon_2,v]+\frac{1}{5}\phi^B[v]\alpha^\alpha[\varepsilon_2]\triangleright_{\alpha BA}\Big)\varepsilon_{3a}\varepsilon_{4b}\Big)\Big)\,,\\
\nonumber
{\rm TI}\Big[\mathcal{F}[\phi_k]-\lambda[\phi_k]\Big] & = & 
\prod_\sigma \prod_{\Delta\in\sigma}\delta\Big(\sum_{\varepsilon,\tau\in\sigma}\varepsilon_a z[\tau,\Delta]W[\varepsilon,\tau]+\frac{1}{15}\sum_{\varepsilon_1,\varepsilon_2\in\sigma}\varepsilon_{1b} \omega^{[cd]}[\varepsilon_2]\triangleright_{[cd]a}{}^{b}W[\varepsilon_1,\varepsilon_2,\Delta]\\
&& \qquad\qquad\qquad +\frac{4\pi i l_p^2}{750}\sum_{v_1,v_2,\varepsilon_1,\varepsilon_2\in\sigma}\bar{\psi}_A[v_1](\gamma_5\gamma^d\psi)^A[v_2]\varepsilon_1^b\varepsilon_2^c\varepsilon_{dbca}W[\varepsilon_1,\varepsilon_2,\Delta]\Big)\,.
\end{eqnarray}
}%
The integration over $\Theta_{abcA}[\sigma]$ and $ \omega^{[ab]}[\varepsilon]$ cannot be explicitly performed because the linear combinations within the Dirac delta functions are complicated. We will perform these integrations using certain approximations in the next Section.

The explicit expression for the $\rm INT[\dots ]$ term, after integration has been performed, is too bulky and it is not convenient to study it all at once. Instead, noting that (\ref{eq:INTtermBeforeIntegration}) consists of six exponent parts,  we will discuss each exponent in turn.

\subsection{Yang-Mills kinetic term}

Let us show that the first exponent corresponds to the kinetic term for the Yang-Mills fields. Namely, if we start from
\begin{equation} \label{eq:FirstExponent}
iS_\text{YM} \equiv -\frac{6i}{5}\sum_\sigma \sum_{\tilde{\Delta},\Delta\in\sigma}C_{\alpha\beta}Q^{\alpha}[\tilde{\Delta},\sigma]W[\Delta,\tilde{\Delta}]\Xi^{\beta}{}_{ab}[\sigma]\Delta^{ab}\,,
\end{equation}
we can read off the explicit expressions for $Q^{\alpha}[\tilde{\Delta},\sigma]$ and $\Xi^{\beta}{}_{ab}[\sigma]$ from the first Dirac delta function in (\ref{eq:TItermBeforeIntegration}) and the second Dirac delta function in (\ref{eq:SCtermBeforeIntegration}), respectively,
{\small
\begin{eqnarray}
\nonumber
Q^\alpha[\Delta,\sigma] & = & -\frac{1}{3} \sum_{\tilde{\Delta},\varepsilon\in\sigma}\alpha^\alpha[\varepsilon]z[\tilde{\Delta},\varepsilon]W[\Delta,\tilde{\Delta}]-\frac{1}{30}\sum_{\varepsilon_1,\varepsilon_2\in\sigma}\alpha^{\beta}[\varepsilon_1]\alpha^{\gamma}[\varepsilon_2]f_{\beta\gamma}{}^{\alpha}W[\Delta,\varepsilon_1,\varepsilon_2]\,, \\
\nonumber
\Xi^\alpha{}_{ab}[\sigma] & = & -\frac{1}{4!V(\sigma)}\Big(\frac{1}{300}\sum_{\varepsilon_1,\varepsilon_2,\varepsilon_3,\Delta\in\sigma}\alpha^{\alpha}[\varepsilon_3]z[\Delta,\varepsilon_3]W[\Delta,\varepsilon_1,\varepsilon_2]\varepsilon_{1a}\varepsilon_{2b}\\
\nonumber
&&\qquad\qquad\qquad+\frac{1}{3000}\sum_{\varepsilon_1,\varepsilon_2,\varepsilon_3,\varepsilon_4\in\sigma}f_{\beta\gamma}{}^{\alpha}\alpha^{\beta}[\varepsilon_3]\alpha^{\gamma}[\varepsilon_4]\varepsilon_{1a}\varepsilon_{2b}W[\varepsilon_1,\varepsilon_2,\varepsilon_3,\varepsilon_4]\Big)\,,
\end{eqnarray}
}%
and we can rewrite them in the form:
{\small
\begin{eqnarray}
\nonumber
Q^\alpha[\Delta,\sigma] & = & -\frac{1}{3} \sum_{\tilde{\Delta}\in\sigma} W[\Delta,\tilde{\Delta}] \Big(\rmd \alpha + \alpha \wedge^{\triangleright} \alpha \Big)^{\alpha}[\tilde{\Delta}] = -\frac{1}{3} \sum_{\tilde{\Delta}\in\sigma} W[\Delta,\tilde{\Delta}] F^{\alpha}[\tilde{\Delta}]\,, \\
\nonumber
\Xi^\alpha{}_{ab}[\sigma] & = & -\frac{1}{4! V(\sigma)} \frac{1}{30}\sum_{\tilde{\Delta},\Delta\in\sigma} W[\Delta,\tilde{\Delta}] \Big(\rmd \alpha + \alpha \wedge^{\triangleright} \alpha \Big)^{\alpha}[\tilde{\Delta}] \Delta_{ab} = -\frac{1}{4!V(\sigma)} \frac{1}{30}\sum_{\tilde{\Delta},\Delta\in\sigma} W[\Delta,\tilde{\Delta}] F^{\alpha}[\tilde{\Delta}] \Delta_{ab}\,.
\end{eqnarray}
}%
Here we can recognize that both depend on the Yang-Mills field strength $2$-form $F^\alpha$.
%
%
Substituting them into (\ref{eq:FirstExponent}), we obtain:
\begin{equation} \label{eq:DiscretizedYangMillsComplicated}
  iS_\text{YM} = -\frac{iC_{\alpha\beta}}{1800}\, \sum_\sigma \sum_{\tilde{\Delta},\Delta, \Delta_1, \Delta_2,\Delta_3\in\sigma} W[\Delta,\tilde{\Delta}]\,  W[\Delta,\Delta_1] W[\Delta_2,\Delta_3] F^\alpha[\Delta_1] F^\beta[\Delta_3] \frac{\tilde{\Delta}^{ab} \Delta_{2ab}}{V(\sigma)} \,.
\end{equation}
We can now compare this expression with the kinetic term for the Yang-Mills field from the ECC theory on the smooth manifold (\ref{eq:ECCaction}),
\begin{equation} \label{eq:SmoothYangMills}
 -i\, C_{\alpha\beta} \ds \int  F^{\alpha}\wedge\dual F^{\beta}\,.
\end{equation}
The general structure is analogous --- both expressions feature the coupling constant factor $-iC_{\alpha\beta}$, the integral over the manifold corresponds to the sum over all $4$-simplices, and the field strength $F^\alpha$ features quadratically. However, (\ref{eq:DiscretizedYangMillsComplicated}) is defined on a piecewise-flat manifold instead of a smooth one, and therefore its details depend a lot on the combinatorial structure of the triangulation. This exemplifies the advantage of the $3BF$ formulation of the theory over the ECC formulation, that was discussed in Section \ref{secII}. Namely, the method of discretizing the $3BF$ action was completely straightforward, taking into account our two main formulas (\ref{eq:DiscretizationMainResult}) and (\ref{eq:DiscretizationSecondMainResult}). These relied on the fact that all fields in the $3BF$ action can be manifestly written as differential forms, and all derivatives that appear in the action are exterior derivatives, at most one per term. In contrast, the ECC action features the Hodge dual operator, which intrinsically depends on the inverse tetrads, which in turn cannot be manifestly written as differential forms, and in addition the ECC action features multiple exterior derivatives in the same term (\ref{eq:SmoothYangMills}). These properties render the ECC action almost impossible to discretize without $3BF$ reformulation as an intermediary. This can be easily seen from the fact that, starting from (\ref{eq:SmoothYangMills}) and given some triangulation, it is far from obvious that its triangulated form should look like (\ref{eq:DiscretizedYangMillsComplicated}). And yet, despite its complicated appearance, (\ref{eq:DiscretizedYangMillsComplicated}) is uniquely determined, according to our formulas (\ref{eq:DiscretizationMainResult}) and (\ref{eq:DiscretizationSecondMainResult}).

Moreover, it is also far from obvious, looking at (\ref{eq:SmoothYangMills}), how is one supposed to implement the Hodge dual operator on a triangulation. But if we apply (\ref{eq:DiscretizationMainResult}), in particular the integral identity (\ref{eq:BetaBetaIntegral}), we can discretize the smooth action (\ref{eq:SmoothYangMills}) into the form
\begin{equation} \label{eq:YMsmoothCorrespondence}
-iC_{\alpha\beta} \ds \int  F^{\alpha}\wedge\dual F^{\beta} = -iC_{\alpha\beta} \frac{1}{30}\sum_\sigma\sum_{\Delta_1,\Delta_2\in\sigma}F^\alpha[\Delta_1] \dual F^\beta[\Delta_2,\sigma]\,W[\Delta_1,\Delta_2]\,.
\end{equation}
This can now be compared with (\ref{eq:DiscretizedYangMillsComplicated}) to read off the definition for the discretized Hodge dual of a $2$-form as:
\begin{equation} \label{eq:HodgeDualTwoFormDisc}
\dual F^\beta[\Delta,\sigma]  = \frac{1}{2} \sum_{\tilde{\Delta}\in\sigma} \Bigg( \frac{1}{30} \sum_{\Delta_1, \Delta_2\in\sigma} W[\Delta,\Delta_1] \frac{\Delta_1^{ab} \Delta_{2ab}}{V(\sigma)} W[\Delta_2,\tilde{\Delta}] \Bigg) F^\beta[\tilde{\Delta}] \,.
\end{equation}
This definition is completely nonobvious, and moreover it is not even clear whether it satisfies the Hodge dual identity $\dual\dual \equiv -1$. Nevertheless, it is uniquely specified from the point of view of $3BF$ theory.

In other words, we have obtained that the first exponent (\ref{eq:FirstExponent}) can be written in the form
\begin{equation}
iS_{YM} = -iC_{\alpha\beta} \frac{1}{30}\sum_\sigma\sum_{\Delta_1,\Delta_2\in\sigma}F^\alpha[\Delta_1] \dual F^\beta[\Delta_2,\sigma]\,W[\Delta_1,\Delta_2]\,,
\end{equation}
which indeed corresponds to the discretization of the kinetic term for the Yang-Mills fields from the ECC theory, via (\ref{eq:YMsmoothCorrespondence}), where the discretization prescription for the Hodge dual operator acting on a $2$-form is given via (\ref{eq:HodgeDualTwoFormDisc}).

\subsection{Regge action term}

Let us now turn to the second exponent in the $\rm INT[\dots ]$ term, and demonstrate that it corresponds to the kinetic term for gravity. Namely, if we start from
\begin{equation} \label{eq:iSgr}
  iS_{\rm GR} \equiv \frac{i}{80\pi l_p^2}\sum_\sigma \sum_{\Delta,\tilde{\Delta}\in\sigma}\varepsilon_{[ab]cd}Q^{[ab]}[\tilde{\Delta},\sigma]W[\Delta,\tilde{\Delta}]\Delta^{cd}\,,
\end{equation}
and introduce a new notation
\begin{equation} \label{eq:DefTildeQ}
\tilde{Q}_{cd}[\Delta,\sigma] \equiv \frac{1}{10} \lc_{[ab]cd} \sum_{\tilde{\Delta}\in \sigma} Q^{[ab]}[\tilde{\Delta},\sigma] \, W[\tilde{\Delta},\Delta]\,,
\end{equation}
we obtain:
\begin{equation}  
iS_{\rm GR} =  \frac{i}{8\pi l_p^2}\sum_{\sigma}\sum_{\Delta\in\sigma}\tilde{Q}_{cd}[\Delta,\sigma]\Delta^{cd}=\frac{i}{8\pi l_p^2}\sum_{\Delta}\sum_{\sigma\ni\Delta}\frac{\tilde{Q}_{cd}[\Delta,\sigma]\Delta^{cd}}{A(\Delta)}A(\Delta) = \frac{i}{8\pi l_p^2}\sum_{\Delta} \delta_\Delta A(\Delta) \equiv i S_{\rm Regge}\,.
\end{equation}
Here we have made an identification
\begin{equation} \label{eq:DeficitAngleDef}
  \delta_\Delta =\sum_{\sigma\ni\Delta}\frac{\tilde{Q}_{cd}[\Delta,\sigma]\Delta^{cd}}{A(\Delta)} \equiv \sum_{\sigma\ni\Delta} \theta_{\Delta,\sigma}\,,
  \qquad \theta_{\Delta,\sigma} \equiv \frac{\tilde{Q}_{cd}[\Delta,\sigma]\Delta^{cd}}{A(\Delta)} \,,
\end{equation}
and interpreted $\delta_\Delta$ as the deficit angle around the triangle $\Delta$. This interpretation means that the second exponent in the $\rm INT[\dots ]$ term is in fact the Regge action for general relativity, and that $\theta_{\Delta,\sigma}$ can be interpreted as a dihedral angle at the triangle $\Delta$ in the $4$-simplex $\sigma$ (see also the next Section for further analysis of $\theta_{\Delta,\sigma}$). A detailed proof that (\ref{eq:DeficitAngleDef}) is indeed the deficit angle is quite involved and therefore out of the scope of this paper. Nevertheless, we can recognize that it has all the hallmark properties that a deficit angle ought to have. Specifically:
\begin{itemize}
\item $\delta_\Delta$ depends only on one triangle $\Delta$, and is expressed as a sum over all $4$-simplices neighboring that triangle of the quantities $\theta_{\Delta,\sigma}$.
\item $\delta_\Delta$ does not depend on the change of the overall scale of the triangulation. Namely, given a change of scale by a factor $\lambda \in \realni^+$, implemented via $\varepsilon \to \lambda \varepsilon$ for all edges $\varepsilon \in T(\cM)$, we have that $A(\Delta) \to \lambda^2 A(\Delta)$ and $\Delta^{cd} \to \lambda^2 \Delta^{cd}$, while $\tilde{Q}_{cd} \to \tilde{Q}_{cd}$, rendering the whole expression (\ref{eq:DeficitAngleDef}) independent of $\lambda$.
\item $\delta_\Delta$ is proportional to the Riemann curvature tensor. Namely, from the second Dirac delta function in (\ref{eq:TItermBeforeIntegration}) we have that
\begin{equation}\label{tildaq}
Q^{[ab]}[\Delta,\sigma]=\frac{1}{3}\sum_{\varepsilon,\tilde{\Delta}\in \sigma}\omega^{[ab]}[\varepsilon]z[\tilde{\Delta},\varepsilon]W[\tilde{\Delta},\Delta]+\frac{1}{30}\sum_{\varepsilon_1,\varepsilon_2\in\sigma}\omega^{[a|c} [\varepsilon_1] \omega_c{}^{|b]}[\varepsilon_2]W[\Delta,\varepsilon_1,\varepsilon_2]\,,
\end{equation}
which can be rewritten in the form
\begin{equation}
Q^{[ab]}[\Delta,\sigma]=\frac{1}{3}\sum_{\tilde{\Delta}\in \sigma} W[\tilde{\Delta},\Delta] \Big( \rmd \omega^{ab} + \omega^{ac} \wedge \omega_c{}^b \Big)[\tilde{\Delta}] =
\frac{1}{3}\sum_{\tilde{\Delta}\in \sigma} W[\tilde{\Delta},\Delta] R^{ab}[\tilde{\Delta}]\,,
\end{equation}
where we can see that $Q^{[ab]}[\Delta,\sigma]$ is proportional to the curvature $2$-form. Since $\tilde{Q}_{cd}[\Delta,\sigma]$ is proportional to $Q^{[ab]}[\Delta,\sigma]$ according to (\ref{eq:DefTildeQ}), we see that $\delta_\Delta$ is proportional to curvature.
\end{itemize}
Therefore, these three properties provide some justification that (\ref{eq:DeficitAngleDef}) is indeed the deficit angle, despite omitting the formal proof.

\subsection{Fermion kinetic term}

Postponing the analysis of the third exponent in (\ref{eq:INTtermBeforeIntegration}) to the next Section, let us focus on the fourth and fifth exponents, taken together:
\begin{equation} \label{eq:FourthFifthExponents}
iS_F \equiv 
\frac{1}{6}\sum_\sigma \varepsilon_{abcd} \frac{1}{25}\sum_{v,\tau\in\sigma} \tau^{abc} \Big[ \bar{Q}_A[\tau,\sigma]\Big(\gamma^d\psi[v]\Big)^A + \bar{\psi}_A[v]\Big(\gamma^d Q[\tau,\sigma]\Big)^A \Big] \,.
\end{equation}
The final two Dirac delta functions in (\ref{eq:TItermBeforeIntegration}) give us:
\begin{eqnarray}
\nonumber
\bar{Q}_A[\tau,\sigma] & = & - \sum_{v\in\sigma}\bar{\psi}_B[v]\Big(\delta^B_Az[\sigma,\tau]+\frac{1}{20}\sum_{\varepsilon\in\sigma}W[\varepsilon,\tau]\Big(\alpha^{\alpha}[\varepsilon]\triangleright_{\alpha A}{}^B+\omega^{[ab]}[\varepsilon]\triangleright_{[ab]A}{}^B\Big)\Big),\\
Q^A[\tau,\sigma] & = & \Big(\delta^A_Bz[\sigma,\tau]+\frac{1}{20}\sum_{\varepsilon\in\sigma}W[\varepsilon,\tau]\Big(\alpha^{\alpha}[\varepsilon]\triangleright_{\alpha B}{}^A+\omega^{[ab]}[\varepsilon]\triangleright_{[ab]B}{}^A\Big)\Big)\sum_{v\in\sigma}\psi^B[v]\,. \label{eq:FermionicQterms}
\end{eqnarray}
Note that $\bar{Q}_A$ has the structure which encodes the operator $\nablar = \rmdr + \alpha \wedge^{\triangleright} + \omega \wedge^{\triangleright}$, while $\bar{Q}_A$ encodes $\nablal = \rmdl + \alpha \wedge^{\triangleright} + \omega \wedge^{\triangleright}$. When substituted into (\ref{eq:FourthFifthExponents}), one obtains the discretized version of the kinetic term for fermion fields.  In the ECC action (\ref{eq:ECCaction}) this term reads
\begin{equation}
\frac{1}{6}\int \varepsilon_{abcd}\,e^a\wedge e^b\wedge e^c\wedge\left[ \bar{\psi}_A(\nablal \gamma^d\psi)^A-\bar{\psi}_A(\gamma^{d} \nablar\psi )^A\right] \,,
\end{equation}
and the correspondence with the discretized version is straightforward.

\subsection{Interaction terms}

Finally, let us briefly examine the last exponent of the $\rm INT[\dots]$ term (\ref{eq:INTtermBeforeIntegration}). It reads:
\begin{eqnarray}
\nonumber
iS_{INT}& = & -\frac{i}{12}\sum_\sigma \frac{1}{3000}\sum_{\varepsilon_1,\varepsilon_2,\varepsilon_3,\varepsilon_4\in\sigma}\varepsilon_{abcd}\varepsilon_1^a\varepsilon_2^b\varepsilon_3^c\varepsilon_4^d\Big(\frac{\Lambda}{8\pi l_p^2}+Y_{ABC}\frac{1}{5^3}\sum_{v_1,v_2,v_3\in\sigma}\bar{\psi}^A[v_1]\psi^B[v_2]\phi^C[v_3]\\
&& +\chi v^4-2\chi v^2\frac{1}{5^2}\sum_{v_1,v_2\in\sigma}\phi_A[v_1]\phi^A[v_2]+\chi\frac{1}{5^4}\sum_{v_1,v_2,v_3,v_4\in\sigma}\phi_A[v_1]\phi^A[v_2]\phi_B[v_3]\phi^B[v_4]\Big)\,,
\end{eqnarray}
and it can be immediately recognized to be the discretization of the corresponding term from the ECC action (\ref{eq:ECCaction}),
\begin{equation}
-\frac{i}{12} \int \varepsilon_{abcd} \, e^a\wedge e^b\wedge e^c\wedge e^d \left[\frac{\Lambda}{8\pi l_p^2}+Y_{ABC}\bar{\psi}^A\psi^B\phi^C+\chi\left(\phi^A\phi_A-v^2\right)^2\right]\,.
\end{equation}
Therefore, it is straightforward to note that the final exponent corresponds to the cosmological constant, Higgs interaction terms and Yukawa couplings between the scalars and fermions.

\subsection{Revisited general structure of the model}

Collecting all results about the structure of the model obtained so far, we can observe that it still has the structure of the form (\ref{oblikF}), where the three terms have now been reduced to the following. The $\rm SC[\dots]$ and $\rm TI[\dots]$ terms have the form (\ref{eq:RemainingDiracDeltaFunctions}), each featuring one remaining Dirac delta function, while the $\rm INT[\dots]$ term (\ref{eq:INTtermBeforeIntegration}) can now be written as:
\begin{equation} \label{eq:ReducedINTterm}
\mathrm{INT}\Big[\lambda[\phi_k]-G[\phi_k]\Big]= \exp \Big(iS_{YM} + iS_\text{Regge}-\frac{i}{5}\sum_\sigma\sum_{\tau\in\sigma}\tilde{Q}^A[\tau,\sigma]\Theta_{abcA}[\sigma]\tau^{abc} + iS_F + iS_{INT}\Big)\,.
\end{equation}
In this form, it becomes very comparable to the ECC action (\ref{eq:ECCaction}), and we can observe that the terms in the action match one to one, with the exception of the unresolved third exponent that should correspond to the kinetic term for the scalar fields, and with the exception of the missing spin-spin coupling term. These two exceptions should be resolved by performing the two remaining integrations over the fields $\Theta_{abcA}[\sigma]$ and $\omega^{[ab]}[\varepsilon]$ to eliminate the remaining Dirac delta functions (\ref{eq:RemainingDiracDeltaFunctions}), which would lead to the precise form of the discretized version of the ECC action. Unfortunately, the integrations over $\Theta_{abcA}[\sigma]$ and $\omega^{[ab]}[\varepsilon]$ are impossible to perform in closed form, so we will discuss them within a certain approximation scheme, namely within the context of the semiclassical limit (see the next Section).

To sum up the results of this Section, the model (\ref{eq:ThirdMainResult}) provides a full definition for the quantization of the constrained $3BF$ theory, corresponding to Einstein-Cartan gravity coupled to the Standard Model of elementary particle physics. It allows one to explicitly implement numerical algorithms that can evaluate expectation values of observables in the model. Moreover, the general structure (\ref{oblikF}) of the model provides additional insight and under some very general assumptions about the underlying triangulation, the model can be significantly simplified by performing explicit integrations over a number of nondynamical fields, drawing it closer to the discretization of the ECC action itself. Along the way, a natural definition of a discretized Hodge dual operator has been introduced, and the deficit angle and Regge action have been identified, coinciding with their definitions from the ordinary Regge calculus.

At this point, one of the main remaining questions is whether our model has the correct semiclassical limit. A preliminary analysis of this question is the topic of the next Section.

\section{\label{secVI}Semiclassical limit}

A semiclassical limit represents an effective classical theory, in an approximation where all quantum effects can be neglected. It is typically studied by evaluating the effective action $\itGamma[\phi]$ of the theory and discussing its asymptotic behavior in the regime where $\itGamma[\phi] \gg \hbar $, where $\hbar$ is the Planck constant, representing the unit of action.
One of the main requirements that every proposed model of quantum gravity must satisfy is to have the correct semiclassical limit, i.e., to reproduce the Einstein's theory of general relativity, and correct dynamics for any possible additional matter fields. In this sense, it is important to study the semiclassical limit for our model (\ref{eq:ThirdMainResult}).

However, the full analysis of the semiclassical limit and the study of the effective action for the theory in the above regime represents a research topic of its own, and is out of the scope of this paper. What we will do instead is to study an approximation scheme for the path integral (\ref{eq:ThirdMainResult}) that will lead us to a simplified path integral, which has already been studied in the literature in the context of the semiclassical limit. In this sense, the analysis of the semiclassical limit of our model presented below can be considered as preliminary, though still providing some convincing evidence that the model can be expected to have correct classical behavior.

Our approximation scheme represents a coarse-graining method applied to our model, based on the following simultaneous double limit:
\begin{equation}\label{dlim}
\lambda_{\phi_k}\gg l_\varepsilon \vargtrsim l_p\,.
\end{equation}
Here, $\lambda_{\phi_k}$ represents an observable scale of wavelengths for all fields $\phi_k$ in the model, $l_\varepsilon$ represents the length scale of the triangulation, while $l_p$ is the Planck length. In other words, the fields vary in magnitude on the scale which is much larger than the triangulation scale, which is itself either larger, or of the same order as the fundamental Planck scale. The fact that the fields are slowly varying allows one to perform some averaging operations within the path integral, described in detail below.

An additional advantage of slowly varying fields is that one can also discuss the smooth limit, in which the path integral defined on a triangulation is substituted with a path integral over a smooth manifold. We will not perform this final smooth limit approximation, since it renders the path integral measure ill-defined. Instead, the smooth limit should be applied only to the effective action, which does not feature a path integral measure, and whose analysis is out of the scope of our paper.

The first averaging operation we will implement is the following:
\begin{equation} \label{eq:DeltaSumToProductSecond}
\delta\Big( \sum_{k\in \sigma } F(k) \Big) \approx \prod_{k\in\sigma} \delta \big(F(k)\big)\,,
\end{equation}
where $k\in\{v,\varepsilon,\Delta,\tau \}$. Here we assume that the quantity $F(k)$ varies very slowly from one $k$-simplex to the next, in accordance with our approximation (\ref{dlim}). Then, the statement of the Dirac delta function on the left hand side is that the sum of quantities $F(k)$ within a $4$-simplex $\sigma$ should be zero, if the corresponding delta function is to give a nonzero contribution. However, since $F(k)$ is very slowly varying, one can approximate it as a constant over the set of all $k$-simplices in $\sigma$, and this constant has an average value of zero. Thus, the left-hand side can then be approximated with the right-hand side, establishing (\ref{eq:DeltaSumToProductSecond}) as an averaging operation.

The second averaging operation is related to the spin connection field $\omega^{[ab]}[\varepsilon]$. According to our approximation, the spin connection also varies slowly, so we can substitute it with its average over each $4$-simplex. Denoting this average as $\Omega^{[ab]}[\sigma]$, we then have that within a given simplex $\omega^{[ab]}[\varepsilon] \approx \Omega^{[ab]}[\sigma]$ (for $\varepsilon\in\sigma$). However, for practical calculations given below, it is more convenient to introduce four different averages $\Omega^{[ab]c}[\sigma]$ instead of one ($c\in\{0,1,2,3\}$), and scale them with four edge vectors $\varepsilon_c$, so that the averaging operation is defined as
\begin{equation} \label{eq:SmallBigOmega}
\omega^{[ab]}[\varepsilon]=\Omega^{[ab]c}[\sigma]\varepsilon_c\,.
\end{equation}
The path integral measure corresponding to this change of variables then changes according to the following rule:
\begin{equation} \label{eq:Jacobian}
\prod_{\varepsilon\in T(\cM)}\prod_{[ab]}d\omega^{[ab]}[\varepsilon] \quad \leftrightarrow \quad \mathcal{N} \prod_{\sigma\in T(\cM)}\big|V(\sigma)\big|^{|[ab]|}\prod_{[ab]}\prod_{c=0}^3 d\Omega[\sigma]^{[ab]c}\,.
\end{equation}
Note that the total number of new fields $\Omega^{[ab]c}[\sigma]$ is $N_\Omega = |\sigma||[ab]||c|$, while the total number of original fields $\omega^{[ab]}[\varepsilon]$ is $N_\omega=|\varepsilon||[ab]|$. Depending on the combinatorial structure of the triangulation, if $N_\omega \geqslant N_\Omega$, we can substitute a subset of original fields $\omega$ with the new fields $\Omega$, while the remainder of the measure term for $\omega$ remains in the path integral. On the other hand, if $N_\omega < N_\Omega$, we can substitute all original fields $\omega$ with the new fields $\Omega$, and redefine the path integral measure by adding the remainder of the fields $\Omega$ to it. This redefinition will change the path integral only by an overall scale factor, since the integrand under the path integral does not depend on these additional fields~$\Omega$. Therefore, within the scope of our approximation scheme, we can modify the path integral measure freely by passing between left and right hand sides of (\ref{eq:Jacobian}).

After introducing both averaging operations, we can proceed to integrate the two remaining Dirac delta functions~(\ref{eq:RemainingDiracDeltaFunctions}). Introducing another auxiliary change of variables
\begin{equation}
  A_{abc}[\sigma]=\frac{1}{2}\big(\Omega_{abc}[\sigma]-\Omega_{acb}[\sigma]\big)\,, \qquad
\Omega_{abc}[\sigma] = A_{abc}[\sigma]-A_{bac}[\sigma]-A_{cab}[\sigma]\,,
\end{equation}
we can apply (\ref{eq:DeltaSumToProductSecond}) and (\ref{eq:SmallBigOmega}) to approximate (\ref{eq:RemainingDiracDeltaFunctions}) into the following form:
{\small
\begin{eqnarray}
\nonumber
\mathrm{SC}\Big[\phi_k\Big] &=&  \prod_{\sigma}
\frac{1}{\big|V(\sigma)\big|^{|A|(|a|-1)|[ab]|}}\delta\Big(\Theta_{abcA}[\sigma]+\frac{1}{6V(\sigma)}\frac{1}{3000}\sum_{v,\varepsilon_1,\varepsilon_2,\varepsilon_3,\varepsilon_4\in\sigma}W[\varepsilon_1,\varepsilon_2,\varepsilon_3,\varepsilon_4]\Big(\vphantom{\frac{1}{5}}\phi_A[v]z[\varepsilon_1,v]\\
&&\qquad\qquad\qquad+\frac{1}{5}\phi^B[v]\alpha^\alpha[\varepsilon_1]\triangleright_{\alpha BA}\Big)\varepsilon_{2c}\varepsilon_{3a}\varepsilon_{4b}\Big)\,, \label{eq:LastSCterm}
\end{eqnarray}
}%
{\small
\begin{eqnarray}
\nonumber
&& {\rm TI}\Big[\mathcal{F}[\phi_k]-\lambda[\phi_k]\Big]= \\
\nonumber
&& \qquad  \prod_\sigma \frac{1}{\big|V(\sigma)\big|^{|[ab]|(\frac{|a|}{2}-1)}}\delta\Big(A_{abc}[\sigma]-\frac{1}{20V(\sigma)}\sum_{\Delta,\varepsilon,\tau\in\sigma}\varepsilon_a z[\tau,\Delta]W[\varepsilon,\tau]\Delta^{ef}\varepsilon_{efbc}-\frac{2\pi i l_p^2}{25}\sum_{v_1,v_2\in\sigma}\bar{\psi}_A[v_1](\gamma_5\gamma^d\psi)^A[v_2]\varepsilon_{dbca}\Big)\,.\\ \label{eq:LastTIterm}
\end{eqnarray}
}%
Note that now the number of Dirac delta functions precisely matches the number of fields $\Theta_{abcA}[\sigma]$ and $A_{abc}[\sigma]$, and the integration can be performed in a straightforward manner. The first integration gives
\begin{equation}
\Theta_{abcA}[\sigma]=-\frac{1}{6V(\sigma)}\frac{1}{20}\sum_{\varepsilon,\tau\in\sigma}W[\varepsilon,\tau]\big( \rmd \phi + \alpha \wedge^\triangleright \phi \big)_A[\varepsilon]\, \tau_{abc} = 
-\frac{1}{6V(\sigma)}\frac{1}{20}\sum_{\varepsilon,\tau\in\sigma}W[\varepsilon,\tau] \, \nabla\phi_A[\varepsilon] \, \tau_{abc} \,.
\end{equation}
Together with the expression for $\tilde{Q}^B[\tau,\sigma]$ obtained by integrating the corresponding Dirac delta function in (\ref{eq:TItermBeforeIntegration}),
\begin{equation} \label{eq:TildeQsolved}
\tilde{Q}^B[\tau,\sigma] = - \sum_{v\in\sigma}\phi^A[v]\Big(\delta^B_Az[\sigma,\tau]+\frac{1}{20}\sum_{\varepsilon\in\sigma}W[\varepsilon,\tau]\alpha^{\alpha}[\varepsilon]\triangleright_{\alpha}{}^B{}_A\Big)
\end{equation}
the result for $\Theta_{abcA}[\sigma]$ can be substituted into the action corresponding to the third exponent of the $\rm INT[\dots ]$ term (\ref{eq:INTtermBeforeIntegration}):
\begin{equation} \label{eq:ThirdExponent}
iS_{SF} = -\frac{i}{5}\sum_\sigma\sum_{\tau\in\sigma}\tilde{Q}^A[\tau,\sigma]\, \Theta_{abcA}[\sigma]\,\tau^{abc}.
\end{equation}
The substitution gives
\begin{equation} \label{eq:ThirdExponent2}
  iS_{SF} = i\sum_\sigma \sum_{\tau\in\sigma}\frac{1}{5} \sum_{v\in\sigma}\phi^A[v]\Big(\delta^B_Az[\sigma,\tau]+\frac{1}{20}\sum_{\varepsilon\in\sigma}W[\varepsilon,\tau]\alpha^{\alpha}[\varepsilon]\triangleright_{\alpha}{}^B{}_A\Big)
\frac{1}{20}  \sum_{\tilde{\varepsilon},\tilde{\tau}\in\sigma} \frac{\tau^{abc}\tilde{\tau}_{abc}}{3!V(\sigma)}W[\tilde{\tau},\tilde{\varepsilon}] \, \nabla\phi_B[\tilde{\varepsilon}] \, .
\end{equation}
The term in the parentheses can be recognized to have the form $\nabla \equiv \rmd + \alpha \wedge^\triangleright$. Moreover, we can observe that this action corresponds to the kinetic term for the scalar field from the ECC theory, in the form
\begin{equation} \label{eq:ECCactionKineticScalar}
 - i \int \left(\nabla\phi\right)^A\wedge \left(\dual\nabla\phi\right)_A  = i\int\phi^A \wedge \big(\nabla \dual \nabla \phi\big)_A + (\text{boundary term})\,.
\end{equation}
Comparing the right hand side of the kinetic term with (\ref{eq:ThirdExponent2}), we can see that the Hodge dual of the $1$-form $\nabla \phi_A$ can be defined on a triangulation as:
\begin{equation} 
\dual \nabla \phi _A [\tau,\sigma] = \frac{1}{20}  \sum_{\tilde{\varepsilon},\tilde{\tau}\in\sigma} \frac{\tau^{abc}\tilde{\tau}_{abc}}{3!V(\sigma)}W[\tilde{\tau},\tilde{\varepsilon}] \, \nabla\phi_A[\tilde{\varepsilon}] \, .
\end{equation}
Given that, the third exponent can be written as
\begin{equation} \label{eq:ThirdExponent3}
  iS_{SF} = i\sum_\sigma \frac{1}{5} \sum_{v\in\sigma}\phi^A[v] \big(\nabla \dual \nabla\phi\big)_A[\sigma] \,.
\end{equation}
We can now see that (\ref{eq:ThirdExponent3}) represents the discretization of the kinetic term (\ref{eq:ECCactionKineticScalar}) of the ECC action (\ref{eq:ECCaction}).

The second integration gives us the field $A_{abc}[\sigma]$ as a sum of two terms. The first term is a function of the edge vectors of $\sigma$, while the second term is a function of the fermion fields. It is convenient to discuss the two contributions separately. To that end, let us write $A_{abc}[\sigma] = \tilde{A}_{abc}[\sigma] + \hat{A}_{abc}[\sigma]$, where we introduce the following notation:
\begin{equation} \label{eq:TildeAndHatConnections}
\tilde{A}_{abc}[\sigma] \equiv \frac{1}{20V(\sigma)}\sum_{\Delta,\varepsilon,\tau\in\sigma}\varepsilon_a z[\tau,\Delta]W[\varepsilon,\tau]\Delta^{ef}\varepsilon_{efbc}\,, \qquad
\hat{A}_{abc}[\sigma] \equiv - \frac{2\pi i l_p^2}{25}\sum_{v_1,v_2\in\sigma}\bar{\psi}_A[v_1](\gamma_5\gamma^d\psi)^A[v_2]\varepsilon_{abcd}\,.
\end{equation}

Focus first on the tilde-term, depending only on geometry. Rewinding the changes of variables, the expression for the field $\tilde{A}_{abc}[\sigma]$ can be rewritten as an expression for the spin connection $\tilde{\omega}^{[ab]}[\varepsilon]$ in the following form:
\begin{equation}
\tilde{\omega}^{[ab]}[\varepsilon]=\frac{1}{V(\sigma)}\frac{1}{20}\sum_{\tilde{\varepsilon},\Delta,\tau\in\sigma}z[\tau,\Delta]W[\tilde{\varepsilon},\tau]\Delta^{ef}\varepsilon_c\tilde{\varepsilon}^q\left(\delta^a_q\varepsilon_{ef}{}^{bc}-\delta^b_q\varepsilon_{ef}{}^{ac}-\delta^c_q\varepsilon_{ef}{}^{ab}\right)\,.
\end{equation}
This can be simplified in a straightforward way to:
\begin{eqnarray}
\tilde{\omega}^{[ab]}[\varepsilon]&=&-\frac{1}{V(\sigma)}\frac{1}{10}\sum_{\tilde{\varepsilon},\Delta,\tau\in\sigma}W[\tilde{\varepsilon},\tau]z[\tau,\Delta]\Delta^{ef}\tilde{\varepsilon}^q\varepsilon_q\varepsilon_{ef}{}^{[ab]}=-\frac{3}{20}\frac{1}{V(\sigma)}\sum_{\tau\in\sigma}\tau^{efq}\varepsilon_q\varepsilon_{ef}{}^{[ab]}\,.
\end{eqnarray}
Substituting this back into equation (\ref{tildaq}) while neglecting the term quadratic in the spin connection, then into (\ref{eq:DefTildeQ}), and finally into the expression for $\theta_{\Delta,\sigma}$ in (\ref{eq:DeficitAngleDef}), we obtain:
\begin{equation}
\theta_{\Delta,\sigma} = \frac{1}{10} \frac{\lc_{[ab]cd} \Delta^{cd}}{A(\Delta)}  \sum_{\Delta_1\in \sigma} W[\Delta_1,\Delta] \frac{1}{3}\sum_{\varepsilon,\tilde{\Delta}\in
  \sigma}z[\tilde{\Delta},\varepsilon]W[\tilde{\Delta},\Delta_1]\Big(-\frac{3}{20}\Big)\frac{1}{V(\sigma)}\sum_{\tau\in\sigma}\tau^{efq}\varepsilon_q\varepsilon_{ef}{}^{[ab]}\,,
\end{equation}
which can be simplified to:
\begin{equation}
\theta_{\Delta,\sigma} = \frac{\Delta_{cd}}{A(\Delta)} \Big( \frac{1}{V(\sigma)}\sum_{\tau\in\sigma}\tau^{cde} \Big) \sum_{\varepsilon\in\sigma} \varepsilon_e \frac{1}{10}\sum_{\tilde{\Delta}\in
  \sigma}z[\tilde{\Delta},\varepsilon] \frac{1}{10} \sum_{\Delta_1\in \sigma}  W[\tilde{\Delta},\Delta_1] \,W[\Delta_1,\Delta] \,.
\end{equation}
At this point we can recognize the following properties of $\theta_{\Delta,\sigma}$:
\begin{itemize}
\item $\theta_{\Delta,\sigma}$ depends only on one triangle $\Delta$ within a given simplex $\sigma$.
\item $\theta_{\Delta,\sigma}$ does not depend on the change of the overall scale of the triangulation. Namely, the term in front of the big parentheses is scale free, since both the numerator and the denominator are proportional to the area of the same triangle. Also, the product of $\tau^{cde}$ and $\varepsilon_e$ scales as the $4$-volume, and is compensated by the term $V(\sigma)$. All other quantities are scale free by definition.
\item $\theta_{\Delta,\sigma}$ features the sum over precisely two tetrahedra. Namely, the term in the parentheses features the sum over all tetrahedra in the $4$-simplex. However, since $\tau^{cde}$ is contracted with $\Delta_{cd}$, this sum has only two terms --- in a given $4$-simplex, any triangle $\Delta$ is common for precisely two tetrahedra. These two tetrahedra are selected via the scalar product with $\Delta_{cd}$, while all other tetrahedra vanish from the sum.
\end{itemize}
These properties all correspond to the dihedral angle between two tetrahedra sharing the common triangle, within a $4$-simplex. Therefore, again despite ommiting the formal proof, these properties provide a certain level of justification to call $\theta_{\Delta,\sigma}$ the dihedral angle, in correspondence with the deficit angle defined in (\ref{eq:DeficitAngleDef}).

Next, focus on the hat-term of the connection (\ref{eq:TildeAndHatConnections}), which depends on fermions. Again rewinding the changes of variables, the expression for the field $\hat{A}_{abc}[\sigma]$ can be rewritten as an expression for the spin connection $\hat{\omega}^{[ab]}[\varepsilon]$ in the following form:
\begin{equation} \label{eq:HatOmega}
\hat{\omega}^{[ab]}[\varepsilon]= -\frac{2\pi i l_p^2}{25}\sum_{v_1,v_2\in\sigma}\bar{\psi}_A[v_1](\gamma_5\gamma^d\psi)^A[v_2]\varepsilon^{abcd} \varepsilon_c \,.
\end{equation}
We want to collect all terms containing $\hat{\omega}^{[ab]}[\varepsilon]$ within $iS_{GR}$ and $iS_F$, and then substitute the above expression into all those terms. The dependence through $iS_{GR}$ is given in (\ref{eq:iSgr}) via (\ref{tildaq}), and just like in the classical theory, the first term in (\ref{tildaq}) corresponds to $\nabla \hat{\omega}$ and transforms into a boundary term, giving no contribution. The second, quadratic term in (\ref{tildaq}) does give a nontrivial contribution. The dependence through $iS_F$ is given in (\ref{eq:FourthFifthExponents}) via (\ref{eq:FermionicQterms}), and both terms in (\ref{eq:FermionicQterms}) contribute. Straightforward substitution of (\ref{eq:HatOmega}) into all those terms gives a result which is too bulky, and we omit it. Instead, looking at the structure of the resulting expressions, we can introduce a convenient change of variables, inspired by the notion of the fermion spin $2$-form (\ref{spinskakoneksija:jna}), as follows:
\begin{equation}
s^a[\Delta,\sigma] \equiv - \frac{1}{2\pi i l_p^2} \hat{A}^{abc}[\sigma] \Delta_{bc}\,, \qquad \dual s^a[\Delta,\sigma] \equiv - \frac{1}{2\pi i l_p^2} \frac{1}{2} \lc_{bcde} \hat{A}^{abc}[\sigma] \Delta^{de}\,.
\end{equation}
Using this, the contributions from $iS_{GR}$ and $iS_F$ terms can be expressed in a much more succinct form. The contribution from $iS_{GR}$ is:
\begin{equation}
iS_{GR}(\hat{\omega}) = i \frac{\pi l_p^2}{4} \sum_{\sigma} \frac{1}{30} \sum_{\Delta,\tilde{\Delta}\in\sigma} W[\Delta,\tilde{\Delta}] \, s^a[\Delta,\sigma] \dual s_a[\tilde{\Delta},\sigma]\,.
 \end{equation}
The contribution from $iS_F$ is identical in form as the contribution from $iS_{GR}$, albeit twice as large, $iS_F(\hat{\omega}) = 2 iS_{GR}(\hat{\omega})$. Taken together, they give rise to the following expression,
\begin{equation}
iS_{GR}(\hat{\omega}) + iS_F(\hat{\omega})  = i \frac{3 \pi l_p^2}{4} \sum_{\sigma} \frac{1}{30} \sum_{\Delta,\tilde{\Delta}\in\sigma} W[\Delta,\tilde{\Delta}] \, s^a[\Delta,\sigma] \dual s_a[\tilde{\Delta},\sigma]\,,
\end{equation}
which represents the discretization of the spin-spin interaction term from the ECC action (\ref{eq:ECCaction}),
\begin{equation}
i\frac{3\pi l_p^2}{4} \int s^{a}\wedge \dual s_{a}\,.
\end{equation}

Finally, after the integrations of all Dirac delta functions in the path integral have been performed, we can observe that the expectation value of the observable (in the given approximation) can be written as
\begin{equation} \label{eq:FinalPathInt}
\langle F \rangle=\frac{\mathcal{N}}{Z}\prod_v\int d\phi[v]d\psi[v]d\bar{\psi}[v]\int\prod_{\varepsilon}d\alpha[\varepsilon]d\varepsilon\,\prod_\sigma \big|V(\sigma)\big|^{-M}\,F(\phi_k)\,{\rm exp}\left(iS_R+iS_{ECC}^{(M)}\right)\,,
\end{equation}
where $S_R$ is the standard Regge action for general relativity, while the action $S_{ECC}^{(M)}$ is given as
\begin{equation} \label{eq:ECCMaction}
S_{ECC}^{(M)} = S_{YM} + S_{SF} + S_F(\tilde{\omega}) + S_{INT} + \big(S_{GR}(\hat{\omega}) + S_F(\hat{\omega}) \big)\,,
\end{equation}
and corresponds to the action for the Standard Model of elementary particles coupled to gravity, with the additional term describing the spin-spin contact interaction. In addition, the constant $M$ is evaluated as
\begin{equation}
M =   |[ab]| \left(|\alpha| + |A| (|a|-1) + \frac{|a|}{2} -1 \right) =  150\,,
\end{equation}
see volume terms in equations (\ref{eq:SCtermBeforeIntegration}), (\ref{eq:LastSCterm}) and (\ref{eq:LastTIterm}). The evaluation is based on the number of generators of the Lorentz group, namely $|[ab]|= 6$, the gauge group $SU(3)\times SU(2) \times U(1)$, which in total gives $|\alpha| = 12$, the spacetime translation group $\realni^4$ with $|a| = 4$, and the group $\kompleksni^4$ for the scalar fields, with $|A| = 4$. The value of $M$ coincides with the one obtained in \cite{Stipsic2025b}.

At this point we can observe that (\ref{eq:FinalPathInt}) represents precisely the discretized version of the path integral for the ECC theory, vindicating the relations (\ref{eq:RelationsBFandECC}) derived in an abstract manner in \cite{Stipsic2025b}.

We finish the analysis of the semiclassical limit by noting that the effective actions for the path integrals of type (\ref{eq:FinalPathInt}) have been studied \cite{sklm1,sklm2}, with the conclusions that they typically have the correct semiclassical behavior. Taken together with our analysis, these results support the conclusion that our $3BF$ model of quantum gravity with matter (\ref{eq:ThirdMainResult}) also has the correct semiclassical limit.

\section{\label{secVII}Conclusions}

\subsection{\label{secVIIa}Summary of the results}

Let us summarize the results of the paper. After the Introduction, in Section \ref{secII} we gave a short review of the classical theory. After reviewing the notion of a $3$-group and its corresponding topological $3BF$ action, we introduced the Standard Model $3BF$ action, as the starting point for the quantization. In addition, we gave a brief review of the ECC action, with an emphasis on the presence of the Hodge dual operator as a stumbling block for the quantization procedure. The $3BF$ action, in contrast, does not feature the Hodge dual, and therefore its fields can be expressed exclusively in terms of differential forms.

Section \ref{secIII} was devoted to the development of the general method to pass from an action defined on a smooth spacetime manifold to an action defined on a piecewise-flat manifold. After studying certain combinatorial aspects of the simplices in the simplicial complex, we derived the prescription (\ref{eq:DiscretizationMainResult}) how to put a given action onto a triangulation. The prescription relies on two key assumptions --- that all fields in the action enter exclusively as differential forms, and that each field remains constant within any given simplex of the triangulation. This enables us to write the action as a sum over $4$-simplices of algebraic contractions of fields on simplices. In particular, $0$-forms live on vertices, $1$-forms live on edges, $2$-forms live on triangles, and so on. Then we extended the prescription to include terms with derivatives, namely (\ref{eq:DiscretizationSecondMainResult}). For this purpose we utilized the Stokes theorem, and the fact that the exterior derivative appears at most once per term in the action. The prescriptions (\ref{eq:DiscretizationMainResult}) and (\ref{eq:DiscretizationSecondMainResult}) represent the first two main results of the paper, establishing the underlying method used to discretize the $3BF$ action.

In Section \ref{secIV} we discussed the discretization of the path integral. One of the main benefits of passing from a smooth manifold to a piecewise-flat manifold lies precisely in the fact that the path integral measure becomes well defined, since we pass from the product of uncountably infinitely many integrals (one per field per spacetime point) to the product of a finite number of integrals, or at most countably infinite number (one per field per simplex in the triangulation). In addition, the underlying $3$-group structure of the $3BF$ action renders each field to be valued in one of the Lie algebras corresponding to the Lie groups within the $3$-group. This is exploited to define the domain of integration for the path integral --- each particular integral in the product is an integral over a corresponding Lie algebra. As a result, the whole path integral is made up of countably many integrals, each for a specific algebra dictated by the $3$-group of the theory. For convenience, at the end we have listed explicit definitions for the Haar measures and other details for each group that appears in the Standard Model $3$-group.

Section \ref{secV} represents the application of the discretization method developed in Sections \ref{secIII} and \ref{secIV} to the Standard Model $3BF$ action (\ref{eq:RealisticAction}) from Section \ref{secII}. The result is an explicit well defined path integral (\ref{eq:ThirdMainResult}) for a model of quantum gravity coupled to matter fields of the Standard Model. It represents the third main result of the paper. In its full generality, the model is specified by (\ref{eq:ThirdMainResult}), but under certain mild additional assumptions, namely that the observable depends only on the dynamic fields (\ref{dinpolja3bf}) and that the triangulation satisfies the property $|\sigma| \leqslant |k|$, the model can be significantly simplified, by explicitly performing a number of integrations within the path integral. This leads to the ``small'' model described using the generic form (\ref{oblikF}) with the three main terms given in (\ref{eq:RemainingDiracDeltaFunctions}) and (\ref{eq:ReducedINTterm}). The set of performed integrations resembles the procedure performed in \cite{Stipsic2025b}, but this time with well defined ordinary integrals instead of relying on abstract path integral calculations. During the integration process, we have obtained the explicit discretization prescriptions for various terms in the ECC action (\ref{eq:ECCaction}), with an emphasis on the highly nontrivial form of the discretized Hodge dual terms.

Finally, in Section \ref{secVI} we gave a preliminary analysis of the semiclassical limit of the model. After defining what the limit actually entails, we introduced two corresponding averaging operations, which enabled us to perform two more sets of integrations and reach the final form of the path integral (\ref{eq:FinalPathInt}). The latter represents the actual quantization of the ECC action, along with its Hodge dual terms. Since the discretized action can be written as the sum of the Regge action for gravity and the discretized Standard Model action for matter fields, the resulting path integral falls into the class of theories for which the semiclassical limit has already been studied \cite{sklm1,sklm2}, supporting the conclusion that both the full model (\ref{eq:ThirdMainResult}) and the reduced model (\ref{eq:FinalPathInt}) have a correct classical limit.

\subsection{\label{secVIIb}Discussion and future lines of investigation}

It should be emphasized that our model (\ref{eq:ThirdMainResult}) represents one of the first models of its kind that features quantum gravity coupled to matter, in particular to the full Standard Model of elementary particle physics. For example, various spinfoam models throughout literature have been developed to quantize the gravitational interaction only, without matter fields, and are not easily extendable to include matter. In contrast, our approach is based on the ideas of higher gauge theory, and the $3$-group structure can encapsulate all known fields in nature, both gravitational and matter, on an equal footing. This advantage was crucial in the development of our model. Moreover, the $3$-group structure may be flexible enough to extend the model with yet more fields. Specifically, the degrees of freedom that ought to describe dark matter are absent from our model. This is because a well accepted model describing the dynamics of dark matter has not been developed yet, so we opted to formulate the model without it. Nevertheless, once the dynamics of dark matter becomes better understood, it is plausible that the Standard Model $3$-group (\ref{eq:StandardModel3GroupDef}) and the $3BF$ action (\ref{eq:RealisticAction}) can be extended to include it, which would then be translated into the quantum version of the theory as well.

On the other hand, a $3$-group can also be chosen in a more constrained fashion, leading to various restrictions on the structure of the theory. This would be a generalization of the idea behind grand unified theories (GUTs), which were historically studied a lot in the context of gauge groups for vector gauge bosons. A GUT-like $3$-group, for example based on the gauge group $SU(5)$, could be defined as
\begin{equation}
G = SO(3,1)\times SU(5) \,, \qquad H = \mathbb{R}^4\,, \qquad L = \mathbb{C}^4\times\mathbb{G}^{64}\times\mathbb{G}^{64}\times\mathbb{G}^{64}\,.
\end{equation}
Today, GUT models of this kind have mostly been experimentally excluded, but a $3$-group still offers alternative possibilities which have not yet been explored. Namely, instead of modifying the group $G$, one could also attempt to modify the groups $H$ and $L$, leading to novel models with possibly interesting properties. For example, a suitable modification of the group $L$ could give rise to a possible explanation of fermion families. Thus, the class of models described by higher gauge theory offers a lot of room for future exploration.

One aspect of our model (\ref{eq:ThirdMainResult}) that was not discussed in the paper is convergence. A model of quantum gravity can be considered well behaved only if it gives finite values for all physically sensible observables. The full analysis of the finiteness of the model is out of the scope of this paper and is a topic for future research. Nevertheless, we can note that (a) the postulated presence of an explicit triangulation in the theory, and (b) the requirement that all fields remain constant within each simplex in the triangulation, together provide a natural small-distance cutoff in the theory. A small-distance cutoff is still not equivalent to a full UV completion --- some fields are integrated over noncompact groups and can therefore have arbitrarily large magnitude, leading to possible divergences even at finite distances --- but it removes the UV divergences at least for the fields integrated over compact groups. On the other hand, large-distance limit and possible IR divergences would probably require a modification of the path integral measure by introducing a suitable dampening term, along the lines of the analysis given in \cite{MikovicVojinovicBook}.

Another aspect of (\ref{eq:ThirdMainResult}) is its natural accessibility to numerical investigations. While analytic solutions of the discretized path integral are hard to come by, it is completely straightforward to implement the whole expression (\ref{eq:ThirdMainResult}) onto a computer, and use various algorithms to evaluate the expectation values of observables numerically. To that end, one may introduce a transformation similar to the Wick rotation, transforming the Fresnel integrals into Gaussian integrals,
\begin{equation}
\int_\realni dx \, e^{ix^2} = \sqrt{\pi} e^{i\frac{\pi}{4}} \qquad \to \qquad \int_\realni dx \, e^{-x^2} = \sqrt{\pi}\,,
\end{equation}
in order to establish better convergence properties for numerical integration. One could also introduce a fully Euclidean theory, based on a $3$-group featuring the Euclidean group $SO(4)$ instead of the Lorentz group $SO(3,1)$. These are all standard techniques to make a model more accessible to numerical methods. The numerical techniques are especially useful in the sense that one can tackle various interesting problems that quantum gravity is expected to address, which depend on matter fields as much as on gravity itself. For example, one can study black hole creation, evaporation, and singularities. Another example would be a cosmological singularity and the low-entropy initial state of matter fields. Yet another example could entail possible violations of the strong equivalence principle, which holds classically (to the best of our knowledge), but quantum correction terms may in principle induce violations of various types. All these open problems fundamentally depend on the presence of matter fields in a tentative quantum gravity model. In this sense, our model lends itself nicely to investigations of these problems.

As a final point, it would be worthwhile to study the semiclassical limit of the model more thoroughly, using the effective action approach \cite{MikovicVojinovicBook}. It is straightforward, although technically maybe a bit tedious, to introduce the effective action equation for the path integral (\ref{eq:ThirdMainResult}), which can then be studied using both analytical and numerical methods. In particular, we expect that in the limit (\ref{dlim}) the effective action will be described, in the leading order, by the Regge action for the gravitational part, and the discretized ECC action (\ref{eq:ECCMaction}) for the matter part. Then, if the scale of observables is much larger than the triangulation scale, one can use the idea of a hydrodynamic approximation to pass from the discrete theory to a smooth theory --- triangulation is approximated by a smooth manifold, the Regge term with the Einstein-Cartan term, and the discretized ECC terms with their smooth counterparts, thus recovering the full classical ECC action on a smooth manifold as the classical limit. This is also an interesting topic for future work.

\acknowledgments


\medskip

This research was supported by the Ministry of Science, Technological Development and Innovations of the Republic of Serbia (MNTRI).

\appendix

\section{\label{AppB}Equations of motion for the Standard Model $3BF$ action}

Let us discuss the equations of motion for the action (\ref{eq:RealisticAction}). It is straightforward to solve the EoMs for all Lagrange multiplier fields, in terms of the dynamical fields and their derivatives (see for example \cite{Radenkovic2019,Stipsic2025a} for details):
\begin{equation}
\begin{array}{rclcrcl}
    M_{\alpha ab}&=& \displaystyle -\frac{1}{48}\varepsilon_{abcd}F_{\alpha}{}^{\mu\nu}e^{c}{}_{\mu}e^d{}_{\nu}\,,&\hspace*{0.5cm}&
    \zeta^{\alpha ab}&=&\ds \frac{1}{4}{C}_{\beta}{}^{\alpha}\varepsilon^{abcd}F^{\beta}{}_{\mu\nu}e_c{}^{\mu}e_d{}^{\nu}\,, \vphantom{\ds\int} \\
    \lambda_{\alpha \mu\nu}&=&\ds -F_{\alpha \mu\nu}\,,&\hspace*{0.5cm}&
    B_{\alpha \mu\nu}&=&\ds -\frac{e}{2}{C}_{\alpha}{}^{\beta}\varepsilon_{\mu\nu\rho\sigma}F_{\beta}{}^{\rho\sigma}\,, \vphantom{\ds\int} \\
    \lambda_{[ab]\mu\nu}&=&\ds R_{[ab]\mu\nu}\,,&\hspace*{0.5cm}&
    B_{[ab]\mu\nu}&=&\ds \frac{1}{8\pi l_p^2}\varepsilon_{[ab]cd}e^c{}_{\mu}e^d{}_{\nu}\,, \vphantom{\ds\int}  \\
    \tilde{\lambda}^A{}_{\mu}&=&\ds \left(\nabla_{\mu}\phi\right)^A\,,&\hspace*{0.5cm}&
    \tilde{\gamma}^A{}_{\mu\nu\rho}&=&\ds -e\varepsilon_{\mu\nu\rho\sigma}\left(\nabla^{\sigma}\phi\right)^A\,, \vphantom{\ds\int} \\
    H^{abcA}&=&\ds \frac{1}{6e}\varepsilon^{\mu\nu\rho\sigma}\left(\nabla_{\mu}\phi\right)^Ae^a{}_{\nu}e^b{}_{\rho}e^c{}_{\sigma}\,,&\hspace*{0.5cm}&
    \Lambda^{abA}{}_{\mu}&=&\ds \frac{1}{6e}g_{\mu\lambda}\varepsilon^{\lambda \nu\rho\sigma}\left(\nabla_{\nu}\phi\right)^Ae^a{}_{\rho}e^b{}_{\sigma}\,, \vphantom{\ds\int} \\
    \gamma^A{}_{\mu\nu\rho}&=&\ds -i\varepsilon_{abcd}e^a{}_{\mu}e^b{}_{\nu}e^c{}_{\rho}\left(\gamma^d\psi\right)^A\,,&\hspace*{0.5cm}&
    \bar{\gamma}_{A\mu\nu\rho}&=&\ds i\varepsilon_{abcd}e^a{}_{\mu}e^b{}_{\nu}e^c{}_{\rho}\left(\bar{\psi}\gamma^d\right)_A\,, \vphantom{\ds\int} \\
    \lambda^A{}_{\mu}&=&\ds \left(\nablar_{\mu}\psi\right)^A\,,&\hspace*{0.5cm}&
    \bar{\lambda}_{A\mu}&=&\ds \left(\bar{\psi}\nablal_{\mu}\right)_A\,, \vphantom{\ds\int} \\
    \beta^a{}_{\mu\nu}&=&\ds 0\,.  \vphantom{\ds\int} 
    \end{array}
\end{equation}

Next focus on the torsion equation. The spin connection $\omega^{[ab]}{}_\mu$ is not equivalent to the Levi-Civita connection, since fermionic fields give rise to nonzero torsion. We therefore rewrite the spin connection as Ricci rotation coefficients $\Delta^{[ab]}{}_\mu$ and contorsion tensor $K^{[ab]}{}_\mu$:
\begin{equation}  \label{spinskakoneksija}
    \omega^{[ab]}{}_\mu = \Delta^{[ab]}{}_\mu+K^{[ab]}{}_\mu\,.
\end{equation}
The Ricci rotation coefficients are given in terms of commutation coefficients,
\begin{equation}
\Delta^{ab}{}_\mu=\frac{1}{2}\left(c^{abc}-c^{bac}-c^{cab}\right)e_{c\mu}\,,
\end{equation}
which are in turn defined as
\begin{equation}
c^{abc}=e^{b\mu} e^{c\nu}\left(\partial_{\mu} e^a{}_\nu-\partial_{\nu} e^a{}_\mu \right)\,.
\end{equation}
The contorsion tensor is given in terms of the components of the torsion:
\begin{equation}\label{kontorzija:def}
K^{ab}{}_\mu = \frac{1}{2} \left( T^{cab} + T^{bac} - T^{abc} \right) e_{c\mu} \,.
\end{equation}
Here $T^{abc} \equiv T^a{}_{\mu\nu} e^{b\mu} e^{c\nu}$, where $T^a{}_{\mu\nu}$ are the components of the torsion $2$-form, defined as:
\begin{equation}
T^a \equiv \nabla e^a = \frac{1}{2} T^a{}_{\mu\nu}\, \rmd x^\mu \wedge \rmd x^\nu\,, \qquad T^a{}_{\mu\nu} \equiv \nabla_\mu e^a{}_\nu - \nabla_\nu e^a{}_\mu\,.
\end{equation}
Given all of the above quantities, one can write the EoM for torsion as:
\begin{equation} \label{eq:EoMforTorsion}
T^a =2\pi i l_p^2\, s^a\,, 
\end{equation}
where the spin $2$-form is given as
\begin{equation} \label{spinskakoneksija:jna}
s^a \equiv \varepsilon^{abcd} \, e_b\wedge e_c \, \bar{\psi}_A(\gamma_5\gamma_d\psi)^A\,.
\end{equation}
We see that the torsion 2-form is proportional to the spin 2-form. Also, using $(\ref{kontorzija:def})$, we obtain the components of the contorsion $1$-form as
\begin{equation}
K^{ab}{}_\mu =-2\pi il_p^2\bar{\psi}_A(\gamma_5\gamma_d\psi)^A\varepsilon^{abcd}e_c{}_{\mu}\,,
\end{equation}
so the relationship between contorsion and torsion simplifies:
\begin{equation}
T^a=K^{ab}\wedge e_b\,.
\end{equation}
In other words, we see that the EoM for torsion (\ref{eq:EoMforTorsion}) is solved explicitly --- torsion is expressed in terms of contorsion, which is expressed in terms of fermion current, while the spin is also expressed in terms of the same fermion current. Therefore, the EoM is satisfied automatically, provided that we write the spin connection in terms of Ricci rotation coefficients and the given contorsion tensor. In this sense, the spin connection is not a dynamical field in the theory, since it can be algebraically determined in terms of other fields, specifically the tetrad field and the fermion fields.

Next, the Einstein field equation can be written in the usual form,
\begin{eqnarray}\label{ajn}
R_{\mu\nu}-\frac{1}{2}g_{\mu\nu}R+\Lambda g_{\mu\nu}=8\pi l_p^2 \, T_{\mu\nu}\,,
\end{eqnarray}
where the stress-energy tensor is given as:
\begin{eqnarray}
\nonumber
T_{\mu\nu}&=&F^{\alpha}{}_{\mu\rho}C_{\alpha}{}^{\beta}F_{\beta\nu}{}^{\rho}-\frac{1}{4}g_{\mu\nu}F^{\alpha}{}_{\rho\sigma}C_{\alpha}{}^{\beta}F_{\beta}{}^{\rho\sigma}\\
\nonumber
&+&\nabla_{\mu}\phi^A\nabla_{\nu}\phi_A-\frac{1}{2}g_{\mu\nu}\left(\nabla_{\rho}\phi^A\nabla^{\rho}\phi_A+2\chi\left(\phi^A\phi_A-v^2\right)^2\right)\\
&+&\frac{i}{2}\left(\bar{\psi}_A ( \nablalr_{\mu}\gamma_d\psi)^A\right)e{}_{\nu}^d-\frac{1}{2}g_{\mu\nu}\left(i\left(\bar{\psi}_A (\nablalr_{\rho}\gamma^d\psi)^A\right)e_d{}^{\rho}-2Y_{ABC}\bar{\psi}^A\psi^B\phi^C\right)\,.
\end{eqnarray}
It consists of three parts, corresponding to the Yang-Mills, scalar and fermion stress-energy, respectively.

Finally, the EoMs for fermion and scalar fields are
\begin{eqnarray}
\left(i(\gamma^\mu  \nablar_{\mu})^A{}_B-Y^A{}_{BC}\phi^C\right)\psi^B=0\,,\\
\bar{\psi}_B\left( i (\nablal_{\mu}\gamma^\mu)^B{}_A+Y^B{}_{AC}\phi^C\right)=0\,,\\
\nabla_{\mu}\nabla^{\mu}\phi^A-4\chi\left(\phi^B\phi_B-v^2\right)\phi^A=0\,,
\end{eqnarray}
while the EoM for Yang-Mills fields is:
\begin{equation}
\nabla_{\mu}F_{\alpha}{}^{\mu\nu}+\frac{1}{2}{C^{-1}}_{\alpha}{}^{\beta}\left(\triangleright_{\beta A B}\left(\phi^A\nabla^{\nu}\phi^B-\phi^B\nabla^{\nu}\phi^A\right)+i\bar{\psi}_A\psi_B\left(\triangleright_{\beta C}{}^A\gamma^{\nu CB}-\gamma^{\nu AC}\triangleright_{\beta C}{}^B\right)\right)=0\,.
\end{equation}
One can observe that all these EoMs correspond precisely to the ECC action (\ref{eq:ECCaction}), demonstrating that it is classically equivalent to the Standard Model $3BF$ action (\ref{eq:RealisticAction}).

\section{\label{appC}Details of the combinatorial analysis}

Let us discuss all seven cases of simplex decompositions in detail, respectively, in order to evaluate the total number of possible choices for each case. This analysis represents the proof of the results given in the Table \ref{TabelaJedan} in the main text.
\begin{itemize}

\item[(1)] The simplest case is of dimension two --- in order to specify a triangle, one chooses two linearly independent vectors along the triangle edges. This can be done in three different ways, while the number of permutations of their order is two, leading to a total of 6 possible choices, as specified in Table \ref{TabelaJedan}.

\item[(2)] Next, we discuss the the case of dimension three. A tetrahedron can be specified by three vectors along its  edges, in two inequivalent ways. See first two diagrams in Figure \ref{SlikaJedan}.

\begin{figure}[H]
\begin{center}
  \begin{tikzpicture}[yscale=0.5]
\draw[very thick] (-1,0) -- (1,0) ;
\draw[very thick] (-1,0) -- (0,1) ;
\draw[very thin] (1,0) -- (0,1) ;
\draw[very thin] (0,1) -- (0,3) ;
\draw[very thick] (-1,0) -- (0,3) ;
\draw[very thin] (1,0) -- (0,3) ;
\end{tikzpicture}
,\ 
\begin{tikzpicture}[yscale=0.5]
\draw[very thin] (-1,0) -- (1,0) ;
\draw[very thin] (-1,0) -- (0,1) ;
\draw[very thick] (1,0) -- (0,1) ;
\draw[very thin] (0,1) -- (0,3) ;
\draw[very thick] (-1,0) -- (0,3) ;
\draw[very thick] (1,0) -- (0,3) ;
\end{tikzpicture}
,\ 
\begin{tikzpicture}[yscale=0.5]
\draw[fill=gray] (-1,0) -- (1,0) -- (0,1) -- (-1,0) ;
\draw[very thin] (-1,0) -- (1,0) ;
\draw[very thin] (-1,0) -- (0,1) ;
\draw[very thin] (1,0) -- (0,1) ;
\draw[very thin] (0,1) -- (0,3) ;
\draw[very thick] (-1,0) -- (0,3) ;
\draw[very thin] (1,0) -- (0,3) ;
\end{tikzpicture}
\end{center}
\caption{\label{SlikaJedan}Tetrahedra discussed in cases (2a), (2b), (3), respectively.}
\end{figure}

\begin{itemize}
\item[(2a)] The first diagram in Figure \ref{SlikaJedan} depicts  the choice of three vectors with a common vertex. These can be chosen in four different ways, corresponding to the choice of the common vertex. In addition, one can order the three vectors in a total of $3!=6$ permutations. This gives a total of 24 possible choices.
\item[(2b)] The second diagram in Figure \ref{SlikaJedan} depicts  the choice of three concatenated vectors. Concatenated vectors can be counted by counting the choices for their vertices. Specifically, the initial vertex can be chosen in four ways, the next one in three, and the final one in two ways, giving a total of 24 apparent choices. However, since there is a symmetry in the direction of concatenation for the vectors (i.e., one can concatenate the vectors from the initial vertex towards the final vertex, or vice versa), the actual number of choices is twice as small, i.e., a total of 12 actual choices. This has to be multiplied by $3!=6$ permutations of vectors, leading to a total of 72 choices.
\end{itemize}
Cases (2a) and (2b) together amount to $24+72=96$ possible choices, as specified in Table \ref{TabelaJedan}.

\item[(3)] In addition to the cases discussed above, a tetrahedron can also be specified by one triangle and one vector that is not coplanar to the triangle (third diagram in Figure \ref{SlikaJedan}). The triangle can be chosen in four different ways, while the vector can be chosen in three different ways, giving a total of 12 possible choices, as specified in Table~\ref{TabelaJedan}.

\item[(4)] Next we discuss the case of dimension four. A $4$-simplex can be specified by four linearly independent vectors in three different ways, depicted in the three diagrams of Figure \ref{SlikaDva}.

\begin{figure}[H]
\begin{center}
\begin{tikzpicture}[scale=1.8,yscale=1.8]
\draw[very thin] (-0.3,0) -- (0.3,0) ;
\draw[very thin] (-0.3,0) -- (0.5,0.3) ;
\draw[very thin] (-0.3,0) -- (0,0.5) ;
\draw[very thick] (-0.3,0) -- (-0.5,0.3) ;
\draw[very thin] (0.5,0.3) -- (0.3,0) ;
\draw[very thin] (0.3,0) -- (0,0.5) ;
\draw[very thick] (-0.5,0.3) -- (0.3,0) ;
\draw[very thin] (0.5,0.3) -- (0,0.5) ;
\draw[very thick] (-0.5,0.3) -- (0.5,0.3) ;
\draw[very thick] (-0.5,0.3) -- (0,0.5) ;
\end{tikzpicture}
,\ 
\begin{tikzpicture}[scale=1.8,yscale=1.8]
\draw[very thick] (-0.3,0) -- (0.3,0) ;
\draw[very thin] (-0.3,0) -- (0.5,0.3) ;
\draw[very thin] (-0.3,0) -- (0,0.5) ;
\draw[very thin] (-0.3,0) -- (-0.5,0.3) ;
\draw[very thin] (0.5,0.3) -- (0.3,0) ;
\draw[very thin] (0.3,0) -- (0,0.5) ;
\draw[very thick] (-0.5,0.3) -- (0.3,0) ;
\draw[very thin] (0.5,0.3) -- (0,0.5) ;
\draw[very thick] (-0.5,0.3) -- (0.5,0.3) ;
\draw[very thick] (-0.5,0.3) -- (0,0.5) ;
\end{tikzpicture}
,\ 
\begin{tikzpicture}[scale=1.8,yscale=1.8]
\draw[very thin] (-0.3,0) -- (0.3,0) ;
\draw[very thin] (-0.3,0) -- (0.5,0.3) ;
\draw[very thin] (-0.3,0) -- (0,0.5) ;
\draw[very thick] (-0.3,0) -- (-0.5,0.3) ;
\draw[very thick] (0.5,0.3) -- (0.3,0) ;
\draw[very thin] (0.3,0) -- (0,0.5) ;
\draw[very thin] (-0.5,0.3) -- (0.3,0) ;
\draw[very thick] (0.5,0.3) -- (0,0.5) ;
\draw[very thin] (-0.5,0.3) -- (0.5,0.3) ;
\draw[very thick] (-0.5,0.3) -- (0,0.5) ;
\end{tikzpicture}
\end{center}
\caption{\label{SlikaDva}$4$-simplices discussed in cases (4a), (4b), (4c), respectively.}
\end{figure}
  
\begin{itemize}
\item[(4a)] The first possible choice corresponds to four vectors sharing a common vertex. The number of possible choices is therefore equal to five, which is the number of vertices in a $4$-simplex, multiplied by $4!=24$ permutations of vectors. This gives a total of 120 possible choices.
\item[(4b)] The second choice corresponds to three vectors sharing a common vertex, while the fourth vector is concatenated. The common vertex can be chosen in five different ways, while the concatenation vertex can be chosen in four ways. In addition, the final vertex can be chosen in three remaining ways, giving a total of $60$ choices. All this is multiplied by $4!=24$ permutations of vectors, giving a total of $60 \cdot 24 = 1440$ possible choices.
\item[(4c)] The third choice corresponds to all four concatenated vectors. Similarly to the case (2b) above, the initial vertex can be chosen in five different ways, the second vertex in four ways, the third vertex in three ways, the fourth vertex in two ways, leaving only a single possibility for the final vertex. In total, there are $120$ apparent choices, which correspond to $60$ actual choices, due to the symmetry of direction of concatenation. Again, this is multiplied by $4!=24$ permutations of vectors, giving once more the total of $60 \cdot 24 = 1440$ possible choices.
\end{itemize}
Cases (4a), (4b) and (4c) together amount to $120+1440+1440 = 3000$ possible choices, as specified in Table \ref{TabelaJedan}.

\item[(5)] In addition to the four linearly independent vectors discussed above, a $4$-simplex can be specified by one triangle and two vectors not coplanar to the triangle, depicted in the three diagrams of Figure \ref{SlikaTri}.

\begin{figure}[H]
\begin{center}
\begin{tikzpicture}[scale=1.8,yscale=1.8]
\draw[fill=gray] (0,0.5) -- (-0.5,0.3) -- (-0.3,0) -- (0,0.5) ;
\draw[very thick] (-0.3,0) -- (0.3,0) ;
\draw[very thin] (-0.3,0) -- (0.5,0.3) ;
\draw[very thin] (-0.3,0) -- (0,0.5) ;
\draw[very thin] (-0.3,0) -- (-0.5,0.3) ;
\draw[very thick] (0.5,0.3) -- (0.3,0) ;
\draw[very thin] (0.3,0) -- (0,0.5) ;
\draw[very thin] (-0.5,0.3) -- (0.3,0) ;
\draw[very thin] (0.5,0.3) -- (0,0.5) ;
\draw[very thin] (-0.5,0.3) -- (0.5,0.3) ;
\draw[very thin] (-0.5,0.3) -- (0,0.5) ;
\end{tikzpicture}
,\ 
\begin{tikzpicture}[scale=1.8,yscale=1.8]
\draw[fill=gray] (0,0.5) -- (-0.5,0.3) -- (-0.3,0) -- (0,0.5) ;
\draw[very thick] (-0.3,0) -- (0.3,0) ;
\draw[very thin] (-0.3,0) -- (0.5,0.3) ;
\draw[very thin] (-0.3,0) -- (0,0.5) ;
\draw[very thin] (-0.3,0) -- (-0.5,0.3) ;
\draw[very thin] (0.5,0.3) -- (0.3,0) ;
\draw[very thin] (0.3,0) -- (0,0.5) ;
\draw[very thin] (-0.5,0.3) -- (0.3,0) ;
\draw[very thick] (0.5,0.3) -- (0,0.5) ;
\draw[very thin] (-0.5,0.3) -- (0.5,0.3) ;
\draw[very thin] (-0.5,0.3) -- (0,0.5) ;
\end{tikzpicture}
,\ 
\begin{tikzpicture}[scale=1.8,yscale=1.8]
\draw[fill=gray] (0,0.5) -- (-0.5,0.3) -- (-0.3,0) -- (0,0.5) ;
\draw[very thick] (-0.3,0) -- (0.3,0) ;
\draw[very thick] (-0.3,0) -- (0.5,0.3) ;
\draw[very thin] (-0.3,0) -- (0,0.5) ;
\draw[very thin] (-0.3,0) -- (-0.5,0.3) ;
\draw[very thin] (0.5,0.3) -- (0.3,0) ;
\draw[very thin] (0.3,0) -- (0,0.5) ;
\draw[very thin] (-0.5,0.3) -- (0.3,0) ;
\draw[very thin] (0.5,0.3) -- (0,0.5) ;
\draw[very thin] (-0.5,0.3) -- (0.5,0.3) ;
\draw[very thin] (-0.5,0.3) -- (0,0.5) ;
\end{tikzpicture}
\end{center}
\caption{\label{SlikaTri}$4$-simplices discussed in cases (5a), (5b), (5c), respectively.}
\end{figure}
 
\begin{itemize}
\item[(5a)] The first choice corresponds to a pair of concatenated vectors, one of which is attached to the triangle. First, the triangle can be chosen in ten possible ways. Then, one of the vectors is attached to the triangle via the attachment vertex, which can be chosen in three ways. This leaves two possible choices for the concatenation vertex, leaving only a single possibility for the final vertex. The two vectors can be arranged into $2!=2$ permutations, so the total amounts  to $10\cdot3\cdot 2 \cdot 2 = 120$ choices.
\item[(5b)] The second choice corresponds to a pair of non-concatenated vectors, both attached to the triangle. Again, the triangle can be chosen in ten possible ways. Then, one vector can be attached to the triangle in three ways, while the other can be attached in two remaining ways. One also multiplies with $2!=2$ permutations of the two vectors, giving the total of $10\cdot 3 \cdot 2 \cdot 2 = 120$ choices.
\item[(5c)] The third choice corresponds to the pair of vectors sharing a common vertex with a triangle. In addition to the ten possible choices for the triangle, the common vertex can be chosen in three ways, together with $2!=2$ permutations of the two vectors. This gives a total of $10\cdot 3\cdot 2 = 60$ choices.
\end{itemize}
Cases (5a), (5b) and (5c) together amount to $120+120+60 = 300$ possible choices, as specified in Table \ref{TabelaJedan}.

\item[(6)] A third way to to specify a $4$-simplex is by one tetrahedron and one vector not co-hyperplanar to the tetrahedron, see first diagram in Figure \ref{SlikaCetiri}. The tetrahedron can be chosen in five different ways, while the vector can then be chosen in four ways, which gives us a total of $5\cdot 4 = 20$ possible choices, as specified in Table \ref{TabelaJedan}.

\begin{figure}[H]
\begin{center}
\begin{tikzpicture}[scale=1.8,yscale=1.8]
\draw[fill=gray] (-0.5,0.3) -- (-0.3,0) -- (0.3,0) -- (0.5,0.3)  -- (-0.5,0.3) ;
\draw[very thin] (-0.3,0) -- (0.3,0) ;
\draw[very thin] (-0.3,0) -- (0.5,0.3) ;
\draw[very thin] (-0.3,0) -- (0,0.5) ;
\draw[very thin] (-0.3,0) -- (-0.5,0.3) ;
\draw[very thin] (0.5,0.3) -- (0.3,0) ;
\draw[very thin] (0.3,0) -- (0,0.5) ;
\draw[very thin] (-0.5,0.3) -- (0.3,0) ;
\draw[very thin] (0.5,0.3) -- (0,0.5) ;
\draw[very thin] (-0.5,0.3) -- (0.5,0.3) ;
\draw[very thick] (-0.5,0.3) -- (0,0.5) ;
\end{tikzpicture}
,\ 
\begin{tikzpicture}[scale=1.8,yscale=1.8]
\draw[fill=gray] (0,0.5) -- (-0.5,0.3) -- (-0.3,0) -- (0,0.5) ;
\draw[fill=gray] (0,0.5) -- (0.5,0.3) -- (0.3,0) -- (0,0.5) ;
\draw[very thin] (-0.3,0) -- (0.3,0) ;
\draw[very thin] (-0.3,0) -- (0.5,0.3) ;
\draw[very thin] (-0.3,0) -- (0,0.5) ;
\draw[very thin] (-0.3,0) -- (-0.5,0.3) ;
\draw[very thin] (0.5,0.3) -- (0.3,0) ;
\draw[very thin] (0.3,0) -- (0,0.5) ;
\draw[very thin] (-0.5,0.3) -- (0.3,0) ;
\draw[very thin] (0.5,0.3) -- (0,0.5) ;
\draw[very thin] (-0.5,0.3) -- (0.5,0.3) ;
\draw[very thin] (-0.5,0.3) -- (0,0.5) ;
\end{tikzpicture}
\end{center}
\caption{\label{SlikaCetiri}$4$-simplices discussed in cases (6) and (7), respectively.}
\end{figure}

\item[(7)] Finally, the fourth way to specify a $4$-simplex is by two triangles sharing a single vertex, see the second diagram in Figure \ref{SlikaCetiri}. The common vertex can be chosen in five ways, while the pair of two triangles can be chosen in three different ways, assuming they do not share any additional vertices. One should also remember to multiply this by $2!=2$ permutations of triangles, giving a total of $5\cdot 3 \cdot 2 = 30$ possible choices, as specified in Table \ref{TabelaJedan}.
\end{itemize}

\section{\label{AppA}Volumes of simplices}

Given a $k$-simplex specified by its edge vectors $\varepsilon_1, \dots, \varepsilon_k$ (where $k \in \{ 0,\dots,4 \}$) in a $4$-dimensional Minkowski spacetime with some chosen coordinate basis $\partial_\mu$, we are interested in expressing the $k$-volume of that simplex in terms of its edge vectors. Using a corresponding Cayley-Menger determinant, one can evaluate the square of this volume as a function of the edge lengths, and this can then be rewritten in terms of edge vectors, as follows:
\begin{eqnarray}
{}^{(0)}V(v)^2&=&-\frac{1}{4!}\frac{1}{(0!)^2}\varepsilon_{\mu\nu\rho\sigma}\,\varepsilon^{\mu\nu\rho\sigma}=1\,,\\
l(\varepsilon)^2&=&-\frac{1}{3!}\frac{1}{(1!)^2}\varepsilon_{\mu\nu\rho\sigma}\, \varepsilon_{\lambda}{}^{\nu\rho\sigma}\,\varepsilon_1^\mu\varepsilon^\lambda_1=\varepsilon_1^{\mu}\varepsilon_1^\nu\eta_{\mu\nu}=\varepsilon_1^2\,,\\
A(\Delta)^2&=&-\frac{1}{2!}\frac{1}{(2!)^2}\varepsilon_{\mu\nu\rho\sigma}\, \varepsilon_{\lambda\xi}{}^{\rho\sigma}\,\varepsilon_1^\mu \varepsilon_1^\lambda \, \varepsilon_2^\nu\varepsilon_2^\xi=\frac{1}{4}\left(\varepsilon_1^2\varepsilon_2^2-(\varepsilon_1^\mu\varepsilon_2^\nu\eta_{\mu\nu})^2\right)\,,\\
{}^{(3)}V(\tau)^2&=&-\frac{1}{1!}\frac{1}{(3!)^2}\varepsilon_{\mu\nu\rho\sigma}\,\varepsilon_{\lambda\xi\theta}{}^{\sigma}\,\varepsilon_1^\mu\varepsilon_1^\lambda \,\varepsilon_2^\nu\varepsilon_2^\xi \, \varepsilon_3^\rho\varepsilon_3^\theta\,,\\
{}^{(4)}V(\sigma)^2&=&-\frac{1}{0!}\frac{1}{(4!)^2}\left(\varepsilon_{\mu\nu\rho\sigma}\,\varepsilon_1^\mu\varepsilon_2^\nu\varepsilon_3^\rho\varepsilon_4^\sigma\right)^2\,.
\end{eqnarray}
Note that, since the values of $l(\varepsilon)$, $A(\Delta)$, ${}^{(3)}V(\tau)$ and ${}^{(4)}V(\sigma)$ are manifestly scalar quantities and thus independent of the choice of the basis $\partial_\mu$, one can freely switch between the components $\varepsilon^\mu$ of edge vectors expressed in this basis, and the components $\varepsilon^a$ of the same vectors expressed in locally inertial coordinates, without changing the values of the volumes.

\end{document}